\documentclass{aa}
\usepackage{graphics}
\usepackage{amssymb}

\newcommand{\pas}{.\hskip-2pt$^{\prime\prime}$}
\begin{document}


\title{A molecular-line study of clumps with embedded high-mass protostar 
candidates}

\author{J. Brand
	  \inst{1} 
	  \and R. Cesaroni
	  \inst{2}
	  \and F. Palla
	  \inst{2}
	  \and S. Molinari
	  \inst{3}
	  \thanks{\emph{Present address:} 
	  Istituto di Fisica dello Spazio Interplanetario, CNR, Via del Fosso 
	  del Cavaliere, I-00133 Rome, Italy} 
	   }

\offprints{J. Brand,\\
{\sl brand@ira.bo.cnr.it}}

\institute{
Istituto di Radioastronomia, CNR, Via Gobetti 101, I-40129 Bologna, Italy 
 \and
Osservatorio Astrofisica di Arcetri, Largo E. Fermi 5, I-50125 Florence, Italy 
 \and
 Infrared Processing and Analysis Center, California Institute of Technology,
MS 100-22, Pasadena, CA 91125, USA 
}

\date{Received /Accepted 29 January 2001}

\abstract{We present molecular line observations made with the IRAM 30-m 
telescope of the immediate surroundings of a
sample of 11 candidate high-mass protostars. These observations are part of an
effort to clarify the evolutionary status of a set of objects which we 
consider to be precursors of UC H{\sc ii} regions. 
\hfill\break\noindent
In a preceding series of papers we have studied a sample of objects, which
on the basis of their IR colours are likely to be associated with compact 
molecular
clouds. The original sample of 260 objects was divided approximately evenly
into a {\sl High} group, with IR colour indices [25--12] $\ge$0.57 and
[60--12] $\ge$1.3, and a {\sl Low} group with complementary colours. 
The FIR luminosity of the {\sl Low}
sources, their distribution in the IR colour-colour diagram, and their lower
detection rate in H$_2$O maser emission compared to the {\sl High} sources, 
led to the hypothesis that the majority of these objects represent an earlier 
stage in the evolution than the members of the
{\sl High} group, which are mostly identifyable with UC H{\sc ii} regions.
Subsequent observations led to the selection of 12 {\sl Low} sources that
have FIR luminosities indicating the presence of B2.5 to O8.5~V$_0$ stars,
are associated with dense gas and dust, have (sub-)mm continuum spectra
indicating temperatures of $\sim$30~K, and have no detectable radio
continuum emission. One of these sources has been proposed by us to be a
good candidate for the high-mass equivalent of a Class~0 object. In the
present paper we present observations of the molecular environment of 11 of
these 12 objects, with the aim to derive the physical parameters of the gas
in which they are embedded, and to find further evidence in support of our 
hypothesis that these sources are the precursors to UC H{\sc ii} regions. We 
find that the data are consistent with such an interpretation.
\hfill\break\noindent
All observed sources are associated with well-defined molecular clumps.
Masses, sizes, and other parameters depend on the tracer used, but typically
the cores have {\it average} diameters of $\sim$0.5--1~pc (with a range of
0.2 to 2.2~pc), and masses of a few tens to a few thousand solar masses.
Compared to a similar analysis of {\sl High} sources, the present sample has
molecular clumps that are more massive, larger, cooler, and less turbulent.
They also tend to have a smaller ratio of virial-to-luminous mass,
indicating they are less dynamically stable than their counterparts in which
the {\sl High} sources are embedded. The large sizes suggest these clumps
should still undergo substantial contraction (their densities are $\sim$10
times smaller than those of the {\sl High} sources). The lower
temperatures and small linewidths are also expected in objects in an earlier
evolutionary state. In various sources indications are found for outflowing
gas, though its detection is hampered by the presence of multiple emission
components in the line spectra. There are also signs of self-absorption,
especially in the spectra of $^{13}$CO and HCO$^+$.
\hfill\break\noindent
We find that the masses of the molecular clumps associated with our objects
increase with L$_{\rm fir}$ ($\rm M_{clump} \propto L_{fir}^{1.17}$), 
and that there is
a (weak) relation between the clump mass and the mass of the embedded
protostellar object $\rm M_{proto} \propto M_{clump}^{0.30}$.
\hfill\break\noindent
The large amount of observational data is necessarily presented in a
compact, reduced form. Yet we supply enough information to allow further
study. These data alone cannot prove or disprove the hypothesis that among
these objects a high-mass protostar is truly present. More observations, at
different wavelenghts and spatial resolutions are needed to provide enough
constraints on the number of possible interpretations. 
\keywords{ISM: clouds -- molecules, Radio lines: ISM}
}

\authorrunning{Brand et al.}
\titlerunning{Molecular clumps around high-mass protostars}
\maketitle

%
%

\section{Introduction \label{intro}}

In spite of the importance of massive stars for the morphology and evolution
of the Galaxy, and galaxies in general, the detailed study of the star 
formation process has
mostly been concentrated on low-mass stars (M$<$1~M$_{\odot}$). The
reasons for this are evident: they are more abundant than high-mass stars,
and the clouds in which they form are closer by, allowing a more detailed
investigation. The study of high-mass (M$\ge 10$~M$_{\odot}$) stars in their 
early evolutionary stages is furthermore complicated by the fact that they 
reach
the ZAMS while still accreting material (and hence still suffering very high
visual extinction), after which they rapidly destroy their natal environment.

\noindent
In recent years however, ever growing efforts have been devoted to the study 
of the formation of high-mass stars, i.e. early type (O--B) stars with mass 
in excess of $\sim$ 10~M$_{\odot}$. In particular, attention has gradually
shifted from the study of H{\sc ii} regions to that of ultracompact (UC)
H{\sc ii} regions, to that of the molecular clumps in which UC~H{\sc ii}
regions are embedded. Going from large, low-density structures to compact
dense cores, corresponds to approaching the very earliest stages of the
evolution of a massive star. Indeed, one of the goals of this type of
research is to identify a massive stellar object in an evolutionary phase
prior to the arrival on the main sequence, when most of its luminosity is 
derived from the release of gravitational energy. Such an
object is defined a ``protostar'', and would observationally be recognized as 
an object of Class~0 (see Andr\'e et al. \cite{andre}).
These are most likely not to be found near UC H{\sc ii}
regions, because an H{\sc ii} region rapidly alters (or even disrupts) the 
molecular cloud in which it forms. To find  the initial conditions
of star formation, one should therefore be looking for the {\it precursors}
of UC H{\sc ii} regions.

\section{The search for massive protostars \label{search}}

The observational approach followed by our group has been illustrated in a 
series
of papers (Palla et al. \cite{palla91}, \cite{palla93}; Molinari et al.
\cite{molinari96}, \cite{molinari98a}, \cite{molinari00}). A sample of 260 
objects was selected
from the IPSC, based on their FIR emission properties (for details see Palla
et al. \cite{palla91}). In particular, the IRAS colour criteria by Richards
et al. (\cite{richards}) were applied, which identify compact molecular
clouds. This sample was then split into two sub-samples, based on their
colour indices. The 125 sources with [25$-$12] $\ge$ 0.57 and [60$-$12]
$\ge$ 1.3 comply 
with the Wood \& Churchwell (\cite{wood}) criteria for identifying 
UC H{\sc ii} regions, and are called ``{\sl High}''; the remaining 135
sources are called ``{\sl Low}''. 

\noindent
Based on their IR properties and the H$_2$O maser occurrence frequency, we
suggested that the {\sl Low} sources are in an earlier evolutionary phase 
than the {\sl High} sources (Palla et al.~\cite{palla91}).
We have then
performed a series of observations with a twofold goal: to confirm that the
{\sl High} sources are indeed UC H{\sc ii} regions, and -- most importantly
-- to clarify the nature of the {\sl Low} sources. This was accomplished by
observations in many different tracers: H$_2$O masers (Medicina telescope),
NH$_3$(1,1) and (2,2) lines (Effelsberg), continuum maps in the IR (ISO
satellite), at centimeter (VLA), and (sub-) millimetre wavelengths (JCMT)
(for a flowchart illustrating this process, see Molinari et al.
\cite{molinari00}, Fig.~1). The main results of these observations can be
summarized as follows:

\smallskip\noindent
$\bullet$\ {\sl High} and {\sl Low} sources have luminosities typical of
high-mass stars, with the latter being only slightly less luminous than the
former;

\smallskip\noindent
$\bullet$\ H$_2$O masers are much more common towards {\sl High} than {\sl
Low} sources (detection rates 26\% and 9\%, respectively);

\smallskip\noindent
$\bullet$\ NH$_3$ emission is detected towards both samples (with a detection
rate of 45\% for {\sl Low} and 80\% for {\sl High} sources), although the 
temperatures derived from the ratio of the (1,1) and
(2,2) lines indicate that {\sl Low} sources are slightly colder than {\sl
High};

\smallskip\noindent
$\bullet$\ Free-free emission is detected towards 43\% of the {\sl High},
but towards only 24\% of the {\sl Low} sources observed with high angular 
resolution with the VLA (at 2 and 6~cm);

\smallskip\noindent
$\bullet$\ For 17 out of 30 (57\%) of the {\sl Low} sources mapped with the
JCMT, the (sub-) millimeter continuum emission arises from a compact
($\sim$30\arcsec) region around the IRAS source; total core masses are 
typically 10$-$200~M$_{\odot}$;

\smallskip\noindent
$\bullet$\ The continuum spectrum of the {\sl Low} sources between 0.35 and
2~mm indicates temperatures of $\sim$30~K, significantly lower than those,
measured towards ``hot cores'': small ($\sim 0.1$~pc), dense ($\sim
10^7$~cm$^{-3}$), hot ($\gtrsim$100~K), and luminous (L$\gtrsim
10^4$~L$_{\odot}$) molecular condensations, which very likely host high-mass 
objects that are already on the main sequence, but that are still too young 
to have developed an UC H{\sc ii} region;

\smallskip\noindent
The main conclusions one can draw are that {\sl High} sources are UC H{\sc ii} 
regions, and that {\sl Low} sources are associated with massive stars (as 
suggested
by their luminosities), but do {\it not} show any of the tracers typical of
the molecular cores hiding a {\it main sequence} early type star (such as
compact free-free emission, high kinetic temperature).

\smallskip\noindent
With the aim of establishing an evolutionary sequence for our sources, we 
have started a mapping survey with the NRAO 12-m at Kitt Peak.
Two phenomena often associated with protostars (of all masses) are
H$_2$O masers and molecular outflows. The association of CO outflows with UC
H{\sc ii} regions has been well established by Shepherd \& Churchwell
(\cite{shepherd}). However, surveys show that H$_2$O masers can appear 
slightly {\it
before} the development of a detectable H{\sc ii} region (Codella et al.
\cite{codella96}, \cite{codella97}), and that they are also closely 
associated with CO outflows
(Felli et al. \cite{felli}; see also Wouterloot et al. \cite{jgaw}, 
for a statistical
analysis of the correlations between FIR, H$_2$O, and CO emission). From maps
of the outflows, one can determine the
dynamical timescale, which provides a lower limit to the protostellar age.
Combining the presence or absence of H$_2$O maser- and radio continuum
emission should provide a rough measure of the age of the source.

\smallskip\noindent
Interferometric (mm-continuum, and -lines) are needed to study the
envelopes and (possible) disks of the embedded objects. Single-dish
observations are important to study the physical and kinematical properties
of the molecular cores, in which the candidate protostellar objects are
embedded. A study of the kinematic properties is especially important:
protostars derive the major part of their luminosity from accretion, and one
therefore expects to see signatures of infall towards these sources (see
e.g. Myers et al. \cite{myers}, and references therein). Although
high-resolution interferometric observations are needed for a complete
investigation, as a first step the resolution provided by a telescope like 
the IRAM 30-m will suffice to study the physical state of the sources, and to
derive a rough estimate for the size and mass of the associated molecular
material. Such observations are reported in the present paper.

\noindent
In Sect.~\ref{sobs} the observations are described; the data are presented
in Sect.~\ref{sdata}. Individual sources are commented on in
Sect.~\ref{sindiv}, and the results are summarized and compared with those 
of {\sl High} sources in Sect.~\ref{discuss}. 

\section{Observations \label{sobs}}

\subsection{IRAM}

Observations were carried out between October 10 -- 12, 1997, with the IRAM
30-m telescope at Pico Veleta (Granada, Spain). We used three SIS receivers 
simultaneously, in combination with the 100~kHz and 1~MHz resolution
filterbanks, and the autocorrelator, split into as many as five parts.
The observed molecules and
frequencies are listed in Table~\ref{freqs}, where we also indicate the
molecules that were observed simultaneously with the same receiver; this was
achieved by tuning the receiver to a frequency intermediate between those of
the two transitions. 

\noindent
Focus, calibration, and receiver alignment were checked by observations of
Venus; the alignment was within 2\arcsec.
During the observations of the program sources, pointing and
calibration were checked by observations of well-known UC H{\sc ii} regions
and continuum sources; the rms pointing accuracy was found to be 3\arcsec, 
while line intensities were reproducible within 10\%--20\%.
All line intensities in this paper are on a main beam brightness temperature 
(T$_{\rm mb}$) scale.

\noindent
Our sample of sources is listed in Table~\ref{sources}. Column~1 gives the
source number from the list of Molinari et al. (\cite{molinari96}); columns~2
and 3 give the equatorial (B1950) coordinates of the central
position of the maps ($\approx$ the position of the sub-mm peak as found by
Molinari et al. \cite{molinari00}); columns~4 and 5 give the galactic
coordinates of the IRAS source, the name of which is in column~6 ; 
column~7 lists the radial velocity of the NH$_3$(1,1) main line 
(from Molinari et al.~\cite{molinari96}). In column~8 we give the kinematic
distances (from Molinari et al.~\cite{molinari00},
\cite{molinari96}), while the luminosity, derived 
from the IRAS fluxes, and (when available) from (sub-)mm data, is given in 
column~9. In columns~10 to 12 
we present information on the presence/absence of H$_2$O maser emission
(from Palla et al. \cite{palla91}), radio continuum (Molinari et al.
\cite{molinari96}), and a mm-detected compact core (Molinari et al.
\cite{molinari00}). 

\smallskip\noindent
We started by making small maps around the sub-mm peak position in
(simultaneously) HCO$^+$, $^{13}$CO, and CS. The initial grid size was
24\arcsec, after which we zoomed in on the peak position on a 12\arcsec\ grid
size. The maps were repeated several times, at least in the inner parts, in
order to get a good signal-to-noise ratio. Total map extent was always
$<$100\arcsec\ in either direction. After having identified the peak position
in this way, C$^{18}$O, C$^{34}$S, CH$_3$C$_2$H, and CH$_3$CN were observed on 
a 3$\times$3 grid, with step size 12\arcsec, around the peak. The peak position 
itself was observed for up to 60~minutes in CH$_3$C$_2$H and 
CH$_3$CN. Observations were primarily done in total
power mode, with an offset position 1800\arcsec\ to the W; observations of
CH$_3$C$_2$H and CH$_3$CN were done in wobbler mode, where the secondary 
mirror was offset by 240\arcsec.

\subsection{KOSMA}

In April 1999, 6 sources from Table~\ref{sources} were searched for
HCO$^+$(4$-$3) emission ($\nu$=356734.253~MHz) with the 3-m KOSMA telescope at
Gornergrat. A liquid He-cooled SIS receiver was used, and the AOS backend
provided a resolution of 0.14~kms$^{-1}$. Single-pointing observations were 
made at the peak positions
previously identified at IRAM. The KOSMA beam size was $\sim$70\arcsec
$\times$78\arcsec; the beam efficiency was 0.75. Each position was observed
for 15 to 30 minutes, resulting in an rms (T$_{\rm mb}$) of 0.05$-$0.08~K. The
spectra suffer from standing waves, and are therefore of rather poor quality.

\section{Presentation of the data 
\label{sdata}}

In this section a general presentation of the available data is made, before
discussing the individual sources in more detail in Sect.~\ref{sindiv}.
Because of the large amount of data, a selection had to be made. In the
following, we try to show a representative sample of the available
observations; individual spectra and maps can be made available upon
request\footnote{Contact J. Brand}.

\subsection{Spectra at the peak positions}

The
spectra of HCO$^+$, CS, $^{13}$CO, C$^{18}$O and C$^{34}$S, taken at the peak
position, are shown in Figs.~\ref{molspec}a$-$k. The relevant parameters of
these lines are collected in Table~\ref{linepars}. In this table we give the
following information: in column~1 the molecular transition; in column~2 the
velocity resolution of the spectrum; the rms noise of the spectrum in
column~3; in columns~4 and 5 the extreme velocities V$_{\rm min}$, 
V$_{\rm max}$ where 
the line intensity drops below the 2$\sigma$ level; in columns~6, 7, and 8
the peak temperature, the velocity of the peak, and the FWHM line width,
respectively. Because of the non-Gaussian nature of many of the line
profiles, these values are {\it read off} from the spectra (as opposed to
being determined from Gaussian fits); in columns~9 and 10 the values of the
integral over the line, between V$_{\rm min}$ and V$_{\rm max}$ (column~9), and
between the velocities where T$_{\rm mb}$=0~K.

\noindent
A comparison between the (IRAM) HCO$^+$(1$-$0) and the (KOSMA)
HCO$^+$(4$-$3) emission is shown in Figs.~\ref{kospec}a, b. The IRAM
spectra were convolved to a 70\arcsec\ beam.

\noindent
The spectra of the CH$_3$C$_2$H(6$-$5) and CH$_3$CN(8$-$7) detections are
presented in Figs.~\ref{molch3c2h}a$-$d and Fig.~\ref{mol98ch3cn},
respectively. Gaussfit parameters of these lines can be found in
Tables~\ref{ch3c2hfits} and \ref{ch3cnfits}. When making the Gaussian fits,
we fixed the velocity separation of the various K-components, and
forced the widths of all K-components in a spectrum to have the same value.
In none of the sources the other transitions of CH$_3$CN were detected, nor 
the CH$_3$OH(v$_{\rm t}$=1) lines, which happen to lie close to the 
C$^{34}$S line. 

\noindent
Density-tracers, such as CS and the rare isotope C$^{34}$S, have been
detected in all sources of our sample, implying that dense cores are indeed
present. The C$^{34}$S lines vary in strength from a few tenths of a K, to
more than 1~K; only in Mol~3 was it barely detected. 

\subsection{Distribution of the integrated emission}

The distributions of the integrated intensities of the emission of
HCO$^+$(1$-$0), $^{13}$CO(2$-$1), CS(3$-$2), C$^{18}$O(2$-$1),
C$^{34}$S(3$-$2) (where strong enough) and CH$_3$C$_2$H(6$-$5) (for
Mol~98; the only source where it has been detected at more than one position) 
are shown in Figs.~\ref{intem}a$-$k. For easy reference we have shown in all
maps the location of the IRAS source (not always coinciding with the map
center), and the location of the peak position 
at which longer integrations were made. From these distributions we derived the
(beam-corrected) source sizes, and a mass estimate of the molecular cores; the
results are collected in Table~\ref{deconsize}. In this Table, column~1
gives the source name, column~2 the kinematic distance. In columns~3 to 6 we
give the core size (in arcseconds and parsec), as determined from the
transitions listed in the column headers. The sizes given here are those of
the diameter of a circle with the same area as that enclosed by
the FWHM contours in Figs.~\ref{intem}a$-$k. If the observed angular
diameter is $\rm \theta_{obs}$, and the beam size at FWHP is 
$\rm \theta_{beam}$, then the beam-corrected size listed in the Table is
$\rm \theta_{cor} = \sqrt{\theta_{obs}^2 - \theta_{beam}^2}$.
Finally, in column~7 we give a
mass estimate of the core, as determined from the $^{13}$CO observations. We
note that this mass represents {\it all} emission above the FWHM contour in
Figs.~\ref{intem}a$-$k, regardless of whether all this emission is associated
with the embedded YSO. For this latter information, and for the assumptions
made in the mass estimate, we refer to Sect.~\ref{sindiv}. The mass
estimates are lower limits, because only emission above the 50\% level was
used, and the FWHM contour is usually not closed inside the mapped area. The
mass range is an order of magnitude, from 130~M$_{\odot}$ (Mol~118) to
2700~M$_{\odot}$ (Mol~3).

\noindent
The FWHM contour of the $^{13}$CO emission extends beyond the limits of the
maps in all but two of the observed sources. For the HCO$^+$ emission we
find the same, for most sources. Only the CS emission is well within the map
boundaries, except for Mol~155, but even there the FWHM contour is almost
closed inside the mapped region (Fig.~\ref{intem}j). 
Unfortunately the
maps in C$^{18}$O and C$^{34}$S are too small to allow a size determination.

\noindent
Fig.~\ref{intem} shows that the peak position of the higher-density
tracers does not always coincide with that of the mm-continuum peak (=
0\arcsec,0\arcsec). However, for those 
sources for which new SCUBA observations (at 450~$\mu$m and/or 850~$\mu$m; 
Molinari et al., in preparation) are available, we find that the 
correspondence between molecular- and sub-mm-peak is good. The reason is that
Molinari et al. (\cite{molinari00}) determined the location of the mm-peak
from 5-point cross observations in most cases, rather than from mapping,
which may have resulted in a not very accurate determination of the mm-peak
in some cases. The IRAS source is usually within a few arcseconds from the
mm-peak position, with the exception of Mols~3 (8\pas 7), 98 (5\pas 9), 155
(11\pas 9), and 160 (8\arcsec). Considering the sizes of the IRAS point
source error ellipses (see e.g. Molinari et al.~\cite{molinari00}), these
deviations are not significant.

\subsection{Line shapes}

At many positions in the maps, but certainly at the peak positions, the
signal-to-noise of the spectra is high enough for a detailed look at the
line shapes. The rare isotopomer C$^{34}$S, with its low abundance and
relatively high critical density, is a tracer of the denser regions of
molecular clouds. Its line profiles can be fitted with a single Gaussian,
with only a few exceptions: Mol~118 and Mol~160, where there are two
components (see Sect.~\ref{sindiv}). In Mol~77 and 98, at low emission
levels, there are deviations from a Gaussian on the blue side (perhaps due
to outflowing gas). Because of this, the velocity of C$^{34}$S can be
assumed to represent the velocity of the high-density gas, and can be used
to identify asymmetries in the line profiles of the other transitions
observed here. For this purpose, in Figs.~\ref{shift}a-d we 
show the spectra of $^{13}$CO(2$-$1), HCO$^+$(1$-$0), CS(3$-$2), and 
C$^{18}$O(2$-$1) aligned with the C$^{34}$S(3$-$2) velocity. 

\noindent
Fig.~\ref{shift}a shows that none of the $^{13}$CO profiles is a simple
Gaussian, except perhaps for Mol~8, although broader emission is visible in
the wings on both sides.  Several of the profiles show dips, especially
Mol~3, 59, and 77, while others (Mol~118, 155, 160) show shoulders. These
deviations may be
due to the superposition at that location of separate velocity components,
or due to self-absorption. The latter can be produced either by an intervening
colder cloud at the same velocity, or by a temperature gradient in the cloud
itself. Profiles with stronger blue peaks can also be the
signature of infalling gas (e.g. Myers et al. \cite{myers}), although it is
unlikely that such a phenomenon is visible in the present observations,
considering the large distances of the sources in our sample, and the
relatively large beam sizes involved. 

\noindent
The asymmetry of the profiles can be
quantified by looking at the ratio of $\int Tdv$ on the blue- and red sides
of the lines (where `blue' and `red' are relative to the C$^{34}$S
velocity). The distribution of this ratio for each transition is shown in
Fig.~\ref{asymm}. The average ratio (indicated in each panel), is $>$1 for
all tracers, indicating a clear blue asymmetry in the sample.
The largest deviations from a Gaussian shape are found for the HCO$^+$
profiles (Fig.~\ref{shift}b). By contrast, the CS spectra are all more or
less symmetric with respect to the C$^{34}$S velocity:
deviations from Gaussian profiles are seen in the CS spectra for Mol~3, 77, 
and 98, which are flat-topped, while there may be
a dip in that for Mol~155. Mol~8, 59, 98, and 160 have ratios $\approx$1. 
The C$^{18}$O profiles are
also more Gaussian in shape than those of $^{13}$CO and HCO$^+$, although
less so than those of CS.

\noindent
The interpretation of these numbers is not straightforward, because most
spectra have multiple emission components (see sect.~\ref{sindiv}), some of
which may not be associated with the embedded IRAS sources (about half of
the objects under investigation are located in the inner Galaxy). 
Some general remarks can be made, however. 
The range in
the ratios for the various tracers is an indication for their kinematical
behaviour. For instance, the profiles of the tracers of the
lower-density gas (like $^{13}$CO or C$^{18}$O) show evidence of a kinematical
behaviour different from that of the higher-density gas (as
represented by the C$^{34}$S lines). On the other hand, the higher-density
tracers (CS) tend to have narrower profiles, which are more symmetric with
respect to the velocity of the high-density gas. The profiles of transitions
like $^{13}$CO, which sample the more diffuse gas in which the high-density
clumps are embedded, are shaped by larger-scale turbulence and by the
superposition of various velocity components, that are not directly related
to the molecular core in which the YSO is embedded. The HCO$^+$ emission
shows the largest average $\int Tdv$ blue/red ratio, as well as the largest
range; this may be related to the fact that HCO$^+$ is also produced in
shocks, causing its kinematics to differ considerably from that of the
C$^{34}$S gas.

\smallskip\noindent
The ratio of $\int Tdv$ CS/C$^{34}$S ranges between 4.7 (Mol~98) and 21.6
(Mol~117); the peak temperature ratio varies from 2.2 (Mol~77) to 15.4 (Mol~3).
The $^{32}$S/$^{34}$S-value for the local ISM $\approx$22 (Wilson \& Rood
\cite{wirood}). The difference is likely due to CS being optically thick.

\smallskip\noindent
Note that if the {\sl Low} objects are really representatives of a very
early evolutionary stage of high-mass stars, they are expected to be 
associated with molecular outflows. These will contribute to the broadening
of the line profiles discussed above, even though the presence of multiple 
emission components tends to obscure the visibility of line wings in the 
spectra. 

\subsection{Boltzmann plots}

Fig.~\ref{mol_boltz} shows the Boltzmann plots constructed from the
CH$_3$C$_2$H(6--5) observations. From weighted least-squares fits to the data 
points, 
temperature and column densities can be derived (see e.g. Kuiper et al. 
\cite{kuiper}; Bergin et al \cite{bergin}). The resulting rotational
temperature T$_{\rm rot}$, asumed to be equal to the gas kinetic temperature 
T$_{\rm kin}$, and column densities, assuming optically thin emission, are 
collected in columns~4 and 6 of Table~\ref{ch3c2h_results}, respectively. 
The uncertainty in T$_{\rm kin}$ in column~5 is derived from that in the slope 
of the fit. An uncertainty for the column density N was obtained in two ways: 
by calculating an N$_{\rm low}$ and N$_{\rm high}$ with the extremes in 
T$_{\rm kin}$, and
taking the average difference between these two values and the N derived for
T$_{\rm kin}$ (given in column~6); and by calculating the average difference in
N as a result of the uncertainty in the slope of the fit. The uncertainties
in N derived in these two ways were added in quadrature, and the resulting
value is listed in column~7.

\noindent
We list the
parameters obtained from both the high- and the low-resolution spectra, for
comparison. Only for Mol~118 is there a significant difference between the
two. The derived T$_{\rm kin}$ are
between 20 and 45~K, in agreement with the dust temperatures (Molinari et
al. \cite{molinari00}). The low temperatures for these {\sl Low} sources are
in marked contrast with those found for members of the {\sl High} group
such as IRAS20126+4104 (Mol~119), where T$_{\rm kin} \sim 200$~K (Cesaroni et
al \cite{cesa97}; from CH$_3$CN observations).

\noindent
The CH$_3$CN(8--7)
data for Mol~98 at (24\arcsec,0\arcsec) yield T$_{\rm kin} \approx$666~K 
(425~K$-$1535~K) from
the high-resolution spectrum, and 1073~K (577~K$-$7599~K) from the
low-resolution spectrum. Though the uncertainties are very large, it does 
show that
the temperature derived from CH$_3$CN is much larger than that, derived from
CH$_3$C$_2$H, indicating that the emission of the former originates from
deeper in the cloud, i.e. from closer to the embedded heating source (i.e.
the YSO).

\section{Comments on individual sources 
\label{sindiv}}

In the following subsections we shall discuss the observed sources in some
detail. If it helps our understanding of the molecular line data, we shall
make use of as yet unpublished observations of $^{12}$CO(2--1) and
C$^{18}$O(2--1) [NRAO 12-m],
3.6-cm radio continuum [VLA D-array], and 450~$\mu$m and/or 850~$\mu$m
[SCUBA]. We try to identify the velocity components of the molecular
emission which are associated with the Molinari sources, and estimate the
masses of these components from the $^{13}$CO data. The first step is to
derive the column densities
of the observed molecule, using the familiar equations (see e.g. Brand \&
Wouterloot \cite{bws151}; Rohlfs \& Wilson \cite{tools}) with the
appropriate constants for the molecule under consideration put in, and
assuming the emission is optically thin. The excitation temperature 
T$_{\rm ex}$ is estimated from the NRAO CO spectra. To get
the column density of H$_2$ we have used the relative abundance
[H$_2$]/[$^{13}$CO] = 5.0$\times 10^5$.
Masses are estimated from the average H$_2$ column density, using all
emission above the FWHM level of the integrated $^{13}$CO emission, and 
correcting the enclosed area for the beam of the observations. 
A correction for He ($\times 1.36$) has been applied as well. Mass estimates 
are lower limits, because not all emission is used, and the FWHM contour is 
not always closed within the area mapped with the 30-m. The results are
collected in Table~\ref{clmass}.

\subsection{\object{Mol~3}}

An optical image of a 10\arcmin $\times$10\arcmin\ region around this
object is shown in Fig.~\ref{optical3} (taken from the Digital Sky
Survey: DSS). Clearly visible are several nebulous patches, that have been
identified by Neckel \& Staude (\cite{neckel84}), and which are known as
GN0042.0$-$1 and GN0042.0$-$2 (Neckel \& Vehrenberg \cite{neckel85}). In
Fig.~\ref{optical3} the easternmost cross marks the position of the sub-mm
peak (Molinari et al. \cite{molinari00}); the
IRAS position is slightly to the NW of this, and almost coincides with the
star which is located at the vertex of one of the nebulous patches.
Neckel \& Staude derive a photometric distance to this star (and to the star
in the nebulosity to the SW) of 1.7~kpc, based on a B5 spectral type for both
stars, and an {\it assumed} luminosity class V.
White \& Gee
(\cite{white}) observed both objects with the VLA D-array. GN0042.0$-$1 was
not detected (at $>$0.3~mJy/beam), while GN0042.0$-$2 was detected at 6~cm.
The westernmost cross in Fig.~\ref{optical3} indicates this radio continuum
peak. The source has a peak flux density of 0.9~mJy/beam 
($\int Fd\nu$=3.5~mJy); the
spectral type derived from the radio data is B1-2, which is a large
discrepancy with the Neckel \& Staude optical data (B5). Hence, their
photometric distance is uncertain, and we will use the kinematic distance.

\noindent
GN0042.0$-$1, the collection of nebulosities nearest to Mol~3, was recently
observed by Molinari et al. (in preparation) with the VLA D-array at 3.6~cm.
They detected 3 faint radio sources, one of which (peak flux density
$\approx$0.19~mJy/beam) is very close to the IRAS source.
A second 3.6~cm source, with similar peak flux, lies at the same
position as the H$_2$O maser and very close ($\sim$8\arcsec SE) to the
850~$\mu$m peak, both measured by Jenness et al. (\cite{jenness}). This sub-mm
peak lies $\sim$ $-$5\arcsec,$-$19\arcsec\ from the sub-mm peak given in
Molinari et al. (\cite{molinari00}; taken as the 0,0 position of the present
maps), who noted however that they missed the
primary peak in their observations (a 5-point cross near the IRAS position
only). Recent, unpublished SCUBA (850~$\mu$m) observations
by Molinari et al. show that the true sub-mm peak
coincides with the Jenness et al. 850~$\mu$m peak.
In Fig.~\ref{intem}a it is seen that the CS peak is slightly offset to the
SE with respect to the communal location of the other molecular peaks, the 
sub-mm peak, the H$_2$O maser, and one of the radio continuum sources.

\smallskip\noindent
A look at Fig.~\ref{molspec}a shows, that at the peak position in Mol~3 the
$^{13}$CO, C$^{18}$O, and HCO$^+$ spectra are double-peaked, while the CS
spectrum is
flat-topped. The velocity of the (weak) C$^{34}$S line falls more or less in
between the dip in the line profiles and the peak of the red component.
Inspection of all profiles in the map, and of position-velocity plots of the
molecular emission shows, that two components can be clearly seen over the
whole mapped region; also the CS profiles are double-peaked away from the
central position. In Fig.~\ref{mol3bluered} we show the integrated areas
over the blue- (at $-$50.85~kms$^{-1}$; left-hand panels) and
the red side of the dip (right-hand panels). The sub-mm peak, the H$_2$O
maser, and one of the 3.6~cm sources lie towards the peaks of the gas
distribution for both components, although the correspondence with the blue
peaks is slightly better (while the red component is more compact). 
In $^{13}$CO, blue component, a separate
clump is visible, which may also be present in the HCO$^+$ and CS maps. 

\subsection{\object{Mol~8}}

The finding chart (at 8000~\AA) published 
by Campbell et al. (\cite{campbell}) shows that some very faint diffuse 
emission is associated 
with the IRAS source, which coincides with a star-like object. This source has
been included in various studies. Ishii et al. (\cite{ishii}) measured its NIR
($1.3 - 4.2~\mu$m) spectrum, and found the 3.1~$\mu$m absorption feature due
to H$_2$O ice. This feature is found in environments protected from UV
radiation, i.e. high-density molecular clouds, and is indeed often detected in 
deeply embedded YSOs (Ishii et al.~\cite{ishii}). Slysh et al. (\cite{slysh}) 
found a strongly circularly
polarized 2.7~Jy OH maser at 1665~MHz. No 6.7~GHz methanol maser was
detected by MacLeod et al. (\cite{macleod}), nor was this object detected by 
Harju et al. (\cite{harju}) in their search for SiO maser emission.

\noindent
Although the spectra taken at the peak position (Fig.~\ref{molspec}b) seem
to show a single emission component, the channel maps in
Fig.~\ref{mol8chan} clearly show the existence of two components, at
$-$25.5 and $-$26.5~kms$^{-1}$ respectively. Although the velocity
difference is small, the two components have a different spatial
distribution and are therefore easily separated by Gaussian fitting to the
line profiles. As an
illustration we show in Fig.~\ref{mol8gauss} the distributions of the
areas under the blue and red components for $^{13}$CO. The redder (northern) 
component is associated with the embedded object.

\noindent
Mol~8 is WB89~621 in the catalogue of Wouterloot \& Brand (\cite{wb89}), who
detected (with the IRAM 30-m) a strong CO(1$-$0) line at the IRAS position,
and found evidence for wings.
Wings are also visible in the present ($^{13}$CO, HCO$^+$, CS) spectra.
Position-velocity plots show that wing emission occurs primarily in an area
within 12\arcsec\ from map center. 
The blue and red lobes are centered on the IRAS/sub-mm 
peak position; the closeness of the peaks of the lobes suggests that the 
bipolar flow is seen nearly pole-on.

\subsection{\object{Mol~59}}

The distance (5.7~kpc) to this object was based on 
V$_{\rm lsr}[NH_3(1,1)]=93.7$~kms$^{-1}$.
This line is however very weak (T$_{\rm mb}=0.34 \pm 0.09$~K), and the
(2,2) line (0.23$\pm 0.10$~K) is found at V$_{\rm lsr}$=97.7~kms$^{-1}$
(Molinari et al. \cite{molinari96}), casting some doubt on the validity of
this distance calculation. The lines measured with the 30-m are however
detected with reasonable signal-to-noise, and the emission is moreover
clearly associated with Mol~59, as can be seen from Fig.~\ref{intem}c. The
average velocity of $\approx$114.5~kms$^{-1}$, corresponds to 
d$_{\rm kin}$=6.6~kpc (using the Brand \& Blitz (\cite{brand}) rotation curve).

\begin{figure}
 \resizebox{\hsize}{!}{\includegraphics{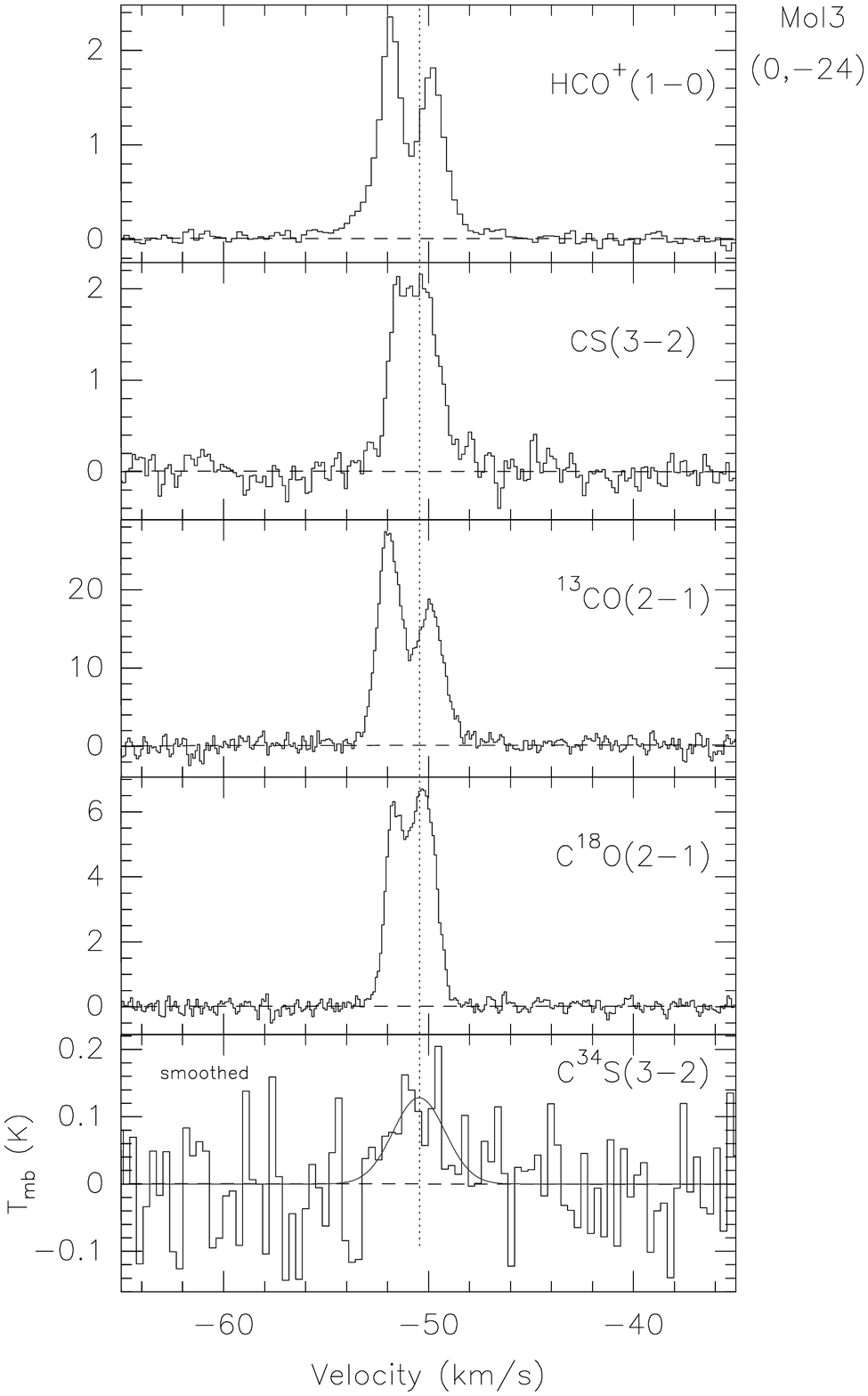}}
 \hfill
 \caption{{\bf a}\ Spectra towards the peak position in the maps for Mol~3. 
Source name, offset position, and molecules/transitions are indicated. 
The offset is relative to the maps' central position, which is the sub-mm
peak (Molinari et al. \cite{molinari00}). 
The main beam
brightness temperature (T$_{\rm mb}$) is plotted against V$_{\rm lsr}$. The 
vertical
dotted line indicates the V$_{\rm lsr}$ of the C$^{34}$S(3$-$2) line, obtained
from a Gaussian fit; the horizontal dashed lines indicate the T$_{\rm mb}$=0~K
level in each spectrum.}
\label{molspec}
\end{figure}

\noindent
This object lies in the galactic plane, $\sim$20\degr\ from the Galactic
Center, and there is emission at various velocities. In the $^{13}$CO
spectra we also found emission at 67~kms$^{-1}$ ($\sim$8~K) for instance. In
the NRAO $^{12}$CO spectra, which have a
larger velocity range, we find emission at practically all velocities
between 0 and 170~kms$^{-1}$. In those spectra, emission is also found at
$\sim$117-128~kms$^{-1}$; in the $^{13}$CO spectra (e.g.
Fig.~\ref{molspec}c) this is visible too, sometimes as a red shoulder to
the line which has its peak at $\sim$115~kms$^{-1}$.  

\smallskip\noindent
The spectra in Fig.~\ref{molspec}c show a depression at 114.3~kms$^{-1}$,
at least in HCO$^+$, $^{13}$CO, and perhaps even in CS. The C$^{34}$S line,
although weak, peaks more or less at the velocity of this depression, as
the CH$_3$C$_2$H and C$^{18}$O lines seem to do. 
All HCO$^+$ and most $^{13}$CO spectra in the mapped region (30
spectra, covering an area of 60\arcsec $\times$48\arcsec) show this dip (the
$^{13}$CO spectra at the edge of the map show asymmetries or shoulders),
whereas CS has been detected at only a few positions near the peak at
(12\arcsec,0\arcsec). This large area, and the relatively large beams
involved in the observations, make self-absorption due to infalling gas an
unlikely interpretation, also because there is little 
consistency between the relative strengths of the blue and red peaks in the
profiles of the different tracers.
The dip could be due to only a temperature gradient
(without infall) in
the cloud, or to an intervening colder cloud at the same velocity,
but we prefer to interpret the spectra as being due to the superposition of
several components, although this does not exclude the presence of
self-absorption at the peak position. We integrated the emission over 5 
velocity
intervals, distinguishing between 5 components: $101-105 (comp.~1), 
105-111 (comp.~2), 111-114.3 (comp.~3), 114.3-117 (comp.~4)$, and 
$117-122 (comp.~5)$~kms$^{-1}$. The first
component is present mostly in the E part of the map, and appears to have its
maximum outside the mapped region. The distribution of the integrated
$^{13}$CO emission from components (2) to (5) is shown in 
Fig.~\ref{mol59areas},
together with that of components $(3)$ and $(4)$ for HCO$^+$ (where the
other components are much fainter or absent). Components $(3)$ and $(4)$ are
dominating the spectra, and are referred to as the `main line blue' and `main
line red', respectively. Component $(2)$ may be outflow emission connected
to component $(3)$, while $(5)$ is most likely not outflowing gas, but
emission from (a) separate component(s) (based on inspection of the 12-m $^{12}$CO
spectrum). From Fig.~\ref{molspec}c it is not evident which of the two
main lines is associated with the sub-mm peak, but Fig.~\ref{mol59areas}
suggests that it is the red component, which is more compact than the
distribution of the blue line, and which has a maximum at the sub-mm peak in
both $^{13}$CO and HCO$^+$.

\subsection{\object{Mol~75}}

Ishii et al. (\cite{ishii}) measured the NIR ($1.3 - 4.2~\mu$m) spectrum of
this source, and found the 3.1~$\mu$m absorption feature due to H$_2$O ice.
As for Mol~3, this is an indication of the presence of high-density material
associated with this object. No 6.7~GHz methanol maser was detected by 
MacLeod et al. (\cite{macleod}).

\noindent
Our observations show that the gas around the IRAS source is
distributed over several clumps. This is particularly so in $^{13}$CO,
where lines are found at V$_{\rm lsr} \approx 41, 49, 53, 56-58$ and
$58-60$~kms$^{-1}$. The latter two components are also found in HCO$^+$ and
CS, suggesting that one or both of these are the associated components.
We note that
the NH$_3$(1,1) line is at 56.8~kms$^{-1}$, and that at the map's peak
position all lines (including CH$_3$C$_2$H) detected with the 30-m, except 
HCO$^+$, have V$_{\rm lsr}
\approx 57$~kms$^{-1}$. The HCO$^+$ line has a dominant component at
$\sim$60~kms$^{-1}$, and the dip seen in Fig.~\ref{molspec}d is at or close
to the velocity where the other lines have their peak (except for CS, which
peaks at the velocity of the blue component of the HCO$^+$ emission). Note
also that the KOSMA HCO$^+$(4--3) spectrum seems to peak at the dip in the 
(1--0) spectrum (Fig.~\ref{kospec}a), but Fig.~\ref{mol75bluered} suggests
the presence of two components.

\noindent
We separated the two main components by integrating over the appropriate
velocity intervals ($55-58.6-61.5, 55-57.6-65$, and $54-58.2-62$ for
$^{13}$CO, HCO$^+$, and CS, respectively). The resulting distributions of
the integrated emission are shown in Fig.~\ref{mol75bluered}. From this
figure it is seen that the blue component peaks at or near the position of
the IRAS source/850~$\mu$m peak, while the red component peaks south
of that. Because the NH$_3$ and CH$_3$C$_2$H lines have the same velocity as 
the blue
component, we assume that this defines the clump in which the FIR source is
embedded.

\noindent
We note that the HCO$^+$ spectra in the central region of the map show a
very broad red wing, up to V$_{\rm lsr} \sim 80$~kms$^{-1}$ at
(0\arcsec,0\arcsec) (i.e. more than 20~kms$^{-1}$ from the center of the
red component; Fig.~\ref{molspec}d). This broad emission is found in all the 
individual observations (ranging in number from 2 to 8) contributing to the 
average spectrum, and
is therefore unlikely to be an artifact. This is strengthened by the fact
that the wing is also present in the HCO$^+$(4--3) spectrum (see
Fig.~\ref{kospec}). Some extended emision on the red
side of the $^{13}$CO spectra is also present, although with a much smaller
extent in velocity (up to 6~kms$^{-1}$ from the line center).

\subsection{\object{Mol~77}}

The $^{13}$CO spectrum at (0\arcsec,0\arcsec) is double-peaked 
(Fig.~\ref{molspec}e). All spectra of this molecule show either two
peaks, or a shoulder (blue or red). The dip (or the start of the shoulder) is
at the velocity of the C$^{34}$S line,
which is also the velocity of the peak of the C$^{18}$O and CH$_3$C$_2$H
lines, and that of
the NH$_3$(1,1) and (2,2) lines (Molinari et al. \cite{molinari96}). At all
observed positions, the HCO$^+$ spectrum shows only very weak emission on the
red side of this dip. In Fig.~\ref{mol77grid} we compare the spectra of
$^{13}$CO and HCO$^+$ with the $^{12}$CO spectra we obtained with the NRAO
12-m (beam FWHP $\sim$29\arcsec). From this comparison it seems likely that
the $^{13}$CO line profiles are the result of the superposition of two
components, rather than being due to self-absorption. In the HCO$^+$
spectra, the bluer component clearly is the dominant one. The $^{12}$CO
spectra show broad emission at velocities $>$80~kms$^{-1}$, some of which is
also seen in $^{13}$CO (see Fig.~\ref{mol77areas}a), and is probably the
result of additional emission components. On the blue side however 
(V$_{\rm lsr} \lesssim 74$~kms$^{-1}$), the $^{12}$CO spectra are much 
steeper. At these velocities
some emission is also present in the spectra of the molecules observed at
IRAM (see Fig.~\ref{molspec}e, and Fig.~\ref{mol77areas}), and may be
outflow emission associated with the component on the blue side of the
dip; an eventual red component of this outflow is obscured
by the emission of the other components redward of the dip. 

\noindent
To isolate the various components, we have integrated the line profiles over
appropriately chosen velocity intervals; the results are shown in
Fig.~\ref{mol77areas}. From the dominance of the 75~kms$^{-1}$ component
of HCO$^+$, and the fact that it peaks at or very near the IRAS source and
sub-mm peak, we conclude that the emission in the interval $\sim
74-76$~kms$^{-1}$ is likely to represent the associated component. However,
also the emission between 76 and 80~kms$^{-1}$ peaks at that location, most
clearly seen in the distribution for the density tracers C$^{18}$O and CS
(Fig.~\ref{mol77areas}c), as does the emission between 71 and 74~kms$^{-1}$.

\subsection{\object{Mol~98}}

Towards
this region we found an H$_2$O maser (Palla et al. \cite{palla91}), while
a 6.7~GHz methanol maser was detected by MacLeod et al. (\cite{macleod}). As
they mention, methanol masers are unique indicators of massive star-forming
regions, because unlike water- and hydroxyl masers, methanol masers have not
been found towards stars of spectral type later than B2.
These latter authors also detected circularly polarized 1665~MHz OH maser
emission. Molinari et al. (\cite{molinari98a}) found no 2 and 6-cm radio
continuum emission associated with this IRAS source. However, in recent VLA-D
observations at 3.6-cm (Molinari et al., in preparation) continuum emission
with a peak flux density $\sim$1~mJy/beam was detected at
$\sim$(+10\arcsec,+14\arcsec) from the
present map center. Recent SCUBA observations (Molinari et al., in
preparation) at 450~$\mu$m show a sub-mm peak at $\sim$(+30\arcsec,+3\arcsec)
from the present map center. We note that this is quite different from the
offset of the sub-mm peak as given in Molinari et al. (\cite{molinari00})
($-$10\arcsec,$-$10\arcsec\ from the IRAS position, i.e.
$-$4\arcsec,$-$10\arcsec\ from the map center), which was however only based 
on a 5-point cross observation, and hence less trustworthy.

\noindent
The present molecular observations show a quite compact core, even in
$^{13}$CO, with a peak $\sim$25\arcsec\ E of the IRAS source
(Fig.~\ref{intem}f). 
The sum of all $^{13}$CO spectra of this source shows two absorption dips,
at V$_{\rm lsr} \sim 62$~kms$^{-1}$ and $\sim 70$~kms$^{-1}$, indicating that
here the offset position ($\Delta \alpha, \Delta \delta =
-1800$\arcsec,0\arcsec\ from the center of the map) was not far enough from 
the molecular cloud in which the
object is embedded. We have tried to correct for this by summing all
$^{13}$CO spectra with narrow (V$_{\rm lsr} \lesssim 60$~kms$^{-1}$) emission,
fitting Gaussians to the dips (which have a strength of $\sim 1-1.5$~K), and
adding those components to all $^{13}$CO spectra in the map. Unfortunately
this does not correct for any absorption that might be present in the main
line, but inspection of the spectra taken at positions without $^{13}$CO
emission ($\Delta \alpha$= 96\arcsec) indicates that any such absorption
will be $\lesssim$2.4~K ($\sim 2\sigma_{\rm rms}$ in the individual spectra at
those positions). 

\noindent
The channel maps for this source indicate that there is only one emission
component associated with the object (the $^{12}$CO spectra in the NRAO 12-m
map, that covers an area of about 300\arcsec $\times$300\arcsec, show many
emission components, but near Mol~98 the dominating one is that between
$\sim 50-60$~kms$^{-1}$); the non-Gaussian profiles seen in
Fig.~\ref{molspec}f (flat top or shoulder) may be caused by saturation or
self-absorption. 

\smallskip\noindent
The spectra in Fig.~\ref{molspec}f show non-Gaussian wings to all
profiles, even that of C$^{34}$S. In Fig.~\ref{mol98wings} we show the
distribution of the integrated emission over the line wings. Note that
HCO$^+$ has a relatively strong blue wing, which extends up to 15~kms$^{-1}$
from the velocity of the bulk of the molecular material. The outflow is
bipolar in all 3 lines shown, and has the center of the line connecting the 
lobes at
offset $\approx$(30\arcsec,5\arcsec), which puts it at the location of the
450~$\mu$m peak.

\subsection{\object{Mol~117}}

On the DSS some faint diffuse emission is
visible at the location of this object. Rather weak NH$_3$(1,1) emission was
detected (Molinari et al. \cite{molinari96}) at V$_{\rm lsr} \approx
-36.4$~kms$^{-1}$, which is also the velocity at which HCO$^+$(4--3) (see
Fig.~\ref{kospec}) is found.
No 6.7~GHz methanol maser was detected by MacLeod et al. (\cite{macleod}).
The molecules observed at IRAM have their peak
emission at V$_{\rm lsr} \approx -35.7$~kms$^{-1}$ (Fig.~\ref{molspec}g and
Table~\ref{linepars}). Our C$^{18}$O(2--1) NRAO 12-m data show (at position
0\arcsec,0\arcsec, and with a 29\arcsec\ beam) a double-peaked profile, with
components at $-$37.9 and $-$35.9~kms$^{-1}$, and a dip at V$_{\rm lsr} \approx
-37$~kms$^{-1}$; the redder component therefore
coincides with the velocities of the peak emission found at IRAM, while the
NH$_3$ velocity lies closer to the dip in the NRAO C$^{18}$O(2--1) spectrum. 
However, also the IRAM spectra show
double-peaked profiles, as illustrated by the $^{13}$CO line at
($-$12\arcsec,0\arcsec) in Fig.~\ref{molspec}g. In fact, at most positions
we find that the line profiles have shoulders or double peaks, with the dip
always at the same velocity. We therefore analyze the data in terms of the
superposition of two emission components. The components are separated by
integrating between $-$40 and $-$36.9~kms$^{-1}$ (blue) and $-$36.9 and
$-$33~kms$^{-1}$ (red) for $^{13}$CO and HCO$^+$, and $-$39.5 and
$-$36.3~kms$^{-1}$ (blue) and $-$36.3 and $-$33~kms$^{-1}$ (red) for CS. The
distributions of the integrated emission are shown in
Fig.~\ref{mol117bluered}. Because the red peak seems to be the dominant
one, we assume this is the associated component. 

\subsection{\object{Mol~118}}

This is the object with the smallest kinematical distance (1.6~kpc) in the
sample. 
The (0,0) position of the maps, and the IRAS source
(at offset $-$3\pas 7,0\arcsec) are located in a dark cloud, which is
surrounded by diffuse emission. 
No 6.7~GHz methanol maser was detected by MacLeod et al. (\cite{macleod}).

\noindent
The spectra shown in Fig.~\ref{molspec}h are non-Gaussian, and as
evidenced by the C$^{34}$S spectrum, and the superimposed $^{13}$CO and
C$^{18}$O spectra at offset (12\arcsec, 12\arcsec), this is due to the
presence of two velocity components. In addition, emission in the wings of
the profiles is visible, which may be due to outflow, or to additional
components. These wings are also visible in the NRAO CO(2--1) spectra.
We separate the two main (`blue' and `red') components by
integrating over the following velocity intervals: 6.5--8.5--9.4~kms$^{-1}$
($^{13}$CO); 6.5--8.4--9.4~kms$^{-1}$ (HCO$^+$); 6.8--8.2--9.2~kms$^{-1}$
(CS). These integration limits omit the wing emission. The distribution of
the emission in each component is shown in Fig.~\ref{mol118bluered}. The
NH$_3$ velocity is 7.8~kms$^{-1}$, that of CH$_3$C$_2$H is 7.9, and thus 
coinciding with the (more intense)
blue component, which we will take as the associated one. 

\subsection{\object{Mol~136}}

Inspection of the data reveal that there is only one emission component
detected towards this source. The line profiles are non-Gaussian, being
skewed towards the blue (see Fig.~\ref{molspec}i). This is also seen in
the $^{12}$CO(1--0) line profile (Wouterloot \& Brand \cite{wb89}; source
WB89~93). Low-level emission is also seen extending over a few kms$^{-1}$
from the central line velocity; the distribution of the integrated emission
in these wings is shown in Fig.~\ref{mol136wings}. The blue emission peaks
near the IRAS source position (which is also the peak of the sub-mm emission), 
while the red component has its maximum more to the West.

\noindent
Although Molinari et al. (\cite{molinari98a}) did not detect associated radio
continuum emission (VLA, 2 and 6-cm), deeper observations at 3.6-cm with the
VLA D-array by Molinari et al. (in preparation), reveal diffuse emission
(peak flux density $\sim$6.2~mJy/beam) just NE of the IRAS source.

\subsection{\object{Mol~155}}

The emission of especially $^{13}$CO and HCO$^+$ is quite extended in this
source (see Fig.~\ref{intem}j). The line profiles are similar to those in
Mol~77 (cf. Figs.~\ref{molspec}e and j): The profiles of $^{13}$CO and
HCO$^+$ show a dip or a shoulder everywhere, while for the latter molecule
the blue component is much stronger than the red one at every observed
position. The dip or shoulder is at V$_{\rm lsr} \sim -51.1, -50.7$ and
$-51.0$~kms$^{-1}$ for $^{13}$CO, HCO$^+$,and CS, respectively. The velocity
of the NH$_3$(1,1) line is at $-$51.5~kms$^{-1}$, thus in the velocity range
of the blue component. We separate the two emission components by
integrating blue- and redwards of the dip. The resulting distributions of
the integrated emission are shown in Fig.~\ref{mol155bluered}. The
dominance of the HCO$^+$ blue component is clearly visible there. We note
that towards this source observations of SiO(2--1) (Harju et al.
\cite{harju}) 
have yielded no detections.

\subsection{\object{Mol~160}}

This source has been observed and discussed extensively by Molinari et al.
(\cite{molinari98b}), who concluded that this is a very good candidate
massive Class~0 object. Inspection of the line profiles at the peak position
(Fig.~\ref{molspec}k; especially C$^{18}$O) strongly suggests the presence
of two velocity components. Individual $^{13}$CO spectra indicate that there
might be more than two components; we also note that the dip, or shoulder,
in these spectra is not always at the same velocity. This may be an
indication of the presence of a velocity gradient in the $^{13}$CO emission,
and separating the contribution of the various components by integrating
over fixed velocity intervals is hazardous. We have therefore performed
Gaussian fits to the profiles of $^{13}$CO, HCO$^+$, and CS (even though for
the latter two we do not see a shift in the position of the dip (if at all)
with position). Based on the V$_{\rm lsr}$ of the NH$_3$(1,1) and
CH$_3$C$_2$H(6--5) lines ($-50.0$ and $-50.45$~kms$^{-1}$ respectively), we
identify the main emission component. After subtraction of the other
Gaussian components, we integrated the spectra; the resulting distributions
of the emission are shown in Fig.~\ref{mol160mainarea}. The emission peaks
lie close to the 3.4-mm continuum peak found with OVRO by Molinari et al.
(\cite{molinari98b}). 

\noindent
Fig.~\ref{mol160rv} shows a series of Right Ascension versus V$_{\rm lsr}$
plots for the $^{13}$CO line, at the Declination offsets labeled in each 
panel. Especially at
positive Declination offsets the distribution of the emission is suggestive
of a velocity gradient, with velocity increasing (becoming redder) from E to
W. The gradient is of the order of 1.4~kms$^{-1}$pc$^{-1}$. The HCO$^+$
data for the main component indicate a gradient of
$\sim$0.8~kms$^{-1}$pc$^{-1}$; no clear velocity gradient is present in the
CS data.

\noindent
The line profiles in Fig.~\ref{molspec}k show some low-level emission in
the wings. For HCO$^+$ and CS
we show the integrated wing emission in Fig.~\ref{mol160wings}. As was found
on a much smaller scale in our OVRO observations of HCO$^+$ and SiO (Molinari 
et al. \cite{molinari98b}), the blue and red lobes almost overlap, and we
are seeing the outflow nearly pole-on. 

\noindent
Finally we note that, although Molinari et al. (\cite{molinari98a}) found no
radio continuum emission (at 2 and 6-cm) towards this object, there is an
11-cm radio continuum source (F3R 3484; integrated flux density 0.16~Jy, peak
flux density 130~mJy) at this location (F\"urst et al. \cite{fuerst}).
Deep VLA-D array observations at 3.6-cm (Molinari et al., in preparation) 
reveal a double-lobed emission structure. The extent of the lobes (at a level 
of $\sim$5\% of the peak value) is $\sim$93\arcsec $\times$127\arcsec\ for the
W lobe ($\int Fd\nu \approx$60~mJy), and $\sim$50\arcsec\ for the E lobe
$\int Fd\nu \approx$17~mJy). The 3.4-mm continuum source (Molinari et
al. \cite{molinari98b}) lies at the sharp E edge of the larger, westernmost
radio lobe, at $\sim 21$\arcsec\ to the NE from its peak. The general
location of the radio lobes coincides with the diffuse emission seen in the
15$\mu$m ISOCAM map (see Molinari et al. \cite{molinari98b}), although the
extent of the radio continuum emission is larger. It is unclear whether this
radio emission is associated with the embedded object.

\section{Discussion and conclusions
\label{discuss}}

The most direct result of this work is that all of the observed sources
are clearly associated with well-defined molecular clumps. It must be stressed
that such clumps are seen in various molecular tracers and are hence real
physical entities. The main goal of our observations
was to obtain a picture of the molecular environment associated with {\sl Low}
sources and look for any evidence supporting the hypothesis that {\sl Low} 
sources are the precursors of {\sl High} sources. In this scenario, the former 
correspond
to YSOs still accreting mass from the surrounding environment, whereas the 
latter are ZAMS early type stars still deeply embedded in their natal cores.
The most direct way to verify this is to make high-angular resolution
observations of the {\sl Low} objects to study their structure and physical
parameters, and indeed we are already performing this type of investigation
on a limited number of objects selected on the basis of the present study.
However, it is also possible to draw some tentative conclusions by comparing
our sample with a sample of {\sl High} sources studied by Cesaroni et al.
(\cite{cesa99}) with the same telescope and in the same lines. For this 
purpose,
we have collected in Table~\ref{tcomp} all the parameters derived from 
both studies:
for each quantity we give the minimum, maximum, and mean values. We note 
that sources Sh-2~233 and NGC\,2024 from the
Cesaroni et al. (\cite{cesa99}) sample have not been considered here, as they 
satisfy neither the requirements of the {\sl High}-, nor of the {\sl Low}
sources.

\smallskip\noindent
In Table~\ref{tcomp}, $\Theta$ is the angular diameter of the clumps after
deconvolution of the beam, $D$ is the corresponding linear diameter,
T$_{\rm mb}$(K) and T$_{\rm b}$(K) are the main beam and intrinsic (after
correcting for the beam filling factor) brightness temperatures,
FWZI the line full width at zero intensity, FWHM the line full width at half
maximum, M$_{\rm vir}$ the virial mass, and M$_{\rm CD}$ the mass obtained
by integrating the line emission over the line profile and over the 
{\it whole} emitting region. For the latter estimate we have assumed LTE at 
30~K and abundances of $1.1\times 10^{-6}$, $10^{-9}$, $10^{-8}$, $4.5\times 
10^{-10}$
respectively for $^{13}$CO, HCO$^+$, CS, and C$^{34}$S (Irvine et al.
\cite{irvine}). Note that the $^{13}$CO abundance is what was used by
Cesaroni et al. (\cite{cesa99}), and is slightly smaller (by a factor of 1.8)
than what we used in Sect.~\ref{sindiv}. In only four {\sl Low} sources 
(Mol~8, 77, 98, and 160) the
C$^{34}$S(3--2) emission was strong enough to be mapped; sizes, masses, and
intrinsic brightness temperature derived from this molecule, reported in 
Table~\ref{tcomp} are derived from these four objects only, and may not be
representative.

\noindent
Inspection of Table~\ref{tcomp} reveals a few interesting differences between
the two samples in spite of the similar bolometric luminosities. Relative to
the clumps associated with {\sl High} sources, those around the {\sl Low} 
sources are
more massive (2--4 times; except for C$^{34}$S), larger ($\sim$3 times), 
less bright (1--3 times), have smaller line widths ($\ga$1.5 times). 
The {\sl Low}
sources also appear to be less dynamically stable than the {\sl High} sources 
since
they typically have a smaller ratio $\rm M_{vir}/M_{CD}$, which is a 
measure of the balance between gravitation and turbulence: the smaller it is,
the weaker is the support against collapse. The results are therefore 
consistent with {\sl Low} sources representing an evolutionary phase prior to 
that of {\sl High} sources. In fact, the
larger diameters of the {\sl Low} clumps suggest that they should still undergo
substantial contraction, as indicated by the lower densities. Moreover, the
lower brightness temperatures may be considered an indication of lower
temperatures and/or optical depths, both expected in an early phase of the
evolution. Finally, the broader lines in the {\sl High} sample might be 
related to
the existence of a larger number of molecular outflows and hence of YSOs
already formed: such outflows could support the surrounding clump from
collapse by injecting high velocity gas, thus accounting for the higher 
ratio $\rm M_{vir}/M_{CD}$.

\smallskip\noindent
We note that the clump mass as reported in Table~\ref{clmass}
increases with the L$_{\rm fir}$ of the IRAS source (see 
Fig.~\ref{masseslumo}a; a least-squares fit to mass
versus luminosity gives a slope of $\sim 1.17 \pm 0.22$ and a corr. coeff.
of $\sim$0.57). Then, assuming that each clump contains a single protostellar
object, we can obtain an estimate of the mass of the central object from
the known value of L$_{\rm fir}$. Using the models of Palla \& Stahler
(\cite{palsta}) for an accretion rate of $10^{-4}$~M$_{\odot}$ /year, we
compute the mass of a protostar that produces a luminosity, L$_{\rm proto}$,
equal to the observed L$_{\rm fir}$. Note that L$_{\rm proto}$ includes a 
component from accretion and one due to gravitational contraction. For the 
range of luminosities considered here, the latter term dominates.
The resulting protostellar masses are plotted against clump masses in 
Fig.~\ref{masseslumo}b. The highest mass that the models of Palla \& Stahler 
(\cite{palsta}) can give is $\sim$17~M$_{\odot}$, hence the two lower limits 
for two of the objects (Mol~3 and 8). The lowest protostellar mass is for 
Mol~118 and amounts to 7.4~M$_{\odot}$. From Fig.~\ref{masseslumo}b we see
that there is a weak dependence on clump mass. A fit to the data points shows 
that $\rm M_{proto} \propto M_{clump}^{0.30 \pm 0.07}$. Note that the 
slope of this relation is similar to that found by Larson (\cite{larson}) in 
his study of young stars and molecular clouds, who found that the maximum mass 
of stars is related to the mass of the cloud as $\rm M_{star} \propto 
M_{cloud}^{0.43}$.

\smallskip\noindent
In conclusion, we believe that our results lend support to the evolutionary
scenario previously proposed by us (Palla et al. \cite{palla91}; Molinari et 
al. \cite{molinari96}; Molinari et al. \cite{molinari98a}), according to which 
the majority of {\sl Low} sources will eventually evolve into {\sl High} 
sources.

\begin{acknowledgements}
We thank Jan Wouterloot for the KOSMA observations. The KOSMA
radio telescope at Gornergrat-S\"ud Observatory is operated by the
University of K\"oln, and supported by the Deutsche Forschungsgemeinschaft
through grant SFB-301, as well as by special funding from the Land
Nordrhein-Westfalen. The Observatory is administered by the Internationale
Stiftung Hochalpine Forschungsstationen Jungfraujoch und Gornergrat, Bern,
Switzerland.
\end{acknowledgements}



\clearpage
\begin{table}
\caption[]{Observed transitions
\label{freqs}
}
\begin{flushleft}
\begin{tabular}{lrccl}
\hline\noalign{\smallskip}
Molecule & Frequency& Vres$^1$ & HPBW & Notes \\
 &\multicolumn{1}{c}{(MHz)} & (kms$^{-1}$) & (\arcsec) & \\
\hline\noalign{\smallskip}
\multicolumn{5}{c}{IRAM 30-m} \\
\hline\noalign{\smallskip}
HCO$^+$(1$-$0)                   & 89188.518 & 0.26& 27 & a\\
$^{13}$CH$_3$CN(5$-$4)           & 89331.297 & 3.36& 27 & a,b \\
CH$_3$C$_2$H(6$-$5)              &102547.984 & 0.22& 23 & b  \\
C$^{34}{\rm S}$(3$-$2)           &144617.147 & 0.16& 17 & c\\
CH$_3$OH(V$_t$=1)                &145103.230 & 2.07& 17 & c\\
CS(3$-$2)                        &146969.049 & 0.16& 16 & d\\
CH$_3$CN(8$-$7)                  &147174.592 & 2.04& 16 & b,d \\
C$^{18}{\rm O}$(2$-$1)           &219560.328 & 0.11& 11 \\
$^{13}$CO(2$-$1)                 &220398.686 & 0.11& 11 & e\\
CH$_3$CN(12$-$11)                &220747.268 & 1.36& 11 & b,e \\
\hline\noalign{\smallskip}
\multicolumn{5}{c}{KOSMA 3-m} \\
\hline\noalign{\smallskip}
HCO$^+$(4$-$3)                   & 356734.253 & 0.14 & 74  & \\
\noalign{\smallskip}
\hline
\multicolumn{4}{l}{$^1$\ Highest velocity resolution available} \\
\multicolumn{4}{l}{a,c,d,e\ Transitions measured in same receiver} \\
\multicolumn{4}{l}{b\ Frequency of the K=0 transition} \\
\noalign{\smallskip}
\end{tabular}
\end{flushleft}
\end{table}


\begin{table*}
\caption[]{Observed Sources 
\label{sources}}
\begin{flushleft}
\begin{tabular}{rccrrcrrllll}
\hline\noalign{\smallskip}
\multicolumn{1}{c}{(1)} & \multicolumn{1}{c}{(2)} & \multicolumn{1}{c}{(3)} &
\multicolumn{1}{c}{(4)} & \multicolumn{1}{c}{(5)} & \multicolumn{1}{c}{(6)} &
\multicolumn{1}{c}{(7)} & \multicolumn{1}{c}{(8)} & \multicolumn{1}{c}{(9)} &
\multicolumn{1}{l}{(10)} & \multicolumn{1}{l}{(11)} & 
\multicolumn{1}{l}{(12)} \\
\multicolumn{1}{r}{Mol}& \multicolumn{1}{c}{$\alpha$(1950)$^{\spadesuit}$}&
\multicolumn{1}{c}{$\delta$(1950)$^{\spadesuit}$}& \multicolumn{1}{c}{$\it{l}$} & 
\multicolumn{1}{c}{$\it{b}$} & \multicolumn{1}{c}{IRAS}&
\multicolumn{1}{c}{V$_{\rm lsr}$(NH$_3$)} & \multicolumn{1}{c}{d$_{\rm kin}$} & 
\multicolumn{1}{c}{L$_{\rm fir}$} & \multicolumn{1}{l}{H$_2$O} & 
\multicolumn{1}{l}{Radio$^1$} & \multicolumn{1}{l}{mm$^2$} \\
\multicolumn{1}{r}{\#} & \multicolumn{1}{c}{({\sl h\ m\ s})}&
\multicolumn{1}{c}{(\degr\ \arcmin\ \arcsec)} & \multicolumn{1}{c}{(\degr)} &
\multicolumn{1}{c}{(\degr)}& \multicolumn{1}{c}{}& 
\multicolumn{1}{c}{(kms$^{-1}$)}&
\multicolumn{1}{c}{(kpc)} & \multicolumn{1}{c}{(10$^4$~L$_\odot$)} &
\multicolumn{1}{l}{} & \multicolumn{1}{l}{}& \multicolumn{1}{l}{} \\
\hline\noalign{\smallskip}
$^{\clubsuit}$3 & 00 42 06.3 & +55 30 50 & 122.015 & $-$7.072 & 00420+5530 & $-$51.2 & 7.7 &
  5.15 & Y & N$^3$ & Npp \\
  8 & 05 13 46.0 & +39 19 10 & 168.061 & +0.821 & 05137+3919 & $-$25.4 & 10.8
  & 3.93$^5$ & Y & N$^3$ & Y\\
 59 & 18 27 49.7 & $-$10 09 19 & 21.561 & $-$0.030 & 18278$-$1009 & +93.7 &
 $^{\dagger}$6.6 & 1.46$^{5,6}$ & Y & N & Y\\
$^{\clubsuit}$75 & 18 51 06.4 & +01 46 40 & 34.821 & +0.351 & 18511+0146 & +56.8 & 3.9 &
 1.30$^5$ & N & U & Y\\
 77 & 18 52 46.2 & +03 01 13 & 36.115 & +0.554 & 18527+0301 & +76.0 & 5.3 &
 0.90$^5$& N & N & Y\\
 98 & 19 09 13.0 & +08 41 27 & 43.035 & $-$0.447 & 19092+0841 & +58.0 & 4.5 &
 0.92$^5$ & Y & U$^3$ & Y\\
$^{\clubsuit}$117 & 20 09 54.3 & +36 40 37 & 74.161 & +1.644 & 20099+3640 & $-$36.4 & 8.7 &
 2.51$^5$ & N & Y & Y\\
$^{\clubsuit}$118 & 20 10 38.3 & +35 45 42 & 73.479 & +1.016 & 20106+3545 & +7.8 & 1.6 &
 0.18 & N & N & Npp\\
136 & 21 30 47.1 & +50 49 01 & 94.262 & $-$0.411 & 21307+5049 & $-$46.7 &6.2 &
 1.16$^5$ & Y & U$^3$ & Y\\
$^{\clubsuit}$155 & 23 14 03.4 & +61 21 17 & 111.871 & +0.820 & 23140+6121 &$-$51.5 &5.2 &
 1.06$^5$ & N$^4$ & Y & Y \\
$^{\clubsuit}$160 & 23 38 31.2 & +60 53 43 & 114.531 & $-$0.543 & 23385+6053 & $-$50.0 &
 4.9 & 1.60$^5$ & Y & N & Y \\ 
\noalign{\smallskip}
\hline
\multicolumn{12}{l}{$^{\spadesuit}$\ map center (sub-mm peak Molinari et
al.~\cite{molinari00})}\\
\multicolumn{12}{l}{$^{\clubsuit}$\ peak position also observed in
HCO$^+$(4$-$3); see Sect.~\ref{sobs}} \\
\multicolumn{12}{l}{$^{\dagger}$\ Derived from the present data; see
Sect.~\ref{sindiv}} \\
\multicolumn{12}{l}{$^1$\ U = unassociated radio source is within
3\arcmin\ radius} \\
\multicolumn{12}{l}{$^2$\ Compact mm-emission; Npp = Detected, but missed
primary peak; see Molinari et al. (\cite{molinari00})} \\
\multicolumn{12}{l}{$^3$\ Recently detected with the VLA-D array at 3.6-cm (see
Molinari et al. \cite{molinari00}); detected signal is compatible with the}\\
\multicolumn{12}{l}{$^{\ }$\ non-detections at 2 and 6-cm reported in Molinari et 
al. (\cite{molinari98a})} \\
\multicolumn{12}{l}{$^4$\ Maser discovered by Valdettaro et al. (\cite{u2})} \\
\multicolumn{12}{l}{$^5$\ Luminosity from Molinari et al. (\cite{molinari00})}\\
\multicolumn{12}{l}{$^6$\ Corrected for revised distance} \\
\end{tabular}
\end{flushleft}
\end{table*}


\clearpage
\begin{table*}
\caption[]{Line parameters at peak position
\label{linepars}}
\begin{flushleft}
\begin{tabular}{lclccrccrr}
\hline\noalign{\smallskip}
Line & Vres & rms & V$_{\rm min}$ & V$_{\rm max}$ & T$_{\rm mb}$ & V$_{\rm lsr}$ &
$\Delta v$ & $\int$T$_{\rm mb}$dv$^1$ & $\int$T$_{\rm mb}$dv$^2$ \\
  & (kms$^{-1}$) & (K) & (kms$^{-1}$) & (kms$^{-1}$) & (K) & (kms$^{-1}$) & 
  (kms$^{-1}$) & (Kkms$^{-1}$) & (Kkms$^{-1}$) \\
\hline\noalign{\smallskip}
\multicolumn{10}{c}{Mol~3 (0\arcsec,$-$24\arcsec)} \\
HCO$^+$(1$-$0)   & 0.26 & 0.045 & $-$53.8 & $-$48.3 & 2.4 & $-$51.9 & 1.3
& 6.2 & 6.6 \\
 & & & & & & & 3.2 & & \\
$^{13}$CO(2$-$1) & 0.11 & 0.78 & $-$53.2 & $-$48.4 & 27.0 & $-$52.0 & 1.5
& 67.3 & 68.8 \\
 & & & & & & & 3.1$^3$ & & \\
CS(3$-$2)        & 0.16 & 0.12 & $-$53.2 & $-$48.5 & 2.1 & $-$50.8 & 2.7 &
5.7 & 5.8 \\
C$^{18}$O(2$-$1) & 0.11 & 0.16 & $-$52.8 & $-$49.0 & 6.7 & $-$50.3 & 2.7 &
16.6 & 16.9 \\
C$^{34}$S(3$-$2) & 0.16 & 0.073 & & & & & & & \\
 smoothed:       & 0.65 & 0.039 & $-$52.0 & $-$49.4 & 0.1 & $-$50.4 & 2.6 &
0.3 & 0.4 \\
\noalign{\smallskip}
\hline
\noalign{\smallskip}
\multicolumn{10}{c}{Mol~8 (0\arcsec,0\arcsec)} \\
HCO$^+$(1$-$0)   & 0.26 & 0.082 & $-$30.7 & $-$21.5 & 1.3 & $-$25.3 & 3.4 &
5.3 & 6.4 \\
$^{13}$CO(2$-$1) & 0.11 & 0.25 & $-$29.6 & $-$20.3 & 16.6 & $-$25.4 & 2.8 &
56.6 & 59.4 \\
CS(3$-$2)        & 0.16 & 0.16 & $-$27.8 & $-$22.7 & 4.1 & $-$25.1 & 2.1 &
10.3 & 10.5 \\
C$^{18}$O(2$-$1) & 0.11 & 0.14 & $-$27.3 & $-$22.6 & 3.5 & $-$25.3 & 2.1 &
7.7 & 7.9 \\
C$^{34}$S(3$-$2) & 0.16 & 0.079 & $-$26.3 & $-$23.9 & 1.2 & $-$25.1 & 1.3 &
1.7 & 2.0 \\
\noalign{\smallskip}
\hline
\noalign{\smallskip}
\multicolumn{10}{c}{Mol~59 (12\arcsec,0\arcsec)} \\
HCO$^+$(1$-$0)   & 0.26 & 0.11 & 111.8 & 118.9 & 1.4 & 115.5 & 1.8 & 4.1 &
4.2 \\
 & & & & & & & 4.2$^3$ & & \\
$^{13}$CO(2$-$1) & 0.11 & 0.49 & 107.2 & 120.1 & 9.1 & 115.1 & 5.5 & 58.3 &
61.3 \\
CS(3$-$2)        & 0.16 & 0.32 & 112.3 & 115.8 & 2.4 & 114.8 & 2.9 & 5.4 &
6.8 \\
C$^{18}$O(2$-$1) & 0.11 & 0.25 & 111.0 & 117.1 & 4.9 & 113.4 & 2.0 & 13.3 &
15.8 \\
C$^{34}$S(3$-$2) & 0.16 & 0.12 & 113.6 & 115.1 & 0.4 & 113.9 & 1.9 & 0.5 &
0.8 \\
\noalign{\smallskip}
\hline
\noalign{\smallskip}
\multicolumn{10}{c}{Mol~75 (0\arcsec,0\arcsec)} \\
HCO$^+$(1$-$0)   & 0.26 & 0.069 & 54.7 & 69.1 & 1.6 & 59.6 & 1.8 & 7.4 & 9.8\\
 & & & & & & & 2.9$^3$ & & \\
$^{13}$CO(2$-$1) & 0.11 & 0.55 & 50.5 & 61.3 & 11.5 & 57.0 & 4.4 & 59.2 &
65.3 \\
CS(3$-$2)        & 0.16 & 0.20 & 53.6 & 61.2 & 5.0 & 56.8 & 2.9 & 17.7 & 17.8\\
C$^{18}$O(2$-$1) & 0.11 & 0.17 & 53.0 & 60.4 & 8.1 & 57.3 & 3.3 & 27.9 &
28.3 \\
C$^{34}$S(3$-$2) & 0.16 & 0.094 & 56.2 & 59.3 & 0.7 & 57.0 & 1.9 & 1.4 & 1.6\\
\noalign{\smallskip}
\hline
\noalign{\smallskip}
\multicolumn{10}{c}{Mol~77 (0\arcsec,0\arcsec)} \\
HCO$^+$(1$-$0)   & 0.26 & 0.090 & 72.1 & 79.1 & 1.4 & 75.2 & 1.8 & 4.0 & 4.3\\
$^{13}$CO(2$-$1) & 0.11 & 0.35 & 71.9 & 82.1 & 10.1 & 75.1 & 4.2 & 42.5 &
42.9 \\
CS(3$-$2)        & 0.16 & 0.25 & 72.8 & 78.0 & 2.6 & 75.5 & 2.9 & 7.4 & 7.8 \\
C$^{18}$O(2$-$1) & 0.11 & 0.22 & 73.8 & 78.3 & 8.5 & 75.7 & 2.4 & 20.8 &
21.1 \\
C$^{34}$S(3$-$2) & 0.16 & 0.073 & 74.1 & 77.5 & 1.0 & 76.4 & 1.6 & 1.9 & 2.4\\
\noalign{\smallskip}
\hline
\multicolumn{10}{l}{$^1$\ Integral taken between V$_{\rm min}$ and V$_{\rm max}$} \\
\multicolumn{10}{l}{$^2$\ Integral taken between velocities where
T$_{\rm mb}$=0~K} \\
\multicolumn{10}{l}{$^3$\ In line profiles with a `dip' which is $<$
0.5T$_{\rm peak}$, the $\Delta v$ on the second line is the FWHM taken over} \\
\multicolumn{10}{l}{$^{\ }$\ the {\it whole} line} \\
\noalign{\smallskip}
\end{tabular}
\end{flushleft}
\end{table*}

\addtocounter{table}{-1}

\clearpage
\begin{table*}
\caption[]{({\it continued})
}
\begin{flushleft}
\begin{tabular}{lclccrccrr}
\hline\noalign{\smallskip}
Line & Vres & rms & V$_{\rm min}$ & V$_{\rm max}$ & T$_{\rm mb}$ & V$_{\rm lsr}$ &
$\Delta v$ & $\int$T$_{\rm mb}$dv$^1$ & $\int$T$_{\rm mb}$dv$^2$ \\
  & (kms$^{-1}$) & (K) & (kms$^{-1}$) & (kms$^{-1}$) & (K) & (kms$^{-1}$) & 
  (kms$^{-1}$) & (Kkms$^{-1}$) & (Kkms$^{-1}$) \\
\hline\noalign{\smallskip}
\multicolumn{10}{c}{Mol~98 (24\arcsec,0\arcsec)} \\
HCO$^+$(1$-$0)   & 0.26 & 0.041 & 45.1 & 61.4 & 2.4 & 56.4 & 5.7 & 14.8 &
15.0 \\
$^{13}$CO(2$-$1) & 0.11 & 0.22 & 53.0 & 63.2 & 17.7 & 57.0 & 4.0 & 74.6 &
75.1 \\
CS(3$-$2)        & 0.16 & 0.096 & 52.2 & 63.8 & 3.1 & 56.9 & 6.1 & 18.9 &
20.2 \\
C$^{18}$O(2$-$1) & 0.11 & 0.13 & 53.7 & 62.3 & 9.0 & 58.3 & 3.6 & 33.7 &
33.9 \\
C$^{34}$S(3$-$2) & 0.16 & 0.070 & 54.0 & 60.5 & 1.0 & 57.7 & 4.1 & 4.0 & 4.3\\
\noalign{\smallskip}
\hline
\noalign{\smallskip}
\multicolumn{10}{c}{Mol~117 (0\arcsec,$-$12\arcsec)} \\
HCO$^+$(1$-$0)   & 0.26 & 0.034 & $-$40.1 & $-$33.5 & 2.4 & $-$35.8 & 2.6 &
6.7 & 6.8 \\
$^{13}$CO(2$-$1) & 0.11 & 0.14 & $-$39.7 & $-$33.3 & 25.3 & $-$35.6 & 2.6 &
66.8 & 66.9 \\
CS(3$-$2)        & 0.16 & 0.085 & $-$38.1 & $-$34.1 & 2.7 & $-$35.7 & 1.8 &
5.1 & 5.4 \\
C$^{18}$O(2$-$1) & 0.11 & 0.12 & $-$38.3 & $-$34.2 & 3.3 & $-$35.8 & 1.8 &
6.6 & 6.7 \\
C$^{34}$S(3$-$2) & 0.16 & 0.057 & $-$36.0 & $-$34.9 & 0.2 & $-$35.5 & 1.0 &
0.24 & 0.25 \\
\noalign{\smallskip}
\hline
\noalign{\smallskip}
\multicolumn{10}{c}{Mol~118 (12\arcsec,0\arcsec)} \\
HCO$^+$(1$-$0)   & 0.26 & 0.048 & 5.2 & 11.2 & 3.7 & 7.7 & 2.4 & 8.6 & 8.7 \\
$^{13}$CO(2$-$1) & 0.11 & 0.39 & 5.4 & 10.7 & 28.0 & 7.6 & 2.5 & 67.7 & 68.4\\
CS(3$-$2)        & 0.16 & 0.13 & 6.2 & 10.0 & 4.0 & 7.7 & 1.9 & 7.7 & 7.9 \\
C$^{18}$O(2$-$1) & 0.11 & 0.19 & 6.2 & 9.6 & 12.4 & 7.7 & 1.7 & 21.2 & 21.5\\
C$^{34}$S(3$-$2) & 0.16 & 0.058 & 6.9 & 9.1 & 0.4 & 7.8 & 1.8 & 0.57 & 0.56 \\
\noalign{\smallskip}
\hline
\noalign{\smallskip}
\multicolumn{10}{c}{Mol~136 (0\arcsec,0\arcsec)} \\
HCO$^+$(1$-$0)   & 0.26 & 0.041 & $-$52.8 & $-$43.0 & 1.5 & $-$46.7 & 3.4 &
5.8 & 6.4 \\
$^{13}$CO(2$-$1) & 0.11 & 0.54 & $-$49.4 & $-$44.1 & 15.7 & $-$46.8 & 2.3 &
40.3 & 41.8 \\
CS(3$-$2)        & 0.16 & 0.098 & $-$49.6 & $-$43.7 & 3.0 & $-$46.8 & 3.2 &
9.3 & 10.1 \\
C$^{18}$O(2$-$1) & 0.11 & 0.11 & $-$49.1 & $-$43.4 & 6.3 & $-$46.3 & 1.8 &
12.7 & 12.7 \\
C$^{34}$S(3$-$2) & 0.16 & 0.054 & $-$47.9 & $-$44.9 & 0.5 & $-$46.2 & 1.9 &
1.0 & 1.1 \\
\noalign{\smallskip}
\hline
\noalign{\smallskip}
\multicolumn{10}{c}{Mol~155 (12\arcsec,$-$12\arcsec)} \\
HCO$^+$(1$-$0)   &0.26 & 0.11 & $-$53.6 & $-$47.6 & 2.0 & $-$52.0 & 1.3 &
4.6 & 4.9 \\
$^{13}$CO(2$-$1) & 0.11 & 0.47 & $-$53.6 & $-$47.2 & 20.9 & $-$51.9 & 3.1 &
61.5 & 62.2 \\
CS(3$-$2)        & 0.16 & 0.21 & $-$52.8 & $-$49.2 & 2.8 & $-$51.6 & 2.1 &
6.2 & 6.6 \\
C$^{18}$O(2$-$1) & 0.11 & 0.14 & $-$53.0 & $-$49.3 & 8.3 & $-$51.3 & 1.7 &
15.3 & 15.5 \\
C$^{34}$S(3$-$2) & 0.16 & 0.075 & $-$51.6 & $-$50.1 & 0.3 & $-$51.4 & 1.5 &
0.38 & 0.61 \\
\noalign{\smallskip}
\hline
\noalign{\smallskip}
\multicolumn{10}{c}{Mol~160 (0\arcsec,0\arcsec)} \\
HCO$^+$(1$-$0)   & 0.26 & 0.044 & $-$56.4 & $-$44.2 & 5.8 & $-$50.3 & 2.3 &
17.5 & 17.6 \\
$^{13}$CO(2$-$1) & 0.11 & 0.32 & $-$54.8 & $-$44.8 & 27.0 & $-$50.2 & 3.7 &
107.2 & 107.7 \\
 & & & & & & & 4.0$^3$ & & \\
CS(3$-$2)        & 0.16 & 0.12 & $-$55.1 & $-$45.2 & 6.0 & $-$50.0 & 2.9 &
21.2 & 22.2 \\
C$^{18}$O(2$-$1) & 0.11 & 0.12 & $-$52.3 & $-$46.4 & 5.6 & $-$50.1 & 2.5 &
18.5 & 18.8 \\
C$^{34}$S(3$-$2) & 0.16 & 0.065 & $-$52.1 & $-$47.0 & 1.1 & $-$50.1 & 2.0 &
2.6 & 3.0 \\
\noalign{\smallskip}
\hline
\noalign{\smallskip}
\end{tabular}
\end{flushleft}
\end{table*}


\clearpage
\begin{table*}
\caption[]{ Gaussfit parameters of CH$_3$C$_2$H(6$-$5) lines 
\label{ch3c2hfits}
}
\begin{flushleft}
\begin{tabular}{rccrccccc}
\hline\noalign{\smallskip}
\multicolumn{5}{c}{} & \multicolumn{4}{c}{$\int$T$_{\rm mb}$dv (Kkms$^{-1}$)} \\
\cline{6-9}
Mol & offset & Vres & V$_{\rm lsr}$ & $\Delta v$ & K=0 & 1 & 2 & 3 \\
\#  & (\arcsec) & (kms$^{-1}$) & (kms$^{-1}$) & (kms$^{-1}$) &  & & & \\
\hline\noalign{\smallskip}
 59 & 12,0& 0.23 & +113.9$\pm$0.1 & 2.98$\pm$0.09 & 0.644$\pm$0.035 & 
 0.490$\pm$0.033 & 0.208$\pm$0.032 & \\
    &     & 2.92 & +114.0$\pm$0.2 & 4.02$\pm$0.16 & 0.619$\pm$0.041 &
 0.479$\pm$0.037 & 0.118$\pm$0.036 & \\
\noalign{\smallskip}
\hline
\noalign{\smallskip}
 75 & 0,0 & 0.23 & +57.28$\pm$0.03 & 2.78$\pm$0.04 & 1.311$\pm$0.031 &
 1.019$\pm$0.030 & 0.325$\pm$0.028 & 0.133$\pm$0.028 \\
    &     & 2.92 & +57.18$\pm$0.11 & 3.62$\pm$0.24 & 1.253$\pm$0.044 & 
 1.025$\pm$0.043 & 0.332$\pm$0.042 & 0.118$\pm$0.041 \\
\noalign{\smallskip}
\hline
\noalign{\smallskip}
 77 & 0,0 & 0.23 & +76.04$\pm$0.04 & 1.52$\pm$0.05 & 0.496$\pm$0.030 & 
 0.464$\pm$0.030 & 0.098$\pm$0.027 & 0.058$\pm$0.028 \\
    &     & 2.92 & +76.04$\pm$0.05 & 2.36$\pm$0.58 & 0.499$\pm$0.226 & 
 0.574$\pm$0.266 & 0.130$\pm$0.056 & 0.041$\pm$0.033 \\
\noalign{\smallskip}
\hline
\noalign{\smallskip}
 98 &24,0 & 0.23 & +57.79$\pm$0.03 & 3.06$\pm$0.04 & 1.681$\pm$0.040 & 
 1.319$\pm$0.036 & 0.691$\pm$0.035 & 0.294$\pm$0.035 \\
    &     & 2.92 & +57.68$\pm$0.07 & 4.33$\pm$0.14 & 1.801$\pm$0.054 & 
 1.428$\pm$0.055 & 0.735$\pm$0.054 & 0.310$\pm$0.051 \\
    &36,0 & 0.23 & +57.48$\pm$0.21 & 3.89$\pm$0.23 & 2.112$\pm$0.256 & 
 1.922$\pm$0.252 & 1.227$\pm$0.247 & 0.563$\pm$0.242 \\
    &     & 2.92 & +57.43$\pm$0.21 & 4.75$\pm$0.26 & 2.431$\pm$0.208 & 
 2.015$\pm$0.208 & 1.303$\pm$0.205 & 0.875$\pm$0.200 \\
   &36,$-$12&2.92& +56.83$\pm$0.41 & 3.77$\pm$0.26 & 2.445$\pm$0.346 & 
 1.002$\pm$0.307 & 0.954$\pm$0.308 & 0.723$\pm$0.339 \\
   & 36,12  &2.92& +56.43$\pm$0.49 & 4.96$\pm$0.48 & 1.278$\pm$0.267 & 
 0.497$\pm$0.250 & 1.071$\pm$0.260 & 0.337$\pm$0.250 \\
   &24,$-$12&2.92& +56.32$\pm$0.61 & 4.45$\pm$0.47 & 1.252$\pm$0.425 & 
 1.196$\pm$0.428 & 1.638$\pm$0.398 & \\
\noalign{\smallskip}
\hline
\noalign{\smallskip}
118 &12,0 & 0.23 & +7.92$\pm$0.04 & 1.55$\pm$0.06 & 0.410$\pm$0.027 & 
0.291$\pm$0.023 & 0.046$\pm$0.022 & \\
    &     & 2.92 & +8.16$\pm$0.28 & 3.96$\pm$0.56 & 0.389$\pm$0.045 & 
 0.213$\pm$0.041 & 0.067$\pm$0.040 & 0.074$\pm$0.040 \\
\noalign{\smallskip}
\hline
\noalign{\smallskip}
160 & 0,0 & 0.23 & $-$50.45$\pm$0.06 & 1.85$\pm$0.08 & 0.277$\pm$0.028 &
0.314$\pm$0.027 & 0.142$\pm$0.025 & 0.048$\pm$0.026 \\
    &     & 2.92 & $-$50.48$\pm$0.25 & 4.17$\pm$0.41 & 0.410$\pm$0.050 &
0.468$\pm$0.054 & 0.185$\pm$0.050 & 0.065$\pm$0.046 \\
\noalign{\smallskip}
\hline
\noalign{\smallskip}
\end{tabular}
\end{flushleft}
\end{table*}


\begin{table*}
\caption[]{ Gaussfit parameters of CH$_3$CN(8$-$7) lines 
\label{ch3cnfits}
}
\begin{flushleft}
\begin{tabular}{rccrccccc}
\hline\noalign{\smallskip}
\multicolumn{5}{c}{} & \multicolumn{4}{c}{$\int$T$_{\rm mb}$dv (Kkms$^{-1}$)} \\ 
\cline{6-9}
Mol & offset & Vres & V$_{\rm lsr}$ & $\Delta v$ & K=0 & 1 & 2 & 3 \\
\#  & (\arcsec) & (kms$^{-1}$) & (kms$^{-1}$) & (kms$^{-1}$) &  & & & \\
\hline\noalign{\smallskip}
 98 &24,0 & 0.64 & +58.31$\pm$0.28 & 4.85$\pm$0.36 & 0.766$\pm$0.114 &
 0.882$\pm$0.117 & 0.511$\pm$0.109 & 0.670$\pm$0.112 \\
    &     & 2.04 & +58.12$\pm$0.40 & 5.65$\pm$0.51 & 0.931$\pm$0.151 &
 0.771$\pm$0.151 & 0.585$\pm$0.135 & 0.738$\pm$0.141 \\
\noalign{\smallskip}
\hline
\noalign{\smallskip}
\end{tabular}
\end{flushleft}
\end{table*}


\begin{table*}
\caption[]{ Deconvolved source equivalent diameter (arcsec, pc) in different 
lines, and mass$^1$
\label{deconsize}
}
\begin{flushleft}
\begin{tabular}{rrrrrrl}
\hline\noalign{\smallskip}
Mol &d & HCO$^+$(1$-$0) & $^{13}$CO(2$-$1) & CS(3$-$2) & C$^{18}$O(2$-$1)$^2$ &
Mass$^1$ \\
\# & (kpc)  & (\arcsec, pc) & (\arcsec, pc) & (\arcsec, pc) & (\arcsec, pc) &
(10$^3$~M$_{\odot}$)\\
\hline\noalign{\smallskip}
 3 &7.72 & 46.2, 1.73 & $\gtrsim$59.3, $\gtrsim$2.22 & 39.3, 1.47 & n.d.&
2.8\\
 8 &10.80& 22.7, 1.19 & $\gtrsim$37.0, $\gtrsim$1.94 & 17.4, 0.91 &
n.d.&1.6 \\
59 &6.57 & $\gtrsim$35.0, $\gtrsim$1.11 & $>$24.7, $>$0.79 & 13.3, 0.42 &
n.d. & 0.25\\
75 &3.86 & $\gtrsim$70.4, $\gtrsim$1.32 & $>$93.4, $>$1.75 & 28.4, 0.53 &
n.d. & 1.3\\
77 & 5.26& $\gtrsim$34.2, $\gtrsim$0.87 & $>$51.7, $>$1.32 & 21.0, 0.54 &
n.d. & 0.34\\
98 & 4.48& $\gtrsim$26.9, $\gtrsim$0.58 & 39.4, 0.86 & 21.3, 0.46 & n.d. &
0.42 \\
117& 8.66& $\gtrsim$40.6, $\gtrsim$1.70 & $\gtrsim$46.2, $\gtrsim$1.94 &
25.3, 1.06 & n.d. &1.7 \\
118& 1.64& $\sim$43.5, $\sim$0.35 & $>$66.3, $>$0.53 & 40.2, 0.32 & n.d. &
0.13\\
136& 6.22& 30.0, 0.90 & 34.1, 1.03 & 17.4, 0.52 & $>$20.9, $>$0.63 &0.37 \\
155& 5.20& $>$80.7, $>$2.03 & $>$83.9, $>$2.12 & $\gtrsim$51.4,
$\gtrsim$1.30 & n.d. &2.2 \\
160& 4.90& 45.7, 1.09 & $>$65.0, $>$1.54 & 29.6, 0.70 & n.d. & 2.0 \\
\noalign{\smallskip}
\hline
\multicolumn{7}{l}{$^1$\ Lower limit for the total molecular mass (including 
correction for He), estimated from}\\
\multicolumn{7}{l}{$^{\ }$\ $^{13}$CO emission above the FWHM level. For 
details of mass estimates, see Sect.~\ref{sindiv}} \\
\multicolumn{7}{l}{$^2$\ n.d. = not determinable: most of half-power contour
outside mapped region} \\
\noalign{\smallskip}
\end{tabular}
\end{flushleft}
\end{table*}


\clearpage
\begin{table}
\caption[]{ Temperature and column densities, derived from the Boltzmann
plots of CH$_3$C$_2$H(6--5) (see Fig.~\ref{mol_boltz})
\label{ch3c2h_results}
}
\begin{flushleft}
\begin{tabular}{rcccccc}
\hline\noalign{\smallskip}
Mol & $\Delta \alpha, \Delta \delta$ & res$^1$ & T$_{\rm k}$ &
$\delta$T$_{\rm k}$ & N &
$\delta$N \\
\# & (\arcsec, \arcsec) & kms$^{-1}$ & K & K & 10$^{14}$~cm$^{-2}$ & 
10$^{13}$~cm$^{-2}$ \\
\hline\noalign{\smallskip}
  59 & 12,0 & 0.23 & 28.6 & 0.0 & 0.353 & 0.002 \\
     &      & 2.92 & 21.8 & 1.2 & 0.269 & 0.254 \\
  75 & 0,0  & 0.23 & 27.3 & 0.5 & 0.692 & 0.207 \\
     &      & 2.92 & 27.7 & 0.6 & 0.686 & 0.252 \\
  77 & 0,0  & 0.23 & 32.1 & 3.0 & 0.334 & 0.507 \\
     &      & 2.92 & 24.0 & 3.4 & 0.267 & 0.761 \\
  98 & 24,0 & 0.23 & 40.4 & 0.3 & 1.351 & 0.139 \\
     &      & 2.92 & 40.0 & 0.4 & 1.424 & 0.202 \\
 118 & 12,0 & 0.23 & 18.4 & 2.0 & 0.151 & 0.284 \\
     &      & 2.92 & 45.9 & 5.3 & 0.446 & 0.814 \\
 160 & 0,0  & 0.23 & 48.4 & 3.6 & 0.326 & 0.387 \\
     &      & 2.92 & 45.9 & 5.3 & 0.446 & 0.814 \\
\noalign{\smallskip}
\hline
\multicolumn{7}{l}{$^1$\ Resolution of the spectrum} \\
\noalign{\smallskip}
\end{tabular}
\end{flushleft}
\end{table}


\begin{table}
\caption[]{ Adopted T$_{\rm ex}$ and estimated masses of clumps associated 
with the IRAS sources
\label{clmass}}
\begin{flushleft}
\begin{tabular}{rcrl}
\hline\noalign{\smallskip}
Mol &T$_{\rm ex}$ & Mass$^1$ \\
\# & (K)  & (M$_{\odot}$) & Associated velocity component \\
\hline\noalign{\smallskip}
 3 & 35 & 2140 & blue (V$_{\rm lsr} < -50.85$) \\
   &    &  534 & red (V$_{\rm lsr} > -50.85$) \\
 8 & 25 &  238 & red (V$_{\rm lsr} \approx -25.5$) \\
59 & 15 &   65 & red (114.3$<$ V$_{\rm lsr} < 117.0$) \\
75 & 25 &  504 & blue (55.0 $<$  V$_{\rm lsr} < 58.6$)\\
77 & 25 &  187 & blue (73.8 $<$  V$_{\rm lsr} < 76.3$)\\
98 & 35 &  424 & (V$_{\rm lsr} \approx 57.0$) \\
117& 40 & 1124 & red ($-36.9 <$  V$_{\rm lsr} < -33.0$)\\
118& 35 &   93 & blue (6.5 $<$  V$_{\rm lsr} < 8.5$)\\
136& 30 &  370 & (V$_{\rm lsr} \approx -46.8$)\\
155& 30 &  968 & blue (V$_{\rm lsr} < -51.1$)\\
160& 35 &  621 & main (V$_{\rm lsr} \approx -50.2$) \\
\noalign{\smallskip}
\hline
\multicolumn{4}{l}{$^1$\ Lower limit, derived from the average H$_2$ column}\\
\multicolumn{4}{l}{\ density, estimated from the $^{13}$CO emission above}\\
\multicolumn{4}{l}{\ the FWHM level, and the beam-corrected area of the}\\
\multicolumn{4}{l}{\ emission.}\\
\noalign{\smallskip}
\end{tabular}
\end{flushleft}
\end{table}


\clearpage
\begin{table}
\caption{Comparison between properties of the {\sl Low} sources observed in 
 this
 study with those of the {\sl High} sources observed by Cesaroni et al. (1999).
 The pair of numbers indicates the minimum and maximum values of the
 corresponding quantity, while the number below each pair is the mean value}
\label{tcomp}
\begin{tabular}{lcc}
\hline\noalign{\smallskip}
	&	Low	&	High	\\
\hline\noalign{\smallskip}
$d$(kpc)	&	1.6--10.8&	0.7--3.5	\\
	&	5.9	&	1.5	\\
Log$[L(L_\odot)]$&	3.3--4.7	&	2.6--4.5	\\
\hline\noalign{\smallskip}
\multicolumn{3}{c}{CS(3--2)} \\
$\Theta(\arcsec)$&	13--51	&	18--38	\\
	&	28	&	29	\\
D(pc)	&	0.32--1.46&	0.13--0.31\\
	&	0.75	&	0.22	\\
$\rm T_{mb}$(K)	&	2.1--6.0	&	3.2--12.9\\
	&	3.4	&	9.4	\\
$\rm T_b$(K)	&	2.5--8.2	&	3.9--18.0\\
	&	5.4	&	13.5	\\
FWZI(km/s)&	3.5--11.6&	18--52	\\
	&	5.9	&	33	\\
FWHM(km/s)&	1.8--6.1	&	2.4--4.4	\\
	&	2.9	&	3.4	\\
$\rm M_{vir}(M_\odot)$&	121--1797&	95--479	\\
	&	629	&	276	\\
$\rm M_{CD}(M_\odot)$	&	7.6--165	&	3.1--55	\\
	&	92.3	&	31.5	\\
$\rm M_{CD}^\ast(M_\odot)$&	16--1167	&	12--177	\\
	&	405	&	81	\\
$\rm M_{vir}/M_{CD}$&	2.6--24.6&	4.8--30.6\\
	&	9.0	&	13.6	\\
$\rm M_{vir}/M_{CD}^\ast$&	0.4--7.6	&	2.3--8.0	\\
	&	2.5	&	4.8	\\
\hline\noalign{\smallskip}
\multicolumn{3}{c}{C$^{34}$S(3--2)} \\
$\Theta(\arcsec)^{\clubsuit}$&	8--17	&	14--34	\\
	&	13	&	23	\\
D(pc)$^{\clubsuit}$	&	0.22--0.9	&	0.12--0.36\\
	&	0.44	&	0.21	\\
$\rm T_{mb}$(K)	&	0.1--1.2	&	0.5--2.5	\\
	&	0.6	&	1.4	\\
$\rm T_b$(K)$^{\clubsuit}$	&	1.3--2.4	&	1.2--3.2	\\
	&	1.8	&	2.4	\\
FWHM(km/s)&	1.0--4.1	&	2.2--4.2	\\
	&	2.0	&	3.1	\\
$\rm M_{vir}(M_\odot)^{\clubsuit}$&	59--160	&	87--370	\\
	&	202	&	227	\\
$\rm M_{CD}(M_\odot)^{\clubsuit}$	&	86--229	&	41--354	\\
	&	143	&	170	\\
$\rm M_{vir}/M_{CD}^{\clubsuit}$&	0.7--3.5	&	0.38--5.0\\
	&	1.5	&	2.1	\\
\noalign{\smallskip}
\hline
\multicolumn{3}{l}{$^\ast$~mass obtained after correcting for line optical
depth:} \\
\multicolumn{3}{l}{\ $\tau_{\rm CS} / [1 - exp(-\tau_{\rm CS})]$, with
$\tau_{\rm CS}$ from T(CS)/T(C$^{34}$S)}\\
\multicolumn{3}{l}{\ and $\tau_{\rm C^{34}S} = \tau_{\rm
CS}$ / f, f=[CS]/[C$^{34}$S] = 22} \\
\multicolumn{3}{l}{$^{\clubsuit}$~Values for {\sl Low} based on Mol~8, 77,
98, 160 only}\\
\noalign{\smallskip}
\end{tabular}
\end{table}
\addtocounter{table}{-1}
\begin{table}
\caption{continued}
\begin{tabular}{lcc}
\hline\noalign{\smallskip}
\multicolumn{3}{c}{$^{13}$CO(2--1)} \\
$\Theta(\arcsec)$&	25--93	&	30--95	\\
	&	55	&	72	\\
D(pc)	&	0.51--2.21&	0.17--0.83\\
	&	1.5	&	0.47	\\
$\rm T_{mb}$(K)	&	9.1--28.0&	14.8--38.2\\
	&	19.0	&	26.6	\\
FWZI(km/s)&	4.8--12.9&	7.6--14.4\\
	&	8.3	&	11.2	\\
FWHM(km/s)&	1.5--5.5	&	2.8--4.4	\\
	&	3.3	&	3.5	\\
$\rm M_{vir}(M_\odot)$&	348--3557&	160--1591\\
	&	1702	&	628	\\
\hline\noalign{\smallskip}
\multicolumn{3}{c}{HCO$^+$(1--0)} \\
$\Theta(\arcsec)$&	23--81	&	26--41	\\
	&	43	&	35	\\
D(pc)	&	0.34--2.03&	0.12--0.63\\
	&	1.17	&	0.30	\\
$\rm T_{mb}$(K)	&	1.3--5.8	&	1.5--13.2\\
	&	2.4	&	7.4	\\
FWZI(km/s)&	5.5--16.3&	34.0--108.3\\
	&	9.1	&	57.4	\\
FWHM(km/s)&	1.3--5.7	&	1.3--4.5	\\
	&	2.5	&	3.2	\\
$\rm M_{vir}(M_\odot)$&	212--1979&	22--1335	\\
	&	757	&	440	\\
$\rm M_{CD}(M_\odot)$	&	97--1838	&	32--360	\\
	&	922 	&	223	\\
$\rm M_{vir}/M_{CD}$&	0.2--2.7	&	0.66--5.8\\
	&	1.1 	&	2.3	\\
\noalign{\smallskip}
\hline
\end{tabular}
\end{table}

%



\clearpage

\addtocounter{figure}{-1}

\begin{figure}
 \resizebox{\hsize}{!}{\includegraphics{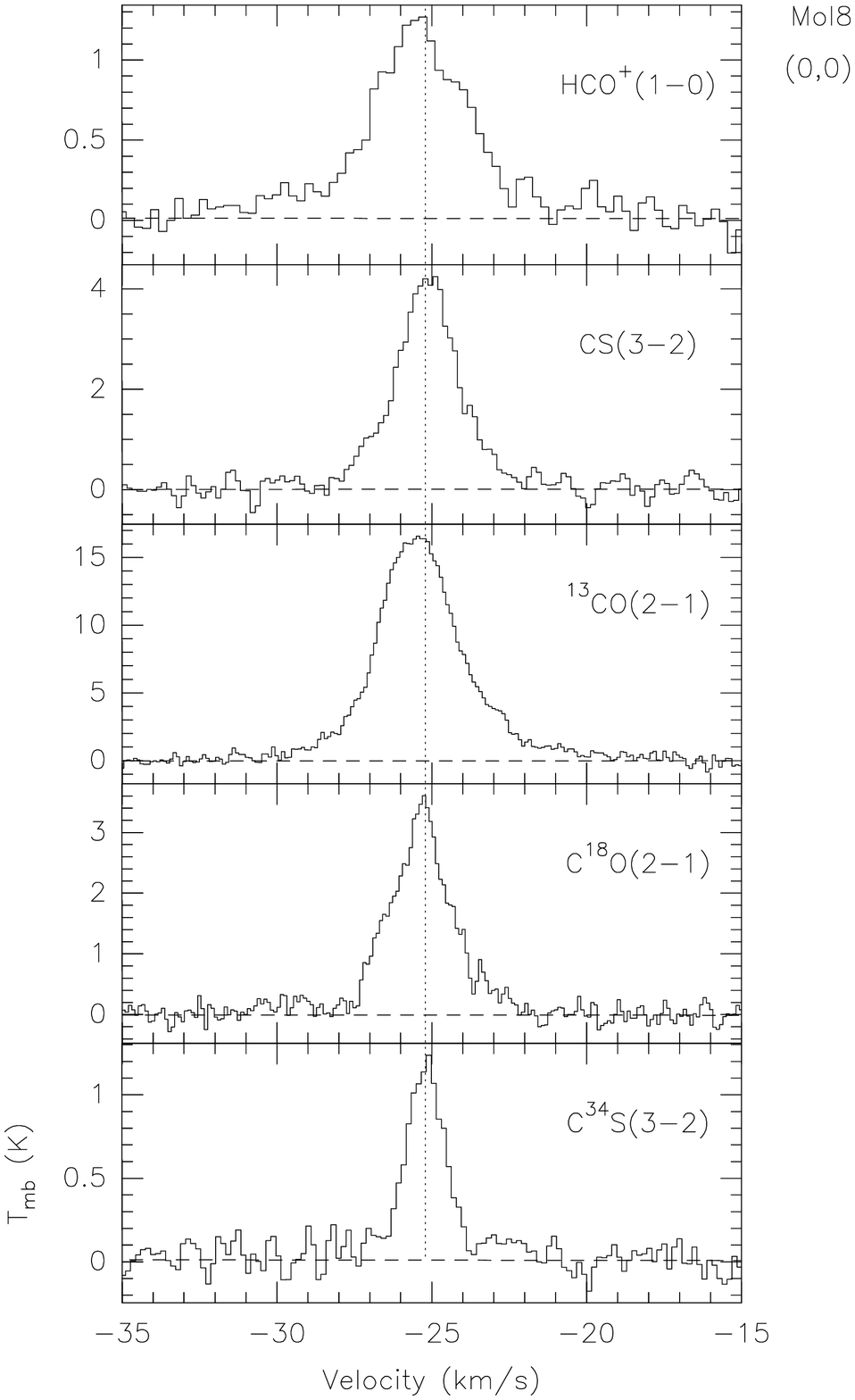}}
 \hfill
 \caption{{\bf b}\ As {\bf a}, for Mol~8}
\end{figure}

\addtocounter{figure}{-1}

\clearpage
\begin{figure}
 \resizebox{\hsize}{!}{\includegraphics{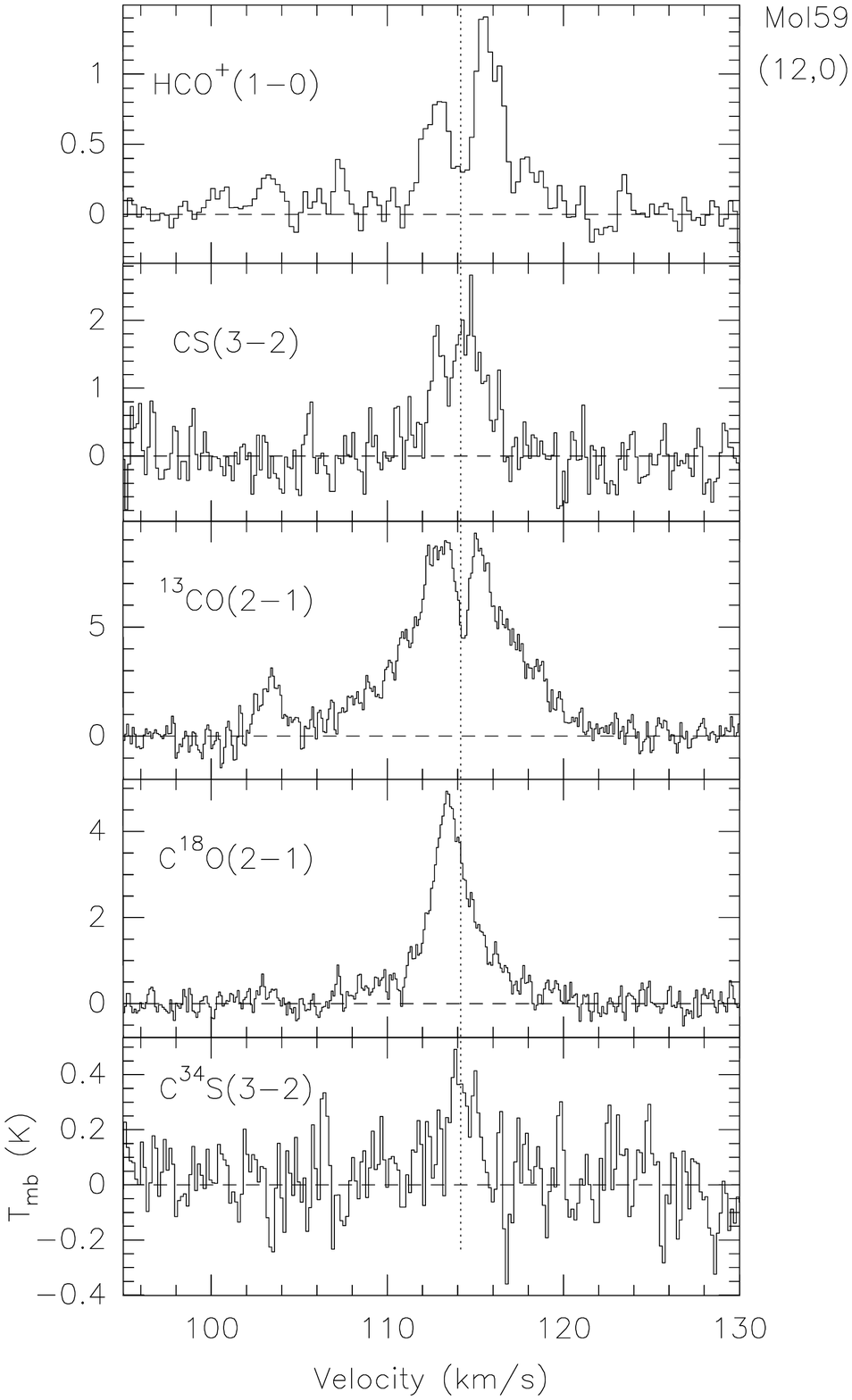}}
 \hfill
 \caption{{\bf c}\ As {\bf a}, for Mol~59}
\end{figure}

\addtocounter{figure}{-1}

\begin{figure}
 \resizebox{\hsize}{!}{\includegraphics{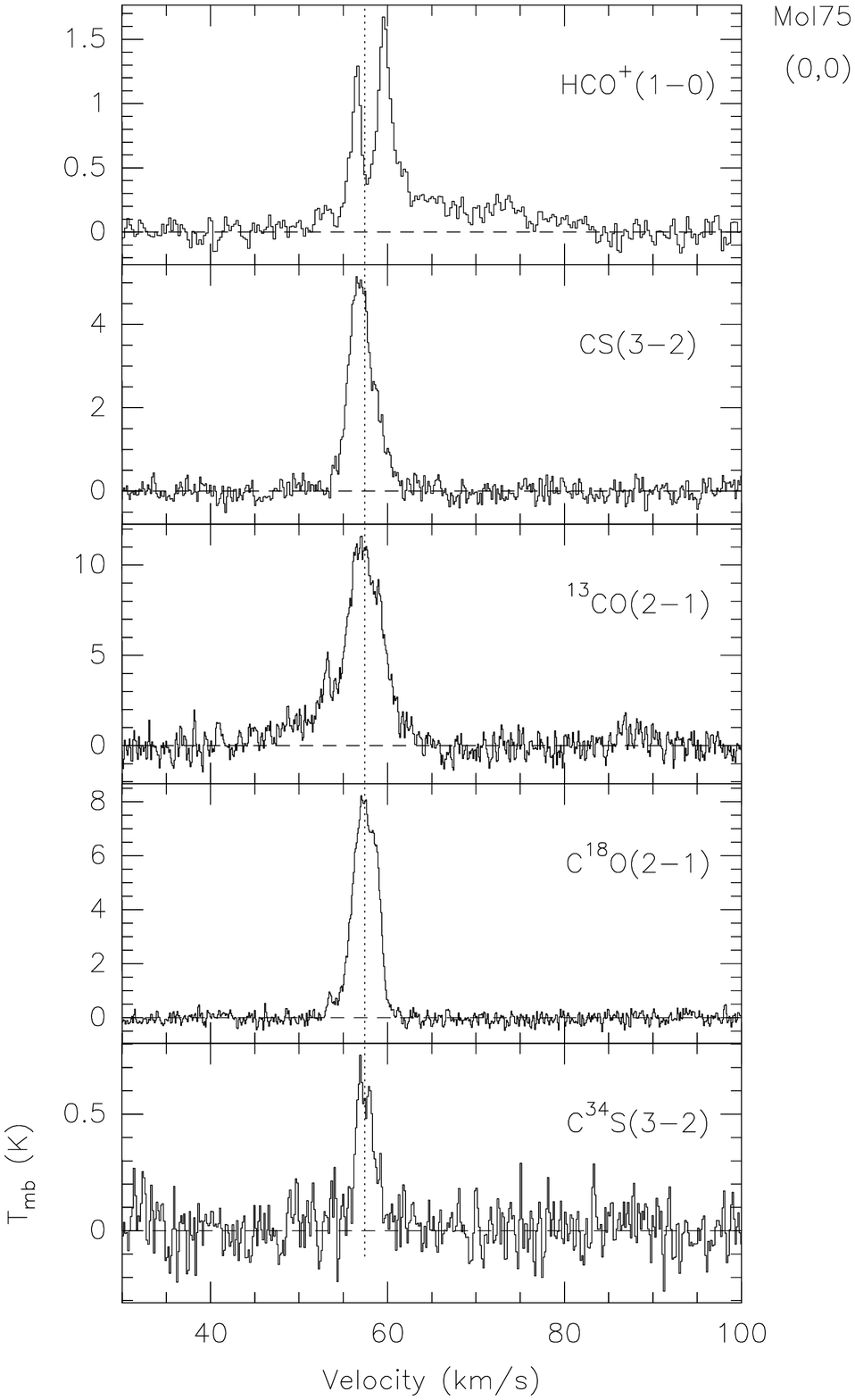}}
 \hfill
 \caption{{\bf d}\ As {\bf a}, for Mol~75}
\end{figure}

\addtocounter{figure}{-1}

\clearpage
\begin{figure}
 \resizebox{\hsize}{!}{\includegraphics{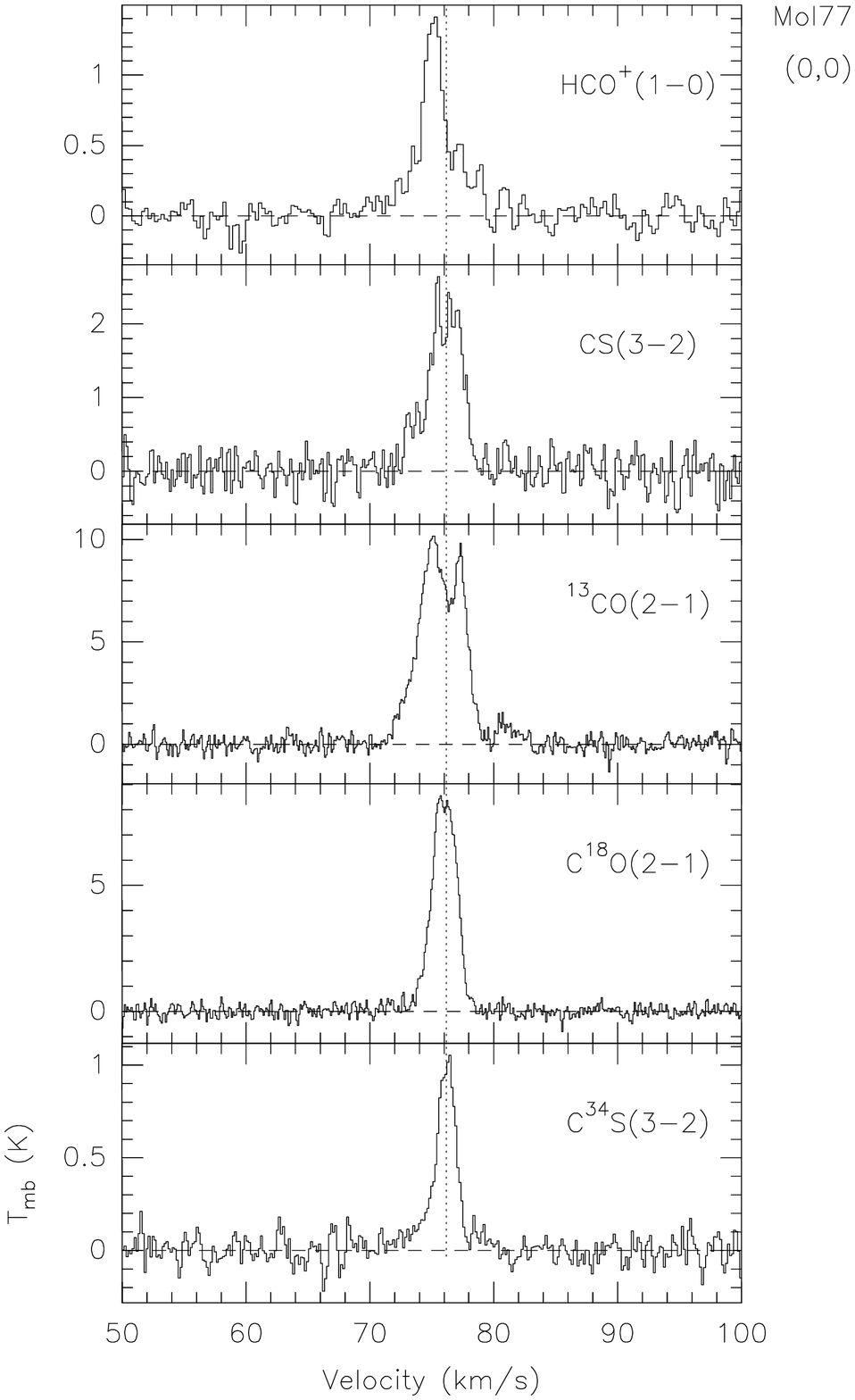}}
 \hfill
 \caption{{\bf e}\ As {\bf a}, for Mol~77}
\end{figure}

\addtocounter{figure}{-1}

\begin{figure}
 \resizebox{\hsize}{!}{\includegraphics{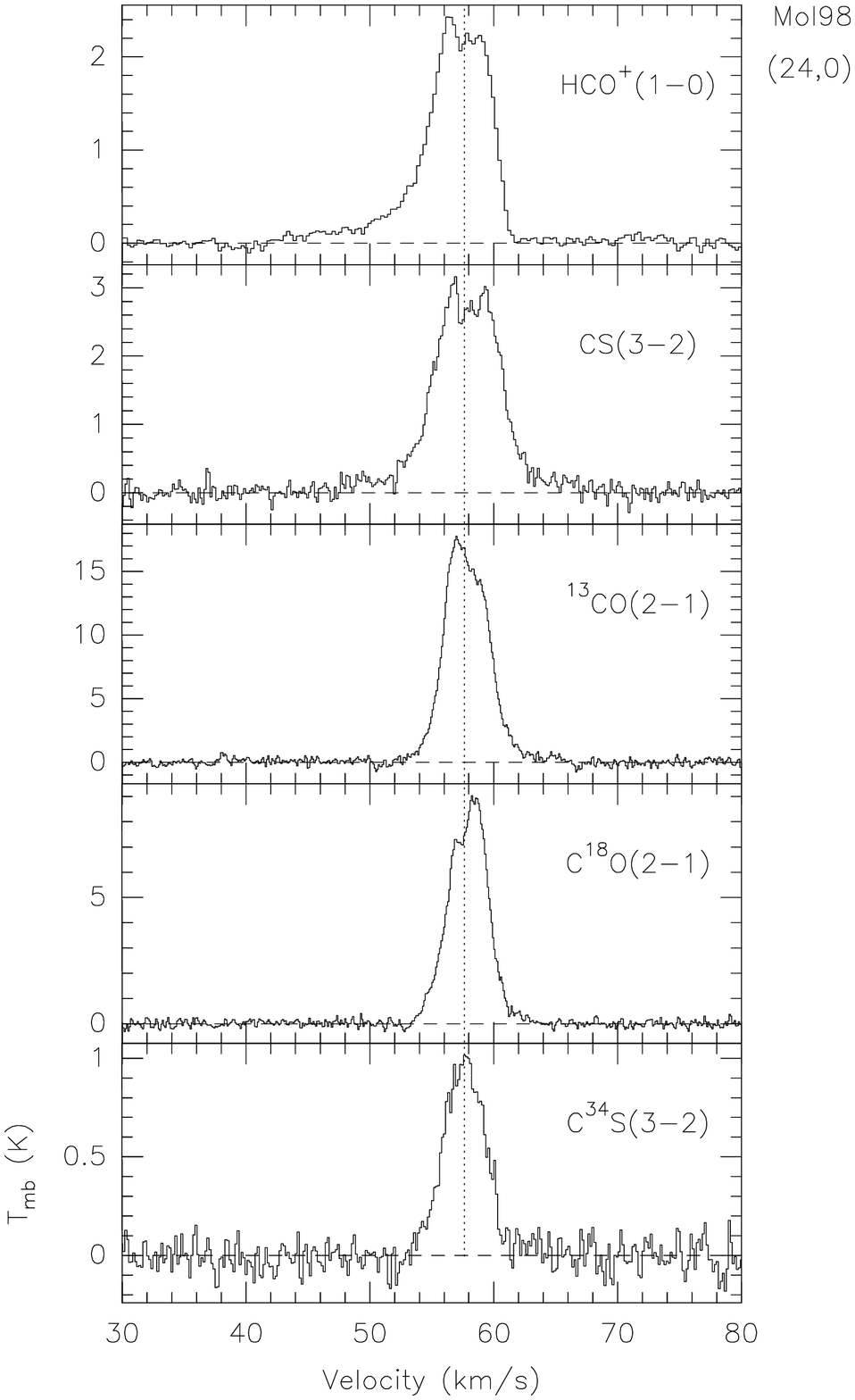}}
 \hfill
 \caption{{\bf f}\ As {\bf a}, for Mol~98}
\end{figure}

\addtocounter{figure}{-1}

\clearpage
\begin{figure}
 \resizebox{\hsize}{!}{\includegraphics{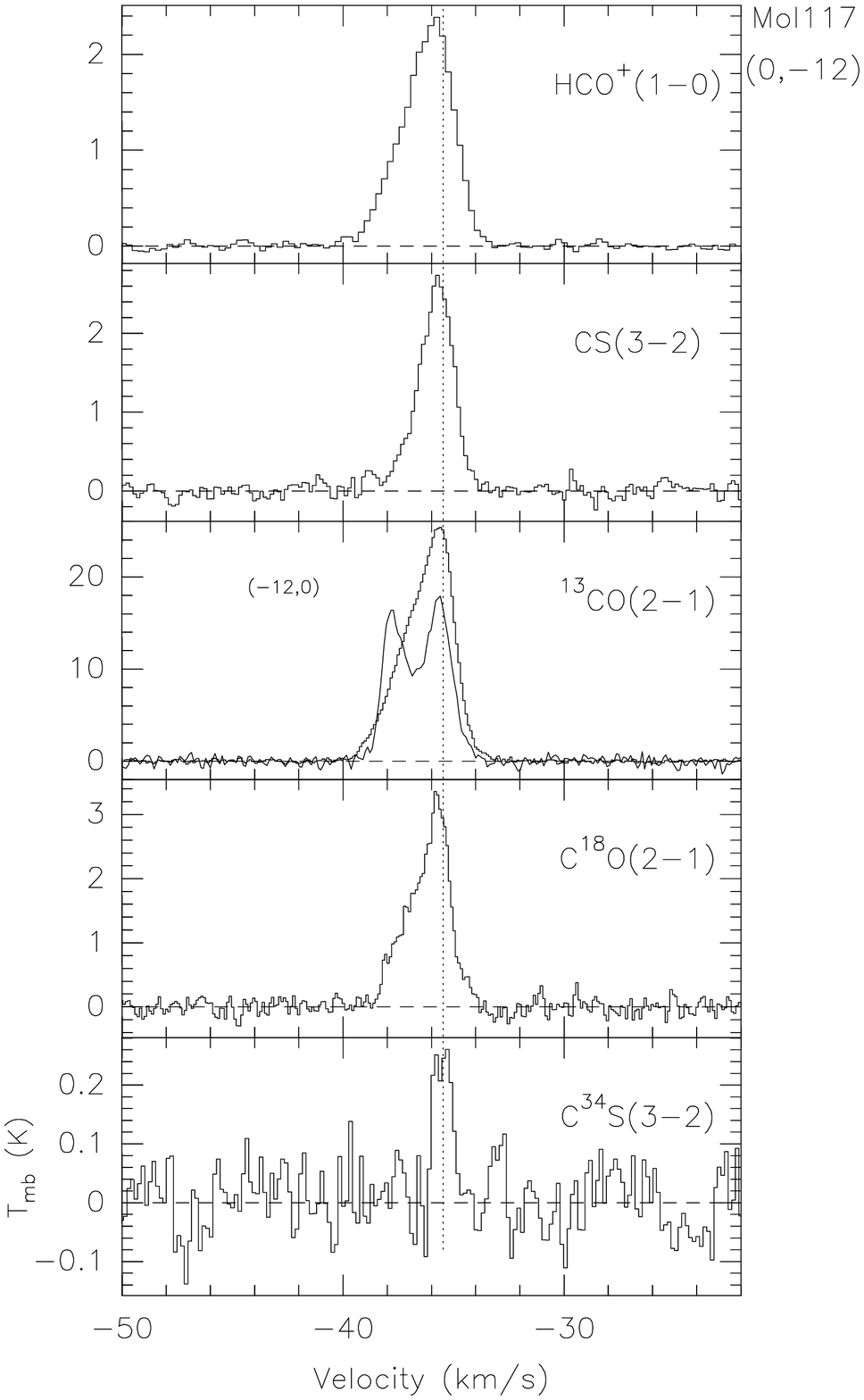}}
 \hfill
 \caption{{\bf g}\ As {\bf a}, for Mol~117. The $^{13}$CO(2$-$1) 
spectrum at (12\arcsec,0\arcsec) is also shown (drawn), for comparison. }
\end{figure}

\addtocounter{figure}{-1}

\begin{figure}
 \resizebox{\hsize}{!}{\includegraphics{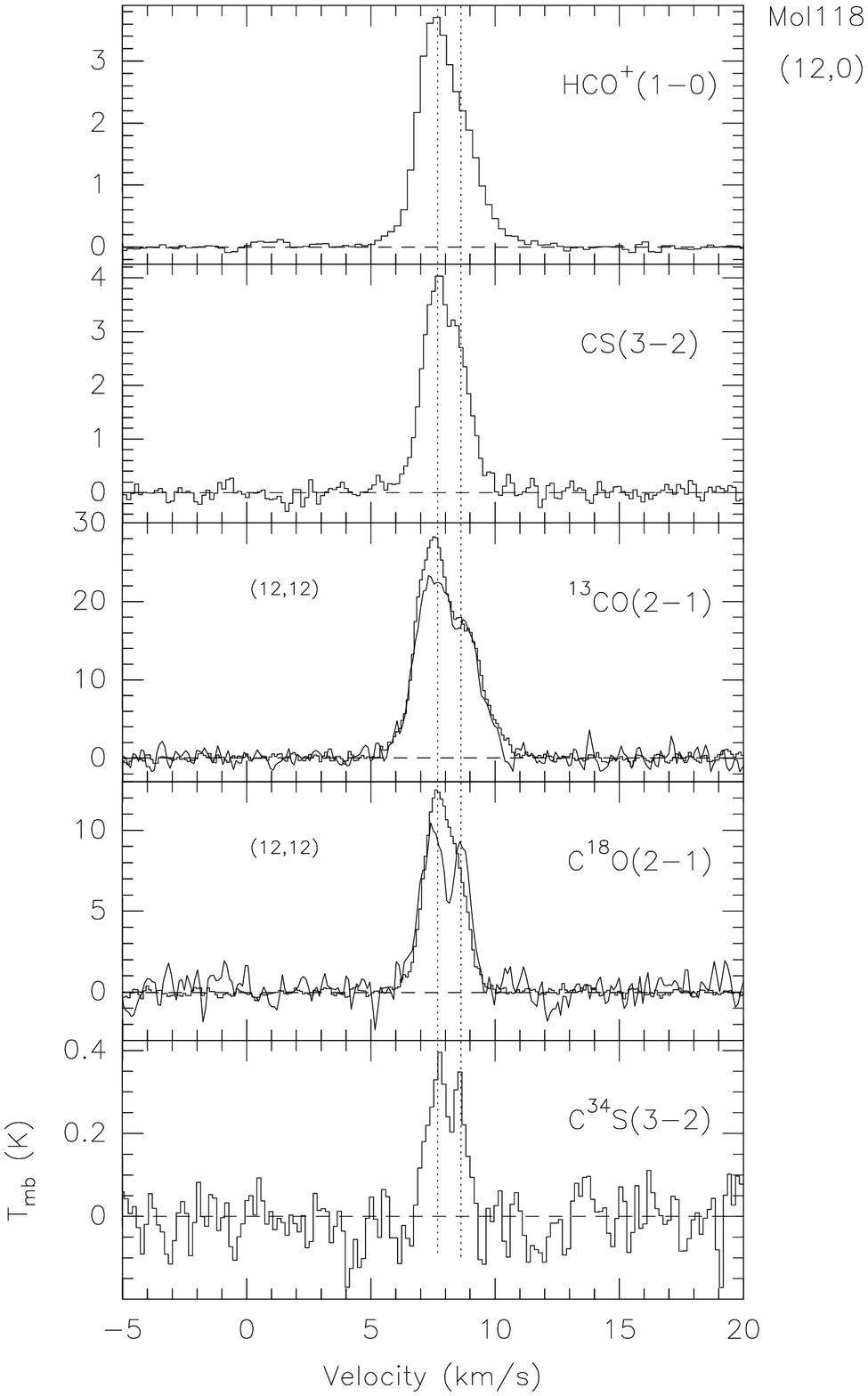}}
 \hfill
 \caption{{\bf h}\ As {\bf a}, for Mol~118. The $^{13}$CO(2$-$1) and 
C$^{18}$O(2$-$1) spectra at (12\arcsec,12\arcsec) are also shown (drawn), for 
comparison. The dotted line with the 
red-shifted velocity, with respect to the V$_{\rm lsr}$ of the C$^{34}$S line, 
indicates the second component present in these observations (see
Sect.~\ref{sindiv}).}
\end{figure}

\addtocounter{figure}{-1}

\clearpage
\begin{figure}
 \resizebox{\hsize}{!}{\includegraphics{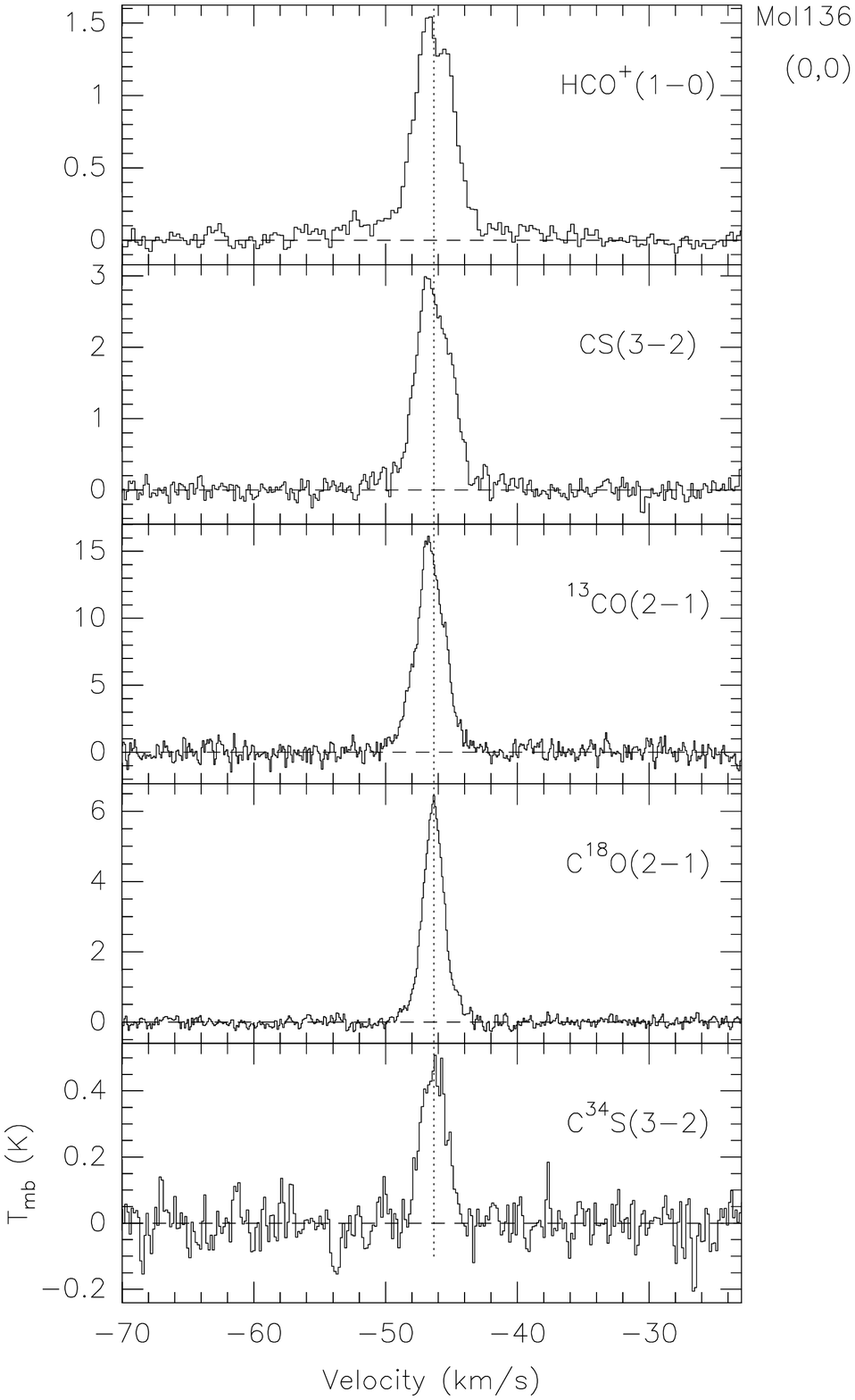}}
 \hfill
 \caption{{\bf i}\ As {\bf a}, for Mol~136}
\end{figure}

\addtocounter{figure}{-1}

\begin{figure}
 \resizebox{\hsize}{!}{\includegraphics{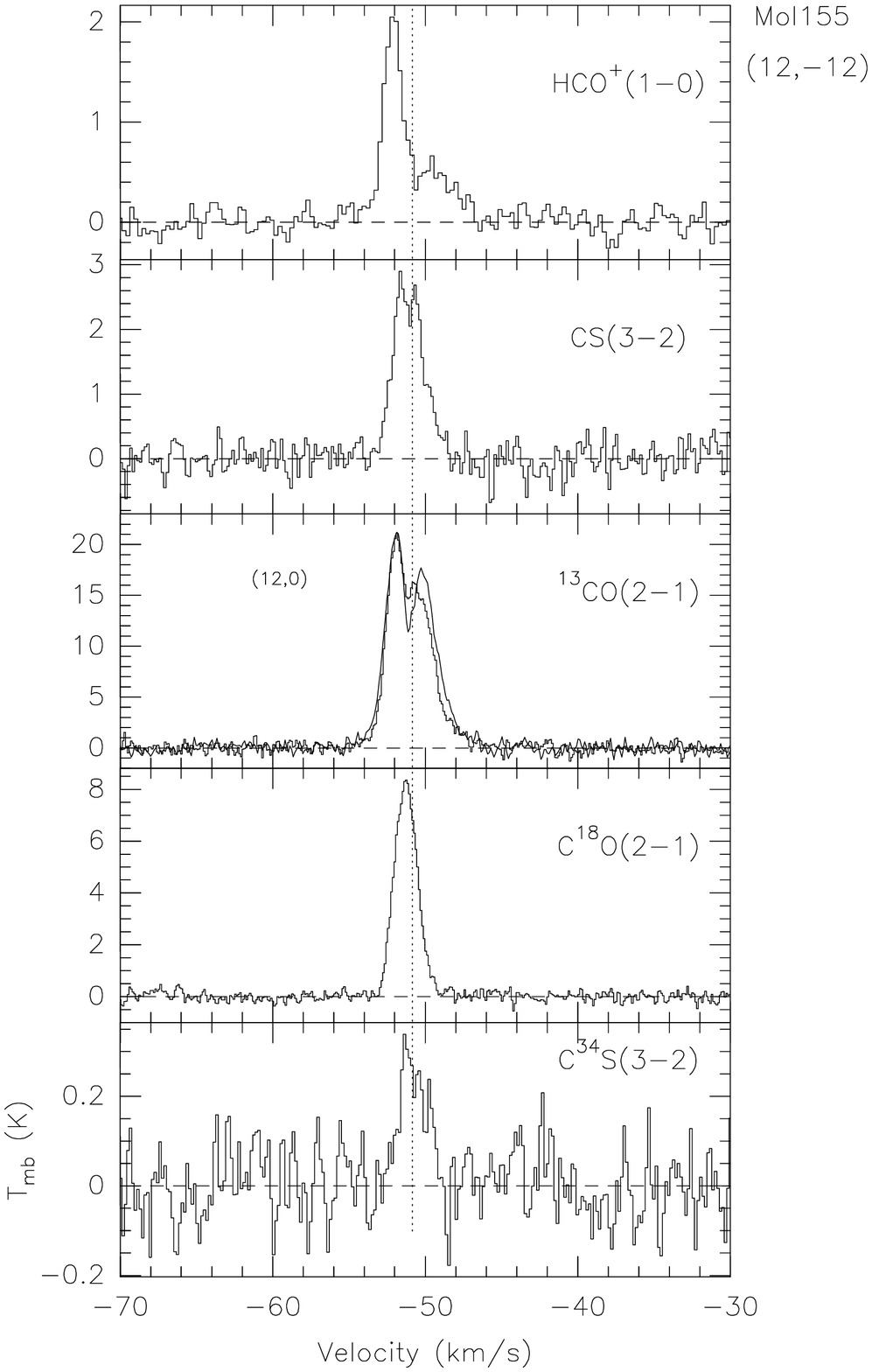}}
 \hfill
 \caption{{\bf j}\ As {\bf a}, for Mol~155. The $^{13}$CO(2$-$1) 
spectrum at (12\arcsec,0\arcsec) is also shown (drawn), for comparison.}
\end{figure}

\addtocounter{figure}{-1}

\clearpage
\begin{figure}
 \resizebox{\hsize}{!}{\includegraphics{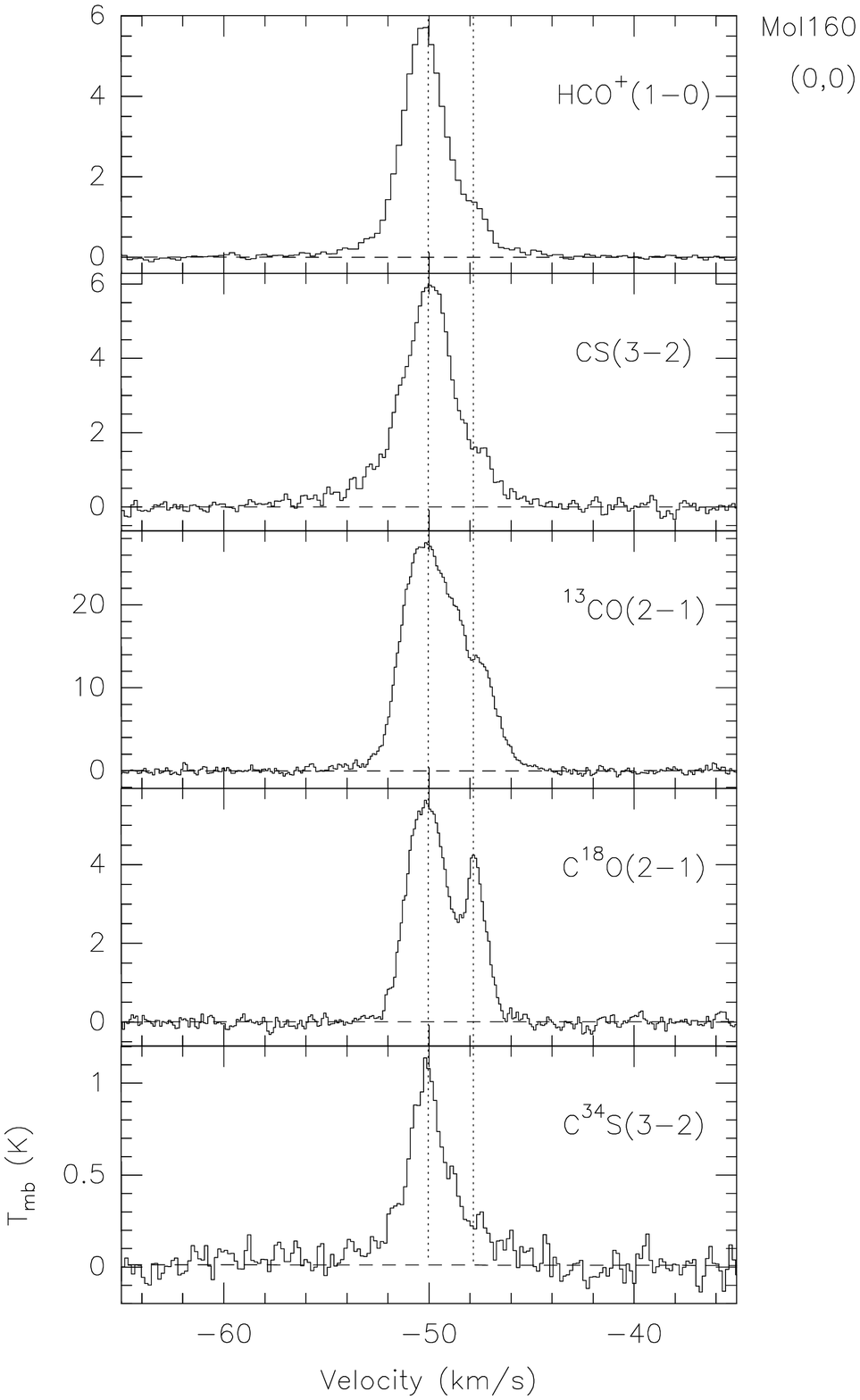}}
 \hfill
 \caption{{\bf k}\ As {\bf a}, for Mol~160. The dotted line with the 
red-shifted velocity, with respect to the V$_{\rm lsr}$ of the C$^{34}$S line, 
indicates the second component present in these observations (see
Sect.~\ref{sindiv}).}
\end{figure}


\clearpage
\begin{figure}
 \resizebox{\hsize}{!}{\includegraphics{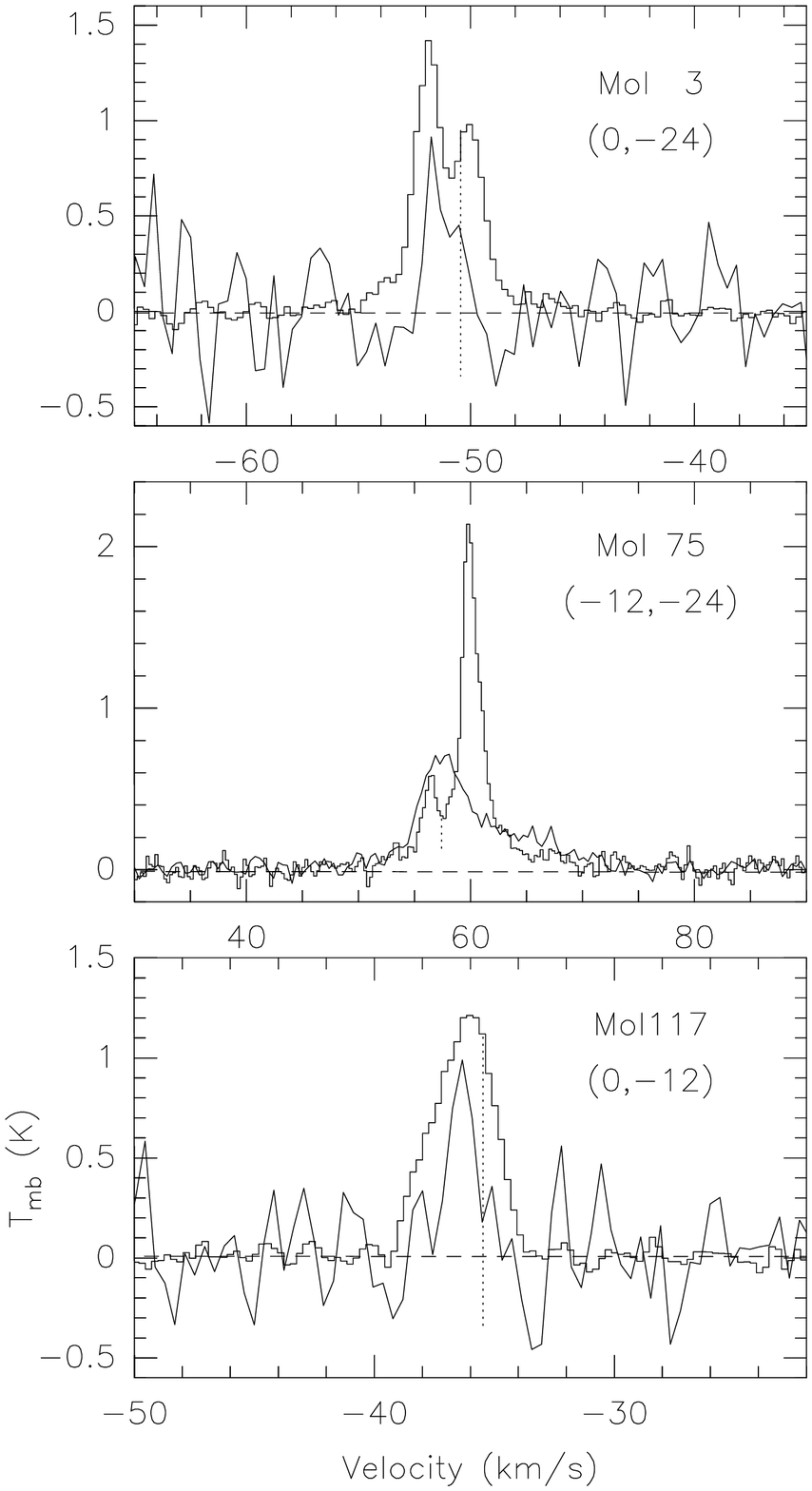}}
 \hfill
 \caption{{\bf a}\ Comparison of the IRAM HCO$^+$(1$-$0) [histogram] spectra 
with the
KOSMA HCO$^+$(4$-$3) [drawn] spectra, at the indicated positions, for Mol~3, 
75, and 117. The dotted vertical lines indicate the V$_{\rm lsr}$ of the 
C$^{34}$S(3$-$2) emission (from Gaussian fits); the dashed horizontal lines 
indicate the 0~K level. The IRAM spectra were convolved to a 70\arcsec\
beam; the KOSMA spectra were smoothed to a resolution of 
0.4~kms$^{-1}$, and multiplied by 7 (Mol~3), and 5 (Mol~117) to make 
comparison easier.}
\label{kospec}
\end{figure}

\addtocounter{figure}{-1}

\begin{figure}
 \resizebox{\hsize}{!}{\includegraphics{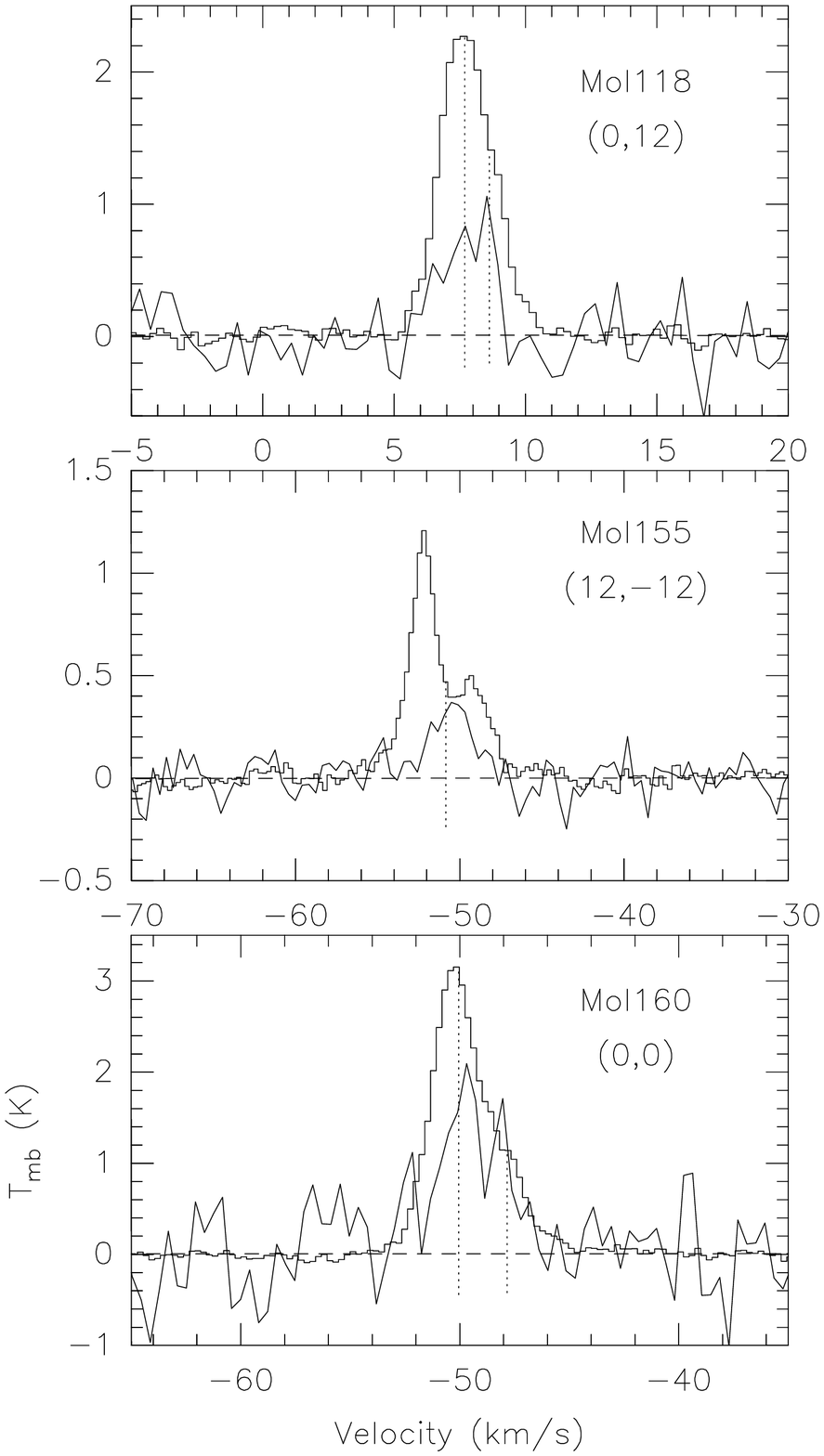}}
 \hfill
 \caption{{\bf b}\ As {\bf a}, for Mol~118, 155, and 160. 
The KOSMA spectra were multiplied by 9 (Mol~118), 2 (Mol~155), and 10
(Mol~160), to make comparison easier.}
\end{figure}


\clearpage
\begin{figure}
 \resizebox{\hsize}{!}{\includegraphics{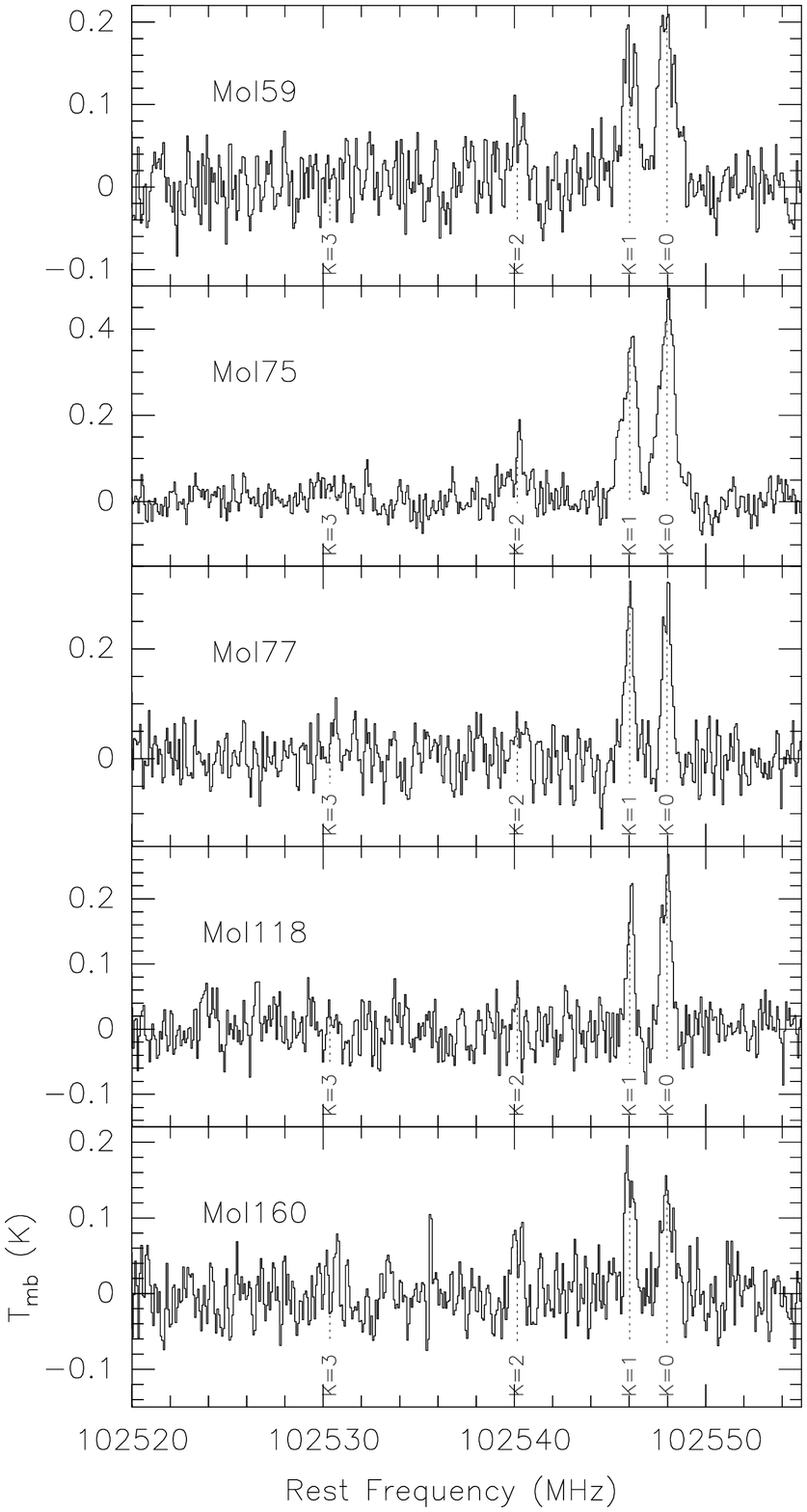}}
 \hfill
 \caption{{\bf a}\ Spectra of the CH$_3$C$_2$H(6$-$5) rotational transition 
towards
the peak position. Only the higher-resolution (0.23~kms$^{-1}$) spectra are
shown. The location of the first few K-components are indicated in the 
spectra.}
\label{molch3c2h}
\end{figure}

\addtocounter{figure}{-1}

\begin{figure}
 \resizebox{\hsize}{!}{\includegraphics{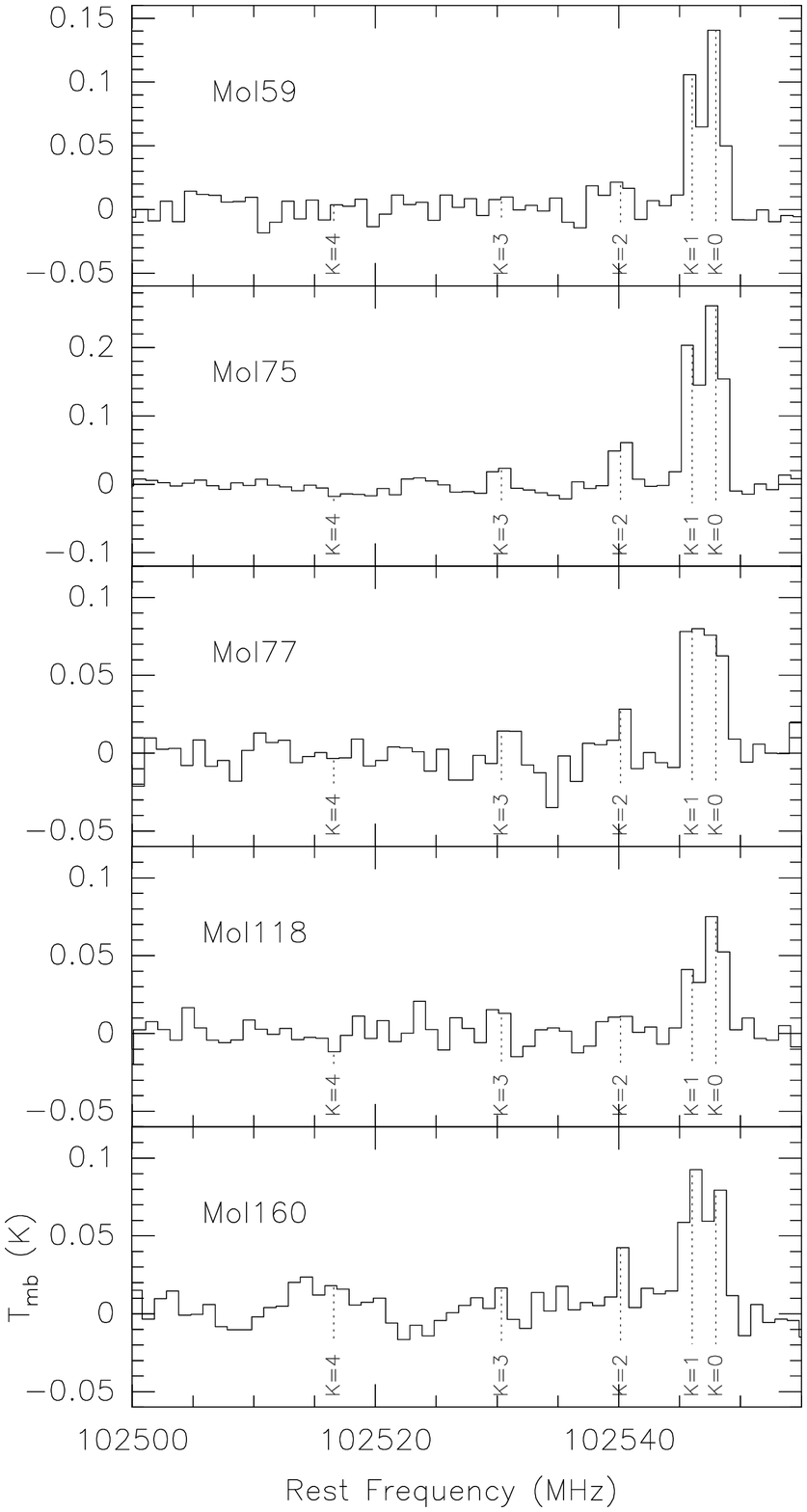}}
 \hfill
 \caption{{\bf b}\ As {\bf a}, but for the low-resolution
(2.92~kms$^{-1}$) spectra.}
\end{figure}

\addtocounter{figure}{-1}

\clearpage
\begin{figure}
 \resizebox{\hsize}{!}{\includegraphics{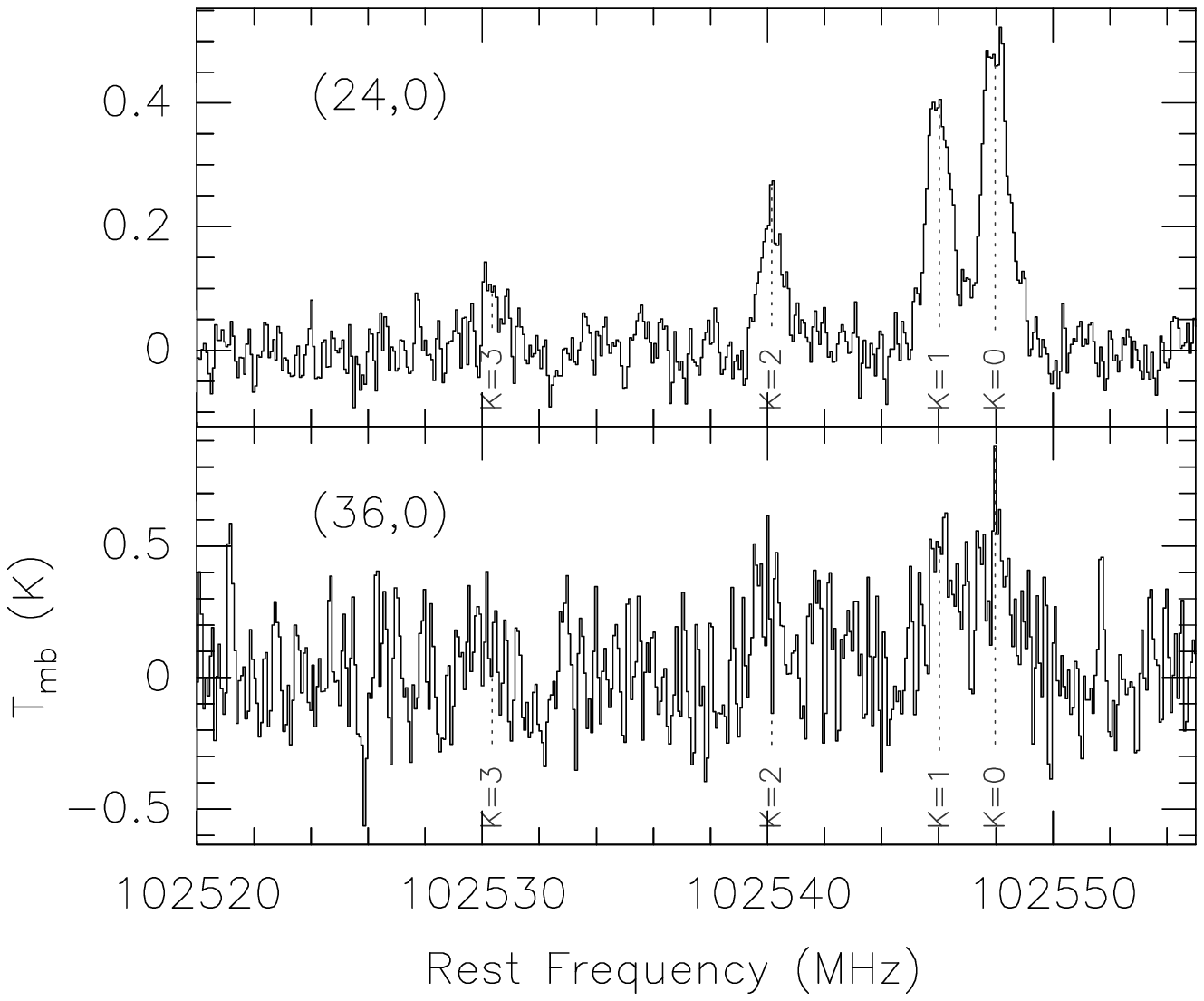}}
 \hfill
 \caption{{\bf c}\ As {\bf a}, but for two positions towards
Mol~98. The velocity resolution of the spectra is 0.23~kms$^{-1}$; the offset 
position at which they were taken are indicated in the panels.}
\end{figure}

\addtocounter{figure}{-1}

\begin{figure}
 \resizebox{\hsize}{!}{\includegraphics{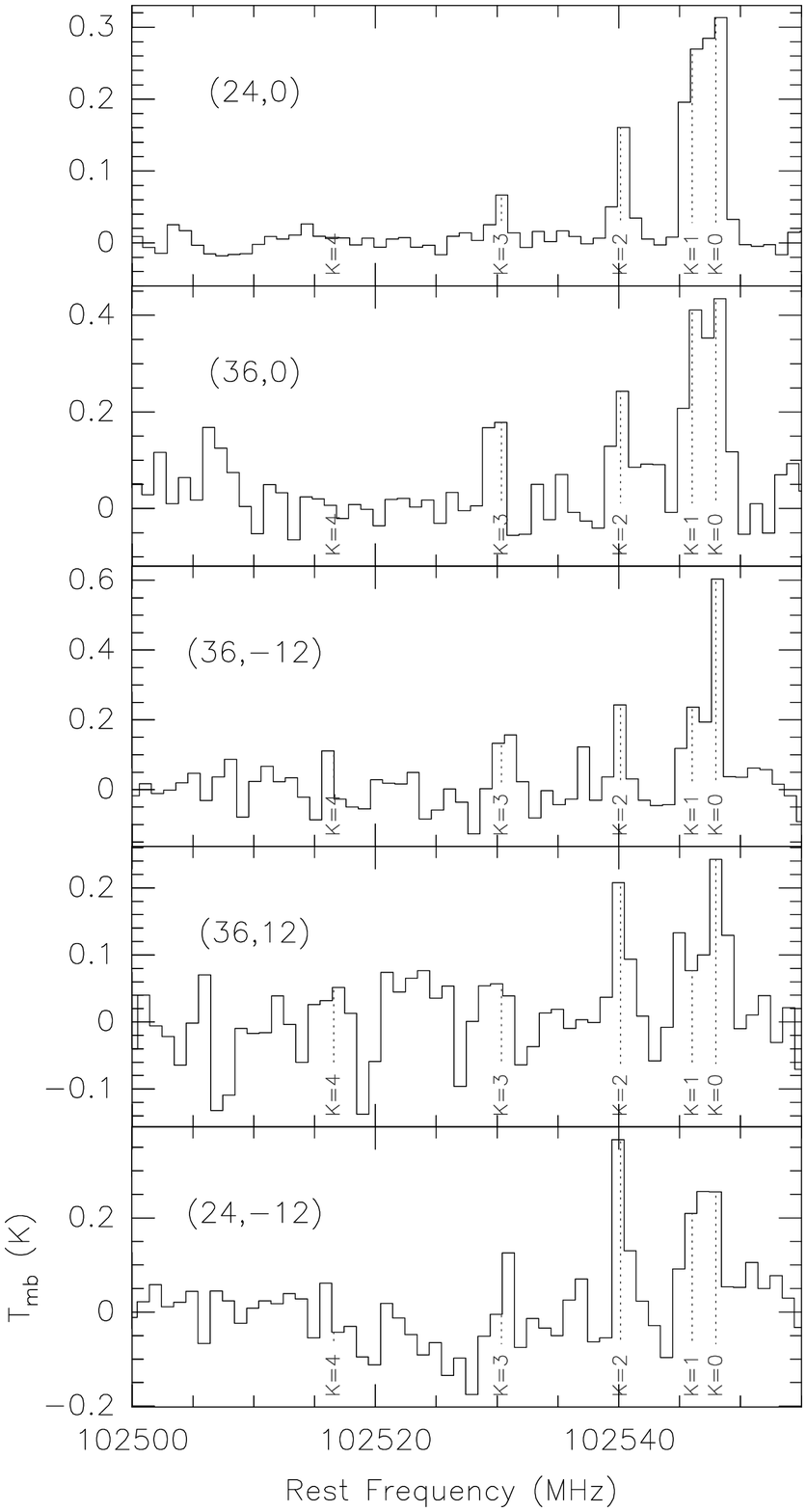}}
 \hfill
 \caption{{\bf d}\ As {\bf a}, but for various positions towards
Mol~98. The velocity resolution of the spectra is 2.92~kms$^{-1}$; the offset 
position at which they were taken are indicated in the panels.}
\end{figure}


\begin{figure}
 \resizebox{\hsize}{!}{\includegraphics{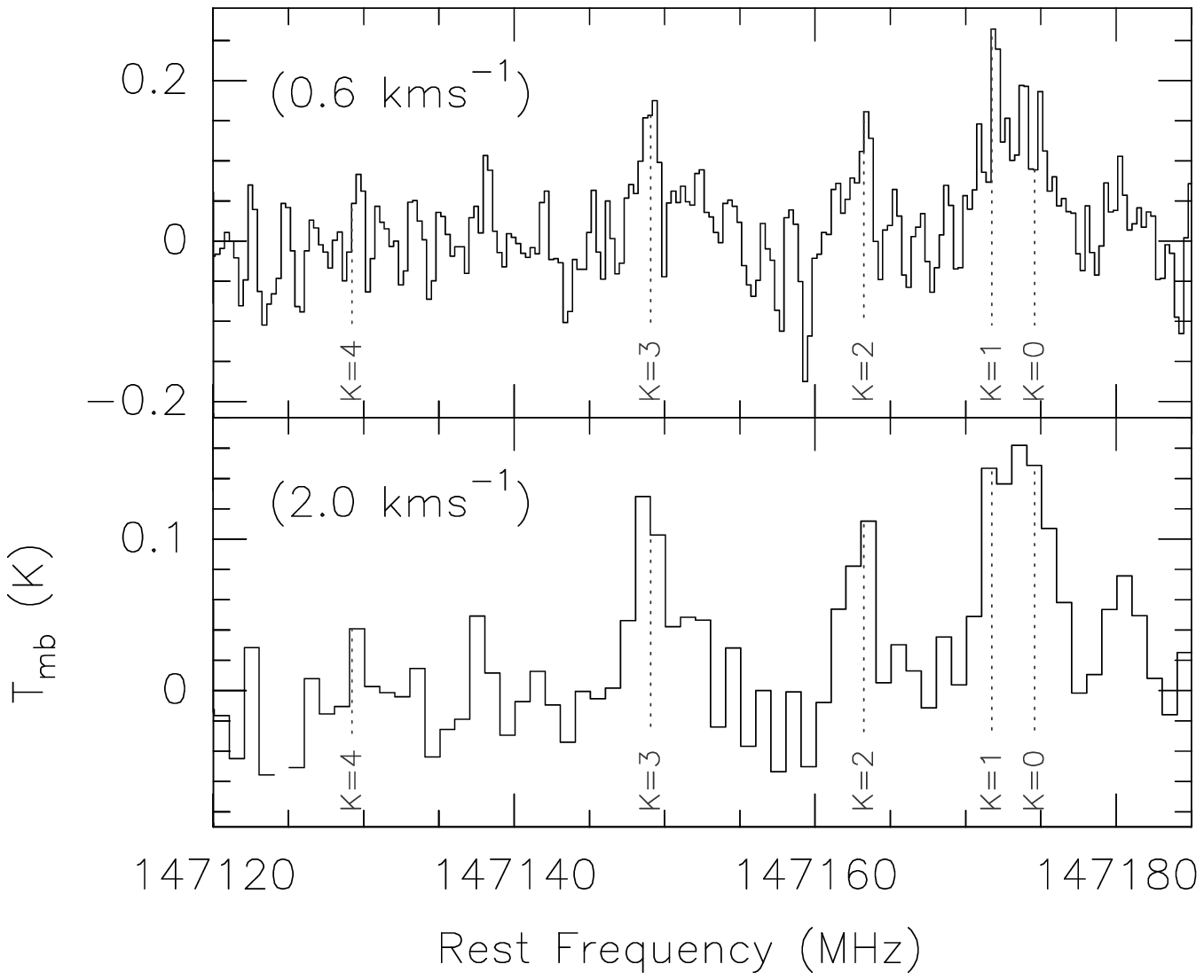}}
 \hfill
 \caption{\ Spectra of the CH$_3$CN(8$-$7) rotational transition 
towards the peak position (24\arcsec,0\arcsec) in Mol~98. The velocity
resolution of the spectra is indicated in the panels. 
The location of the first few K-components are indicated in the spectra.}
\label{mol98ch3cn}
\end{figure}


\clearpage
\begin{figure}
 \resizebox{\hsize}{!}{\includegraphics{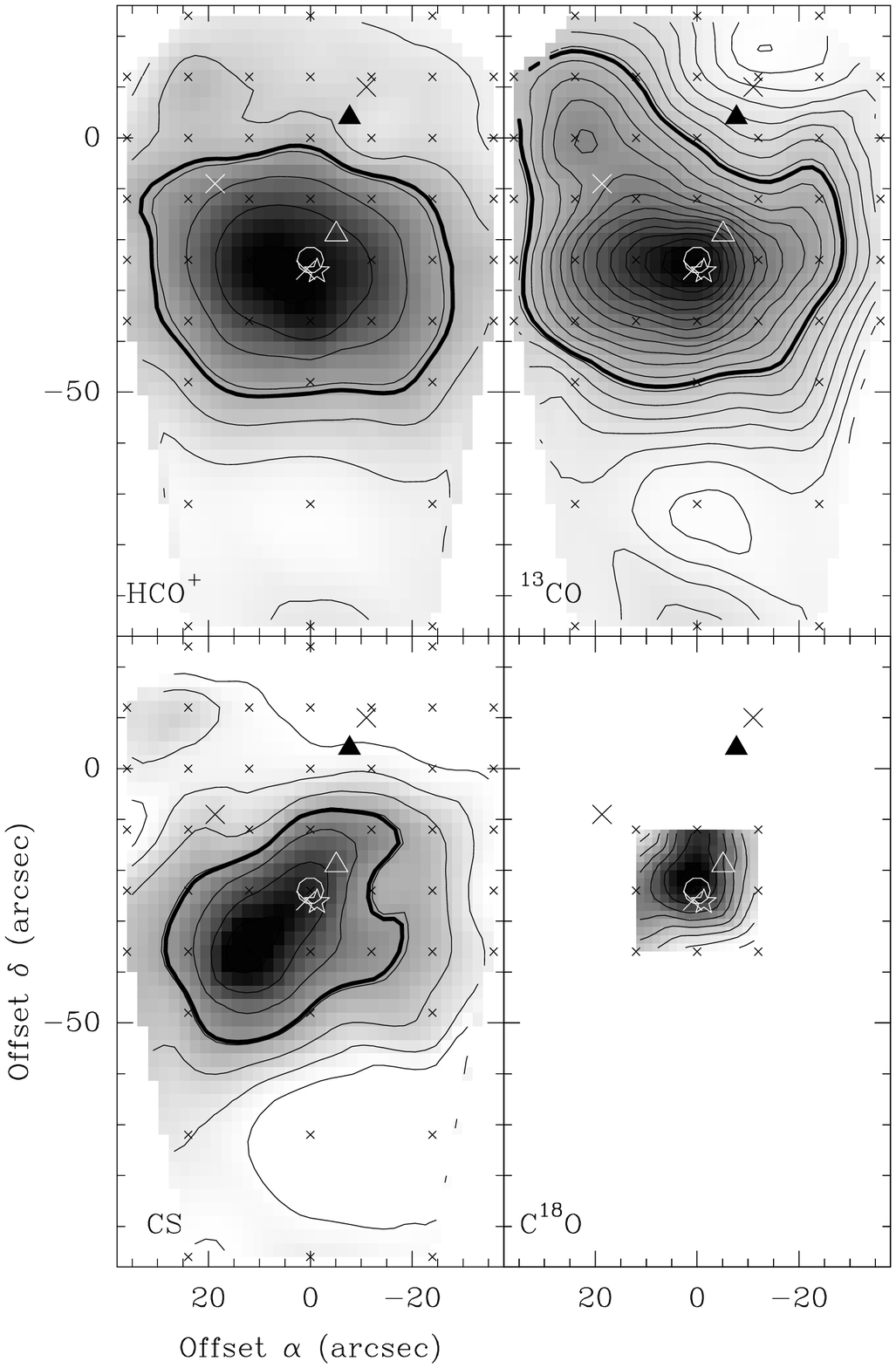}}
 \hfill
 \caption{Maps towards all sources of the integrated emission in the
lines of HCO$^+$(1$-$0), $^{13}$CO(2$-$1), CS(3$-$2), and C$^{18}$O(2$-$1).
Where C$^{34}$S(3$-$2) was strong enough, a panel with a map of that
molecule is shown as well. Thick contours indicate the
50\% level in each map. The (0,0) position is the position of the sub-mm
peak (Molinari et al. \cite{molinari00}).
The small crosses mark the observed positions, the filled
triangle indicates the IRAS point source position, and the circle marks the
position where we made longer integrations at the frequency of the
CH$_3$C$_2$H(6$-$5) and CH$_3$CN lines. 
\hfill\break\noindent
{\bf a}\ Data for Mol~3. The open triangle is the peak of the 850~$\mu$m
emission, while the star indicates the position of the H$_2$O maser (Jenness
et al. \cite{jenness}). The big crosses indicate the positions of the peaks of
the radio continuum emission (VLA D-array, 3.6~cm; Molinari et al. in
preparation). For details, see Sect.~\ref{sindiv}. Contour values
(low(step)high, in Kkms$^{-1}$) are: 0.5(1)6.5 (HCO$^+$); 13(3)69
($^{13}$CO); 0.5(1)6.5 (CS); 9(1)17 (C$^{18}$O).
}
\label{intem}
\end{figure}

\addtocounter{figure}{-1}

\begin{figure}
 \resizebox{\hsize}{!}{\includegraphics{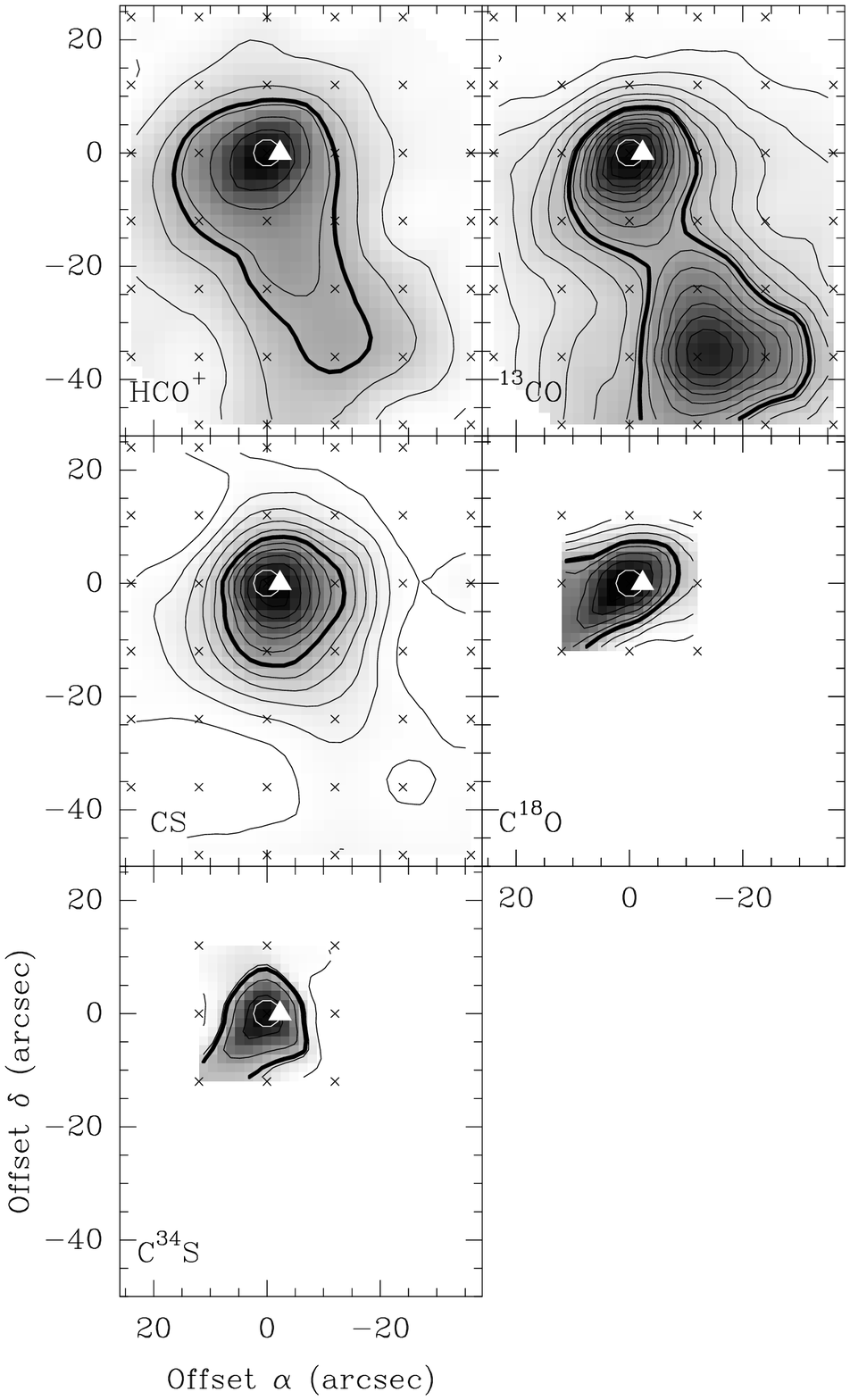}}
 \hfill
 \caption{{\bf b}\ Same as {\bf a}, for Mol~8. Contour values
(low(step)high, in Kkms$^{-1}$) are: 0.5(1)6.5 (HCO$^+$); 4.5(4)60.5
($^{13}$CO); 0.5(1)11 (CS); 1(1)8 (C$^{18}$O); 0.4(0.4)2 (C$^{34}$S).} 
\end{figure}

\addtocounter{figure}{-1}

\begin{figure}
 \resizebox{\hsize}{!}{\includegraphics{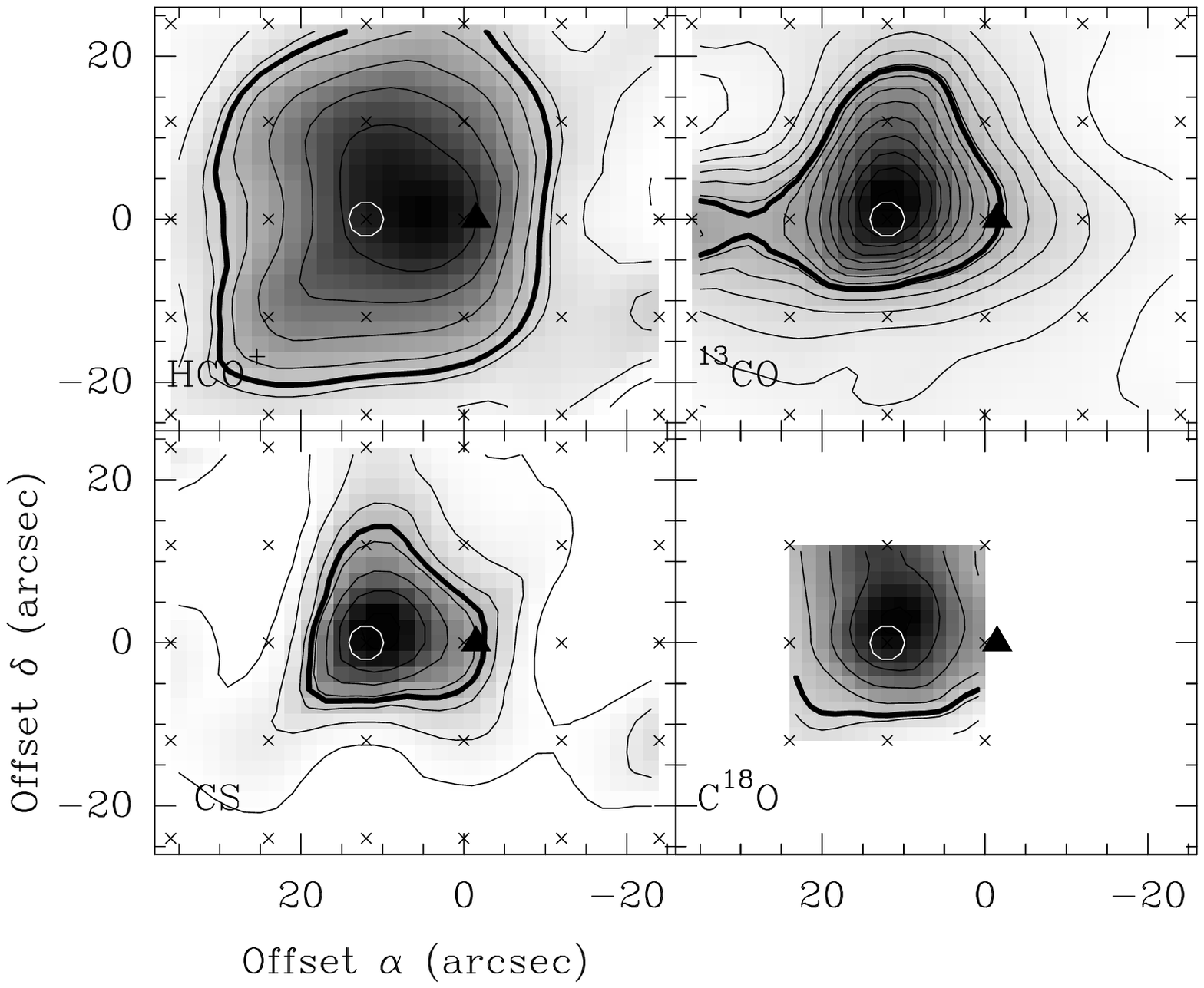}}
 \hfill
 \caption{{\bf c}\ Same as {\bf a}, for Mol~59. Contour values
(low(step)high, in Kkms$^{-1}$) are: 1(0.5)4.5 (HCO$^+$); 6(4)62
($^{13}$CO); 1(1)7 (CS); 3(2)15 (C$^{18}$O).}
\end{figure}

\addtocounter{figure}{-1}

\begin{figure}
 \resizebox{\hsize}{!}{\includegraphics{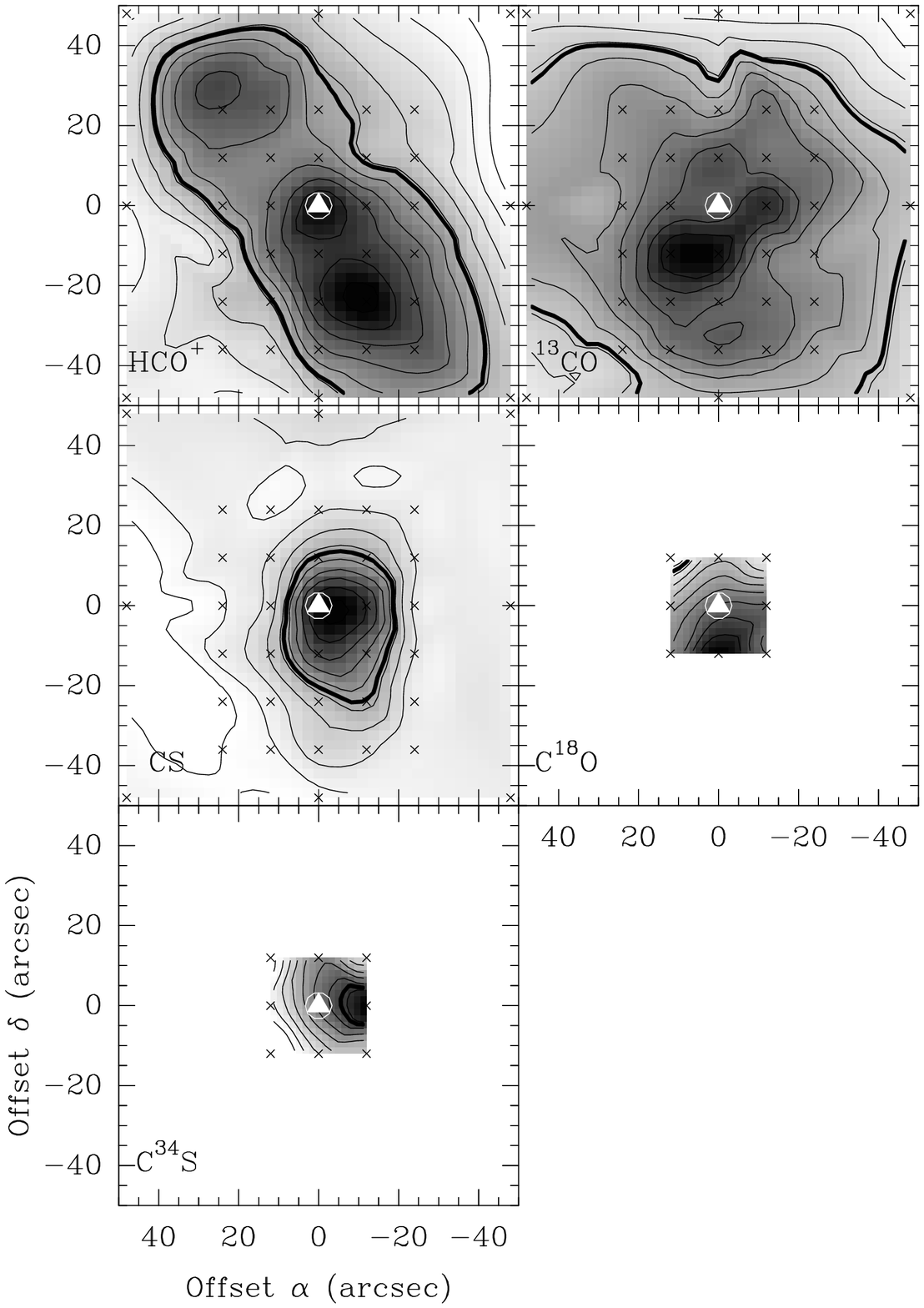}}
 \hfill
 \caption{{\bf d}\ Same as {\bf a}, for Mol~75. Contour values
(low(step)high, in Kkms$^{-1}$) are: 1(1)10 (HCO$^+$); 15(6)81
($^{13}$CO); 1(2)19 (CS); 14(2)34 (C$^{18}$O); 0.3(0.3)2.7 (C$^{34}$S).}
\end{figure}

\addtocounter{figure}{-1}

\begin{figure}
 \resizebox{\hsize}{!}{\includegraphics{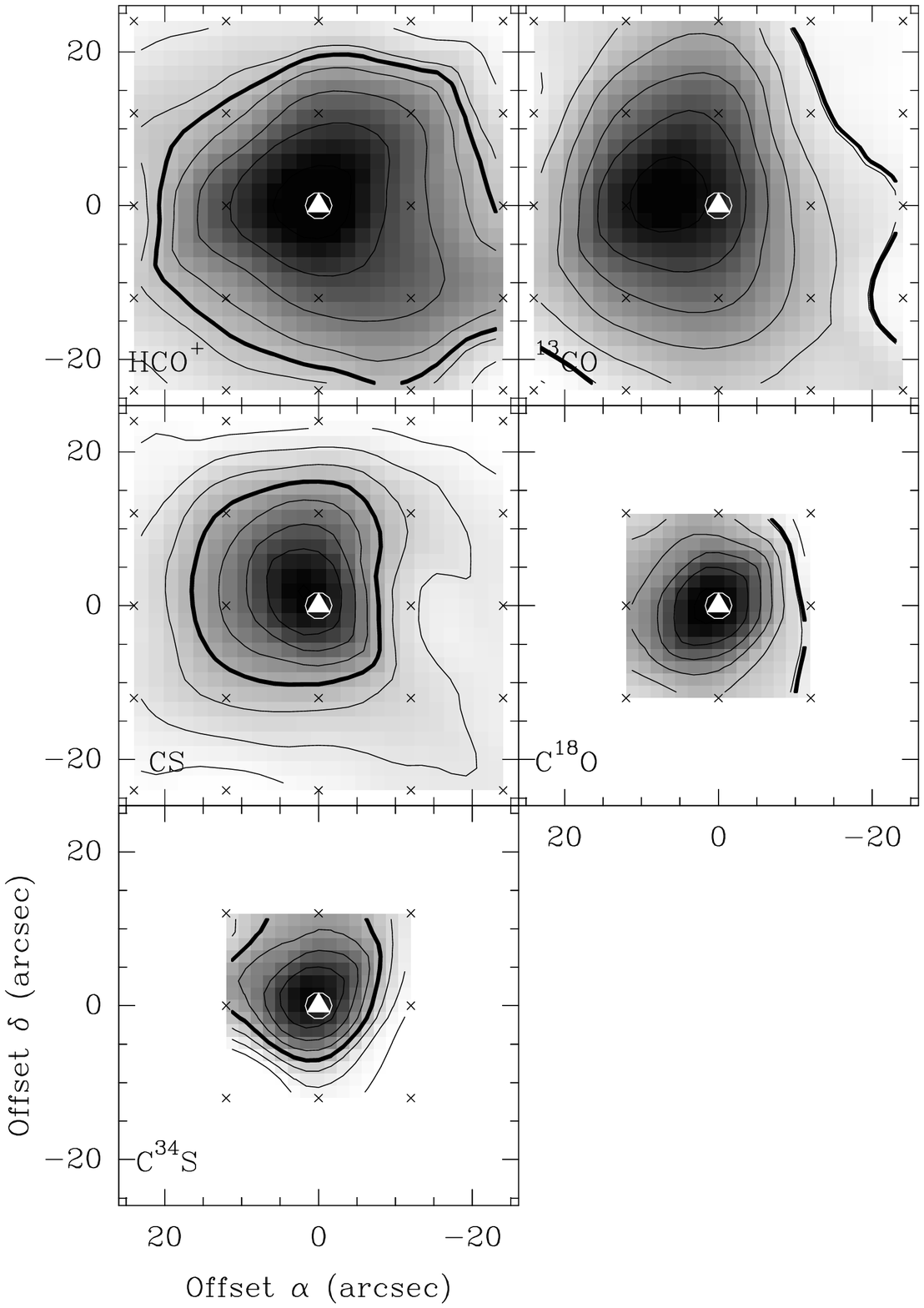}}
 \hfill
 \caption{{\bf e}\ Same as {\bf a}, for Mol~77. Contour values
(low(step)high, in Kkms$^{-1}$) are: 1(0.5)4 (HCO$^+$); 18(4)42
($^{13}$CO); 1(1)8 (CS); 7(2)21 (C$^{18}$O); 0.3(0.3)2.4 (C$^{34}$S).}
\end{figure}

\addtocounter{figure}{-1}

\begin{figure*}
 \resizebox{\hsize}{!}{\rotatebox{-90}{\includegraphics{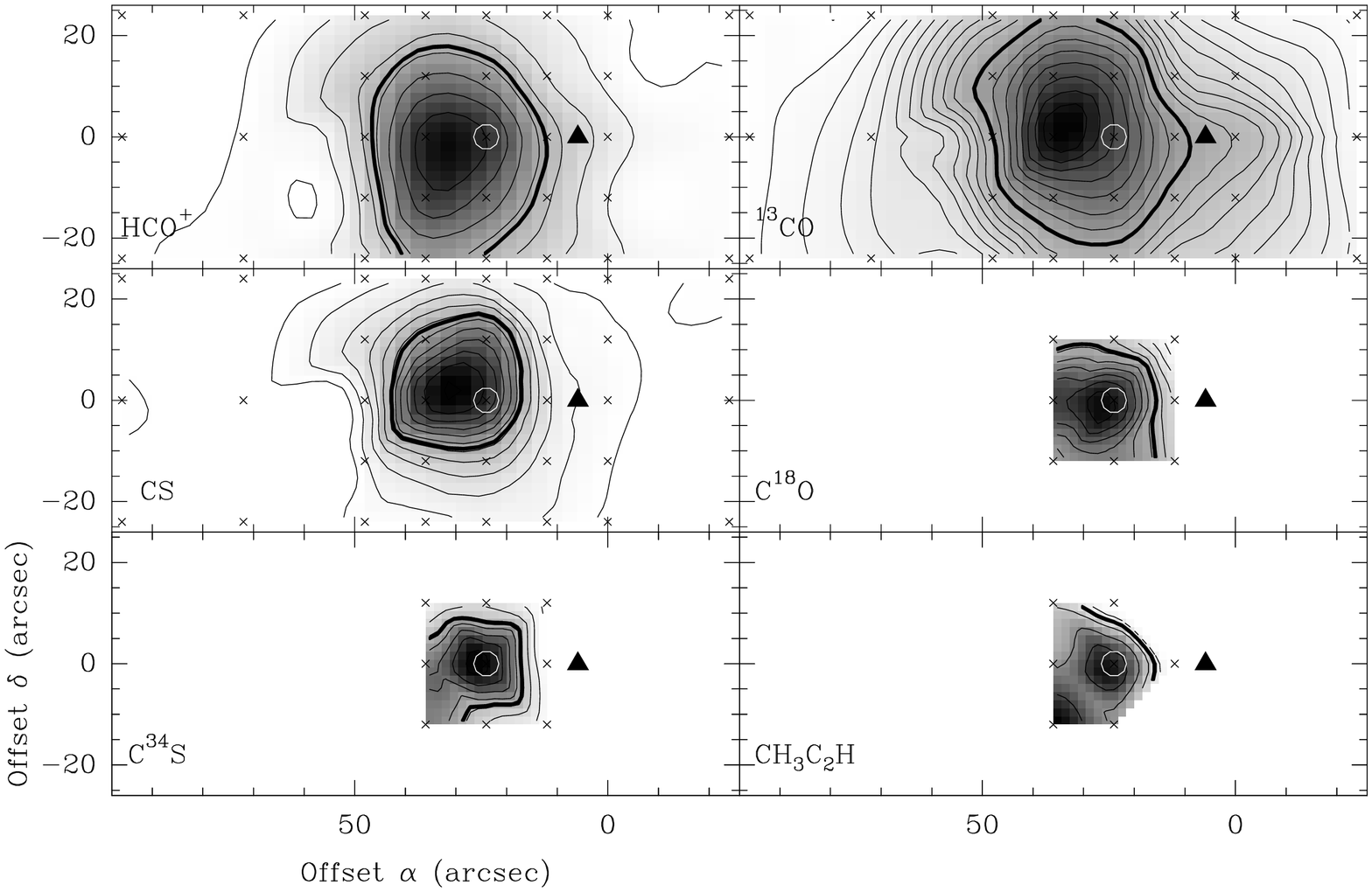}}}
 \hfill
 \caption{{\bf f}\ Same as {\bf a}, for Mol~98. Added here is a
panel showing the distribution of the integrated emission over the K=0,1
components of CH$_3$C$_2$H(6$-$5). Contour values
(low(step)high, in Kkms$^{-1}$) are: 1(2)17 (HCO$^+$); 1(4)41(6)95
($^{13}$CO); 1(2)23 (CS); 3(3)35 (C$^{18}$O); 0.5(0.5)4 (C$^{34}$S);
1(0.5)3.5 (CH$_3$C$_2$H).}
\end{figure*}

\addtocounter{figure}{-1}

\begin{figure}
 \resizebox{\hsize}{!}{\includegraphics{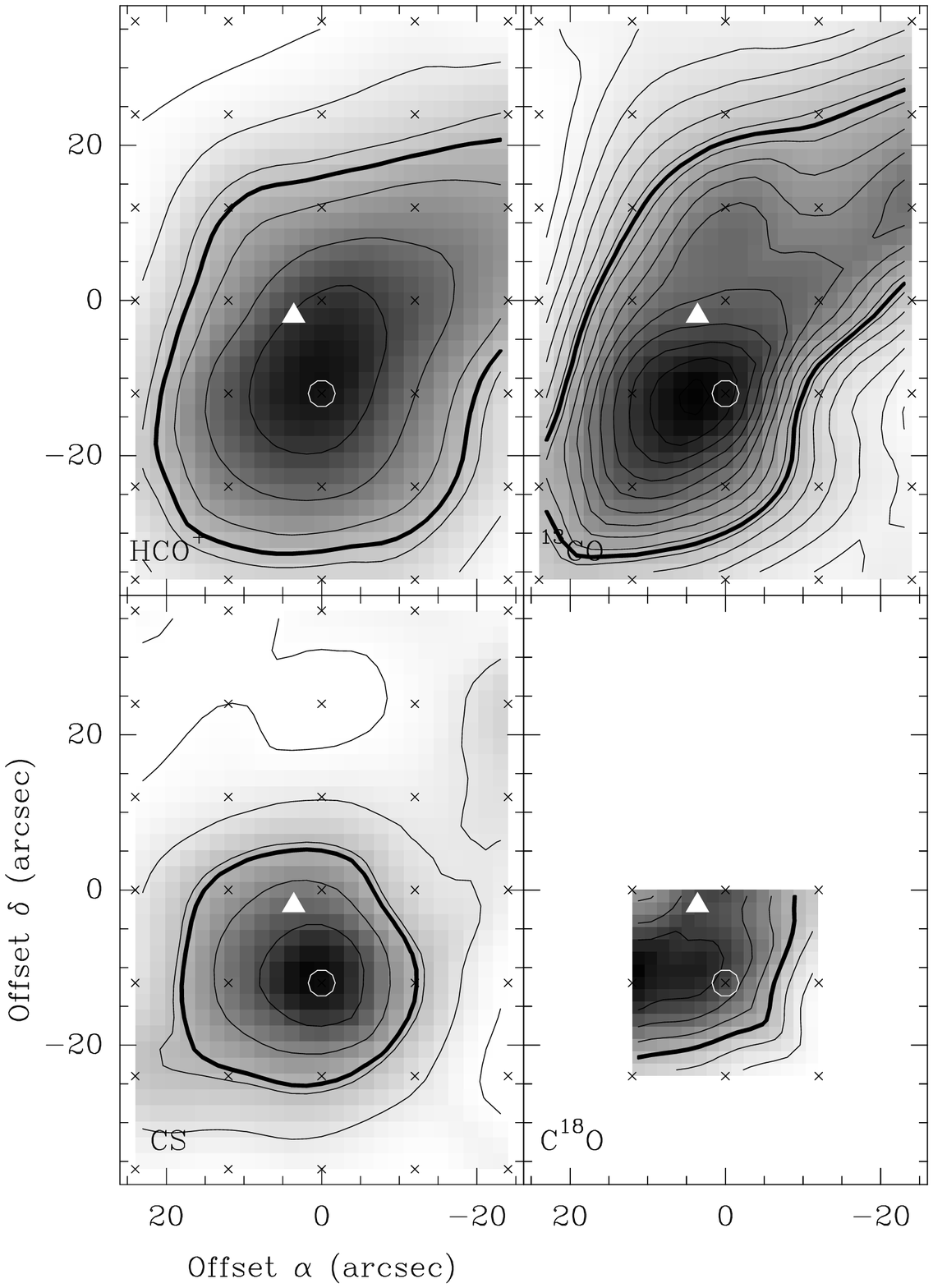}}
 \hfill
 \caption{{\bf g}\ Same as {\bf a}, for Mol~117. Contour values
(low(step)high, in Kkms$^{-1}$) are: 1(1)7 (HCO$^+$); 5(4)70
($^{13}$CO); 0.5(1)5.5 (CS); 1(1)8 (C$^{18}$O).}
\end{figure}

\addtocounter{figure}{-1}

\begin{figure*}
 \resizebox{\hsize}{!}{\rotatebox{-90}{\includegraphics{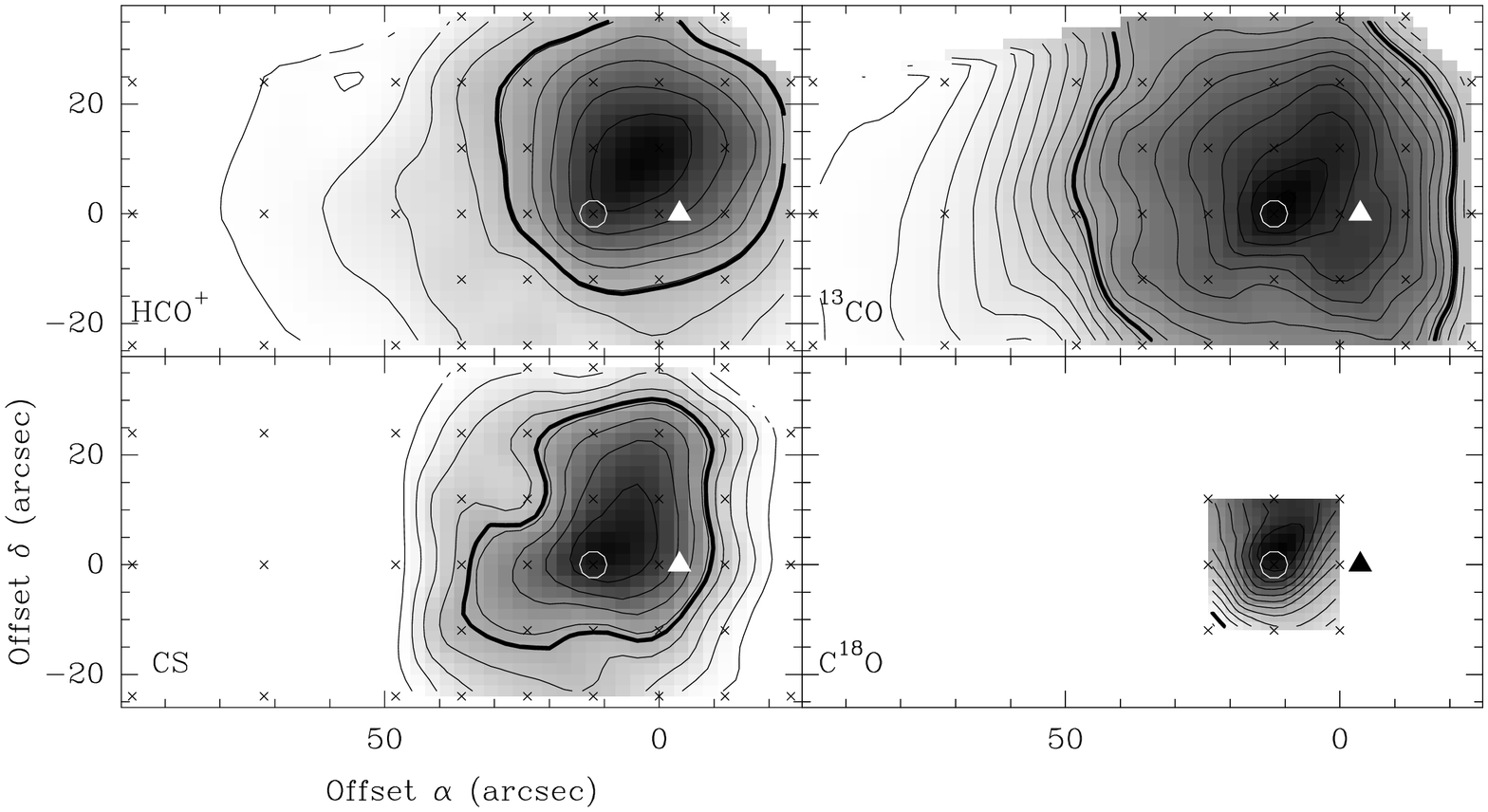}}}
 \hfill
 \caption{{\bf h}\ Same as {\bf a}, for Mol~118. Contour values
(low(step)high, in Kkms$^{-1}$) are: 1(1)10 (HCO$^+$); 1(4)71
($^{13}$CO); 0.5(1)8.5 (CS); 10(1)22 (C$^{18}$O).}
\end{figure*}

\addtocounter{figure}{-1}

\begin{figure*}
 \resizebox{\hsize}{!}{\rotatebox{-90}{\includegraphics{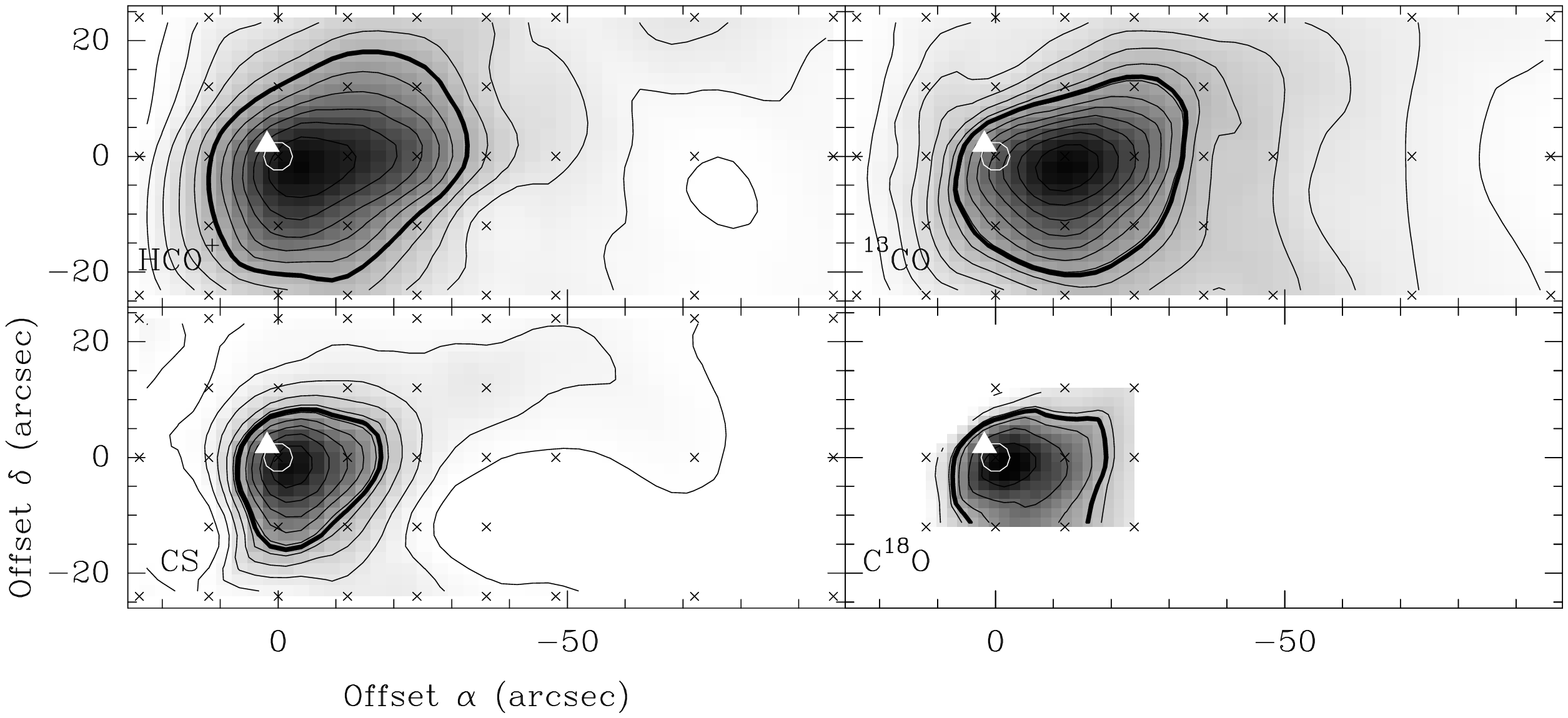}}}
 \hfill
 \caption{{\bf i}\ Same as {\bf a}, for Mol~136. Contour values
(low(step)high, in Kkms$^{-1}$) are: 0.5(0.5)6 (HCO$^+$); 1(4)57
($^{13}$CO); 0.5(1)10.5 (CS); 2(2)13 (C$^{18}$O).}
\end{figure*}

\addtocounter{figure}{-1}

\begin{figure}
 \resizebox{\hsize}{!}{\includegraphics{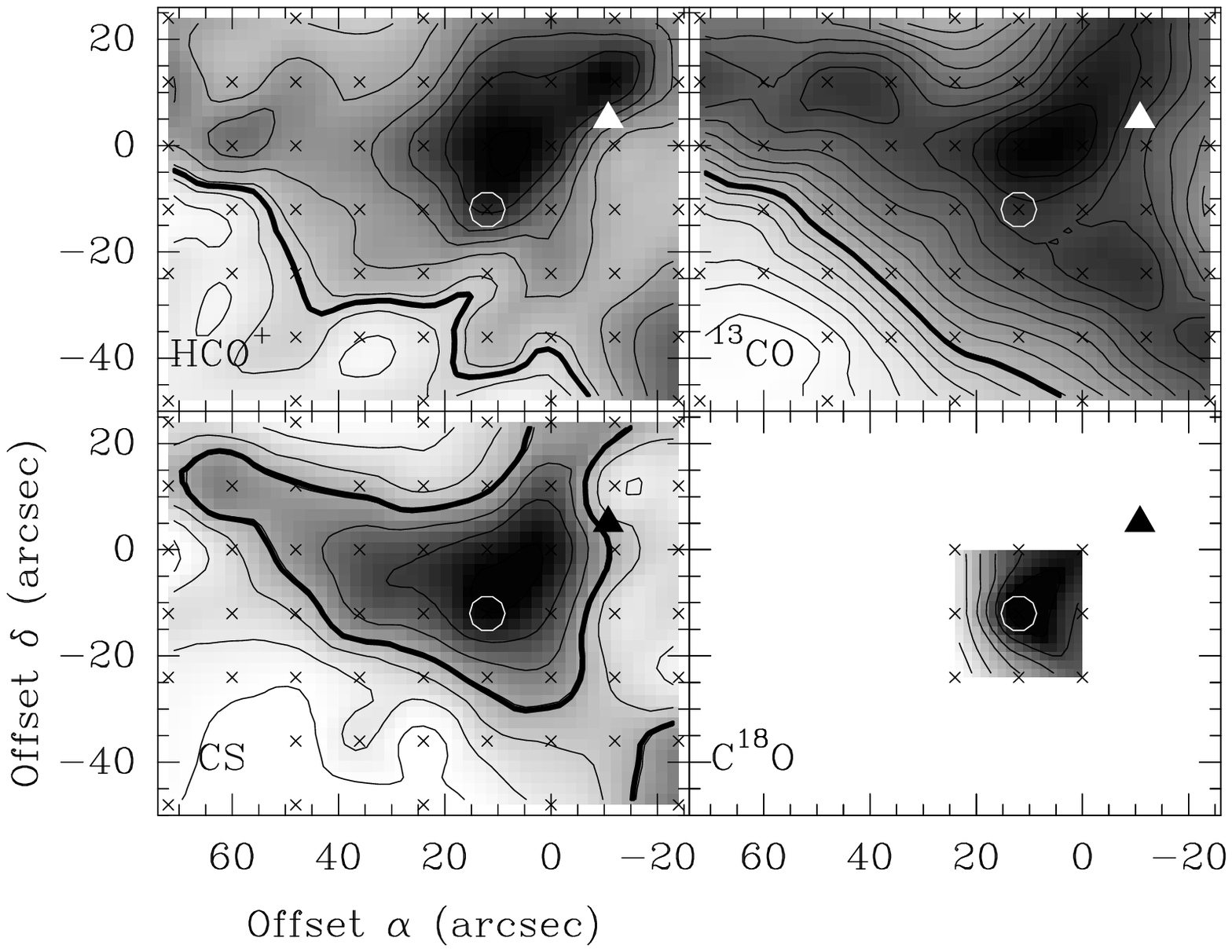}}
 \hfill
 \caption{{\bf j}\ Same as {\bf a}, for Mol~155. Contour values
(low(step)high, in Kkms$^{-1}$) are: 1.5(1)5.5 (HCO$^+$); 8(4)72
($^{13}$CO); 0.5(1)6.5 (CS); 9(1)15 (C$^{18}$O).}
\end{figure}

\addtocounter{figure}{-1}

\begin{figure}
 \resizebox{\hsize}{!}{\includegraphics{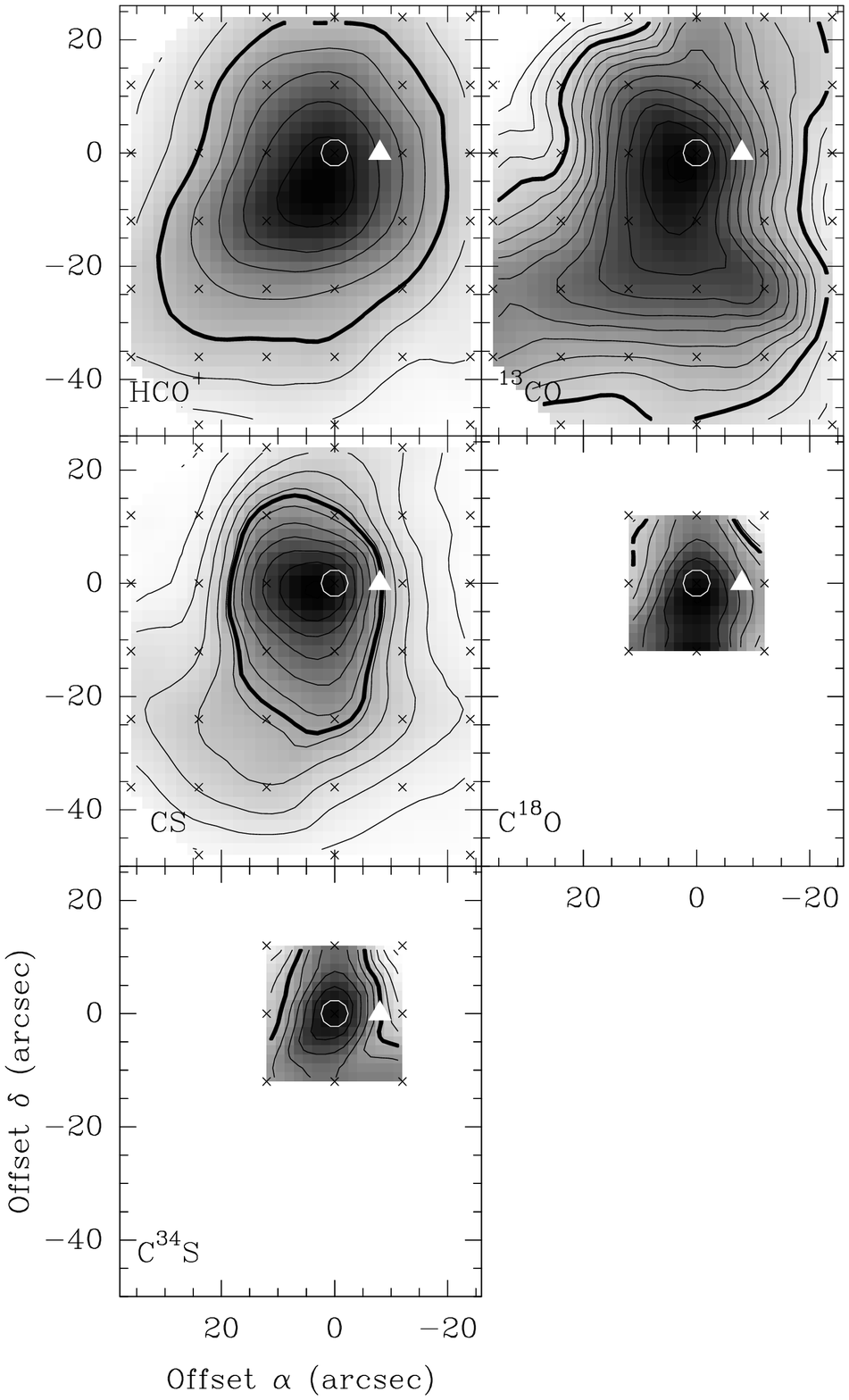}}
 \hfill
 \caption{{\bf k}\ Same as {\bf a}, for Mol~160. Contour values
(low(step)high, in Kkms$^{-1}$) are: 3(2)18 (HCO$^+$); 31(6)110
($^{13}$CO); 0.5(2)22.5 (CS); 5(2)19 (C$^{18}$O); 0.3(0.3)3 (C$^{34}$S).}
\end{figure}



\clearpage
\begin{figure}
 \resizebox{\hsize}{!}{\includegraphics{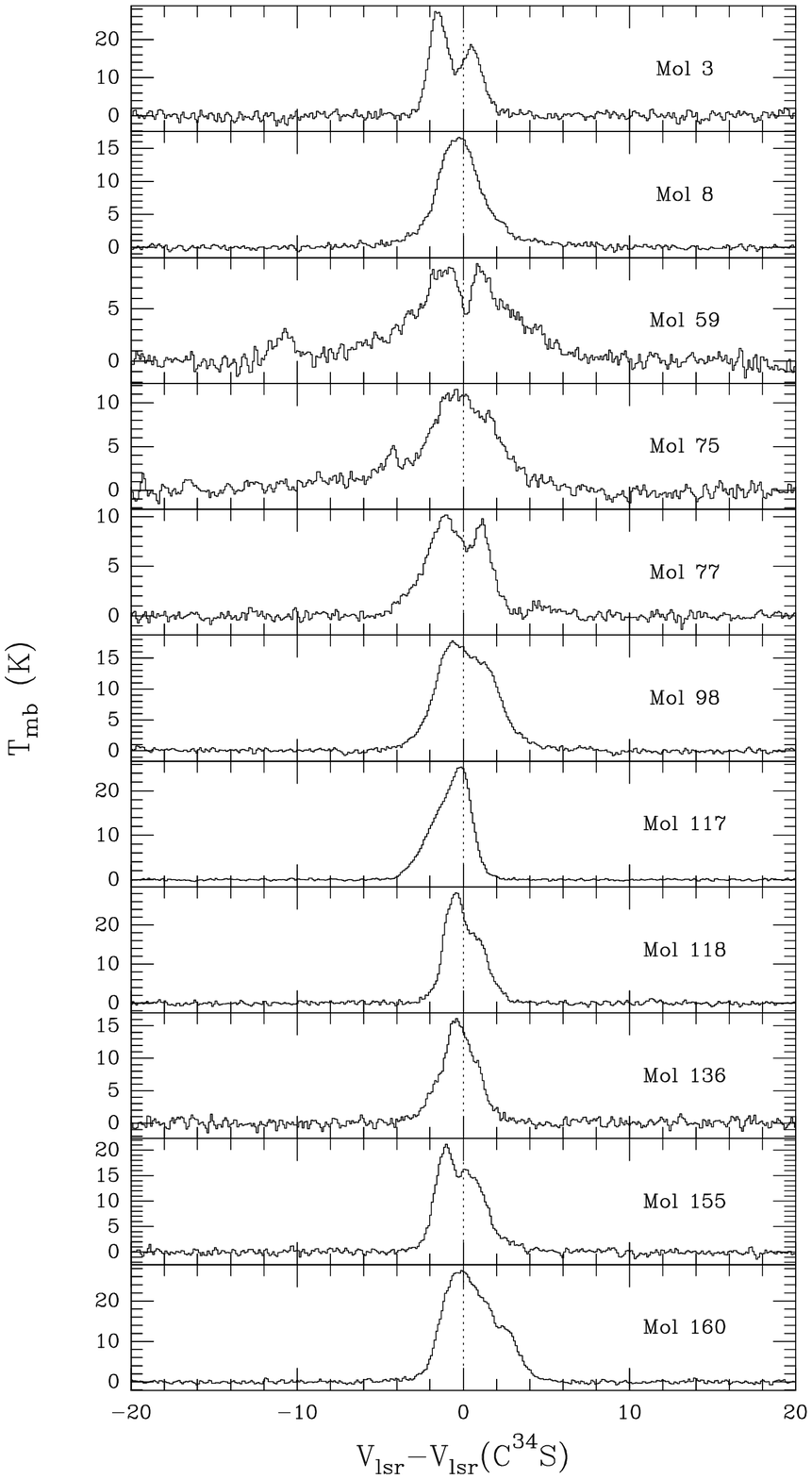}}
 \hfill
 \caption{{\bf a}\ Spectra of the $^{13}$CO(2$-$1) emission towards the peak
position in each source (as indicated in Fig.~\ref{molspec}). The
velocity is indicated with respect to that of the C$^{34}$S(3$-2$) line,
taken to respresent the velocity of the bulk of the high-density gas.}
\label{shift}
\end{figure}

\addtocounter{figure}{-1}

\begin{figure}
 \resizebox{\hsize}{!}{\includegraphics{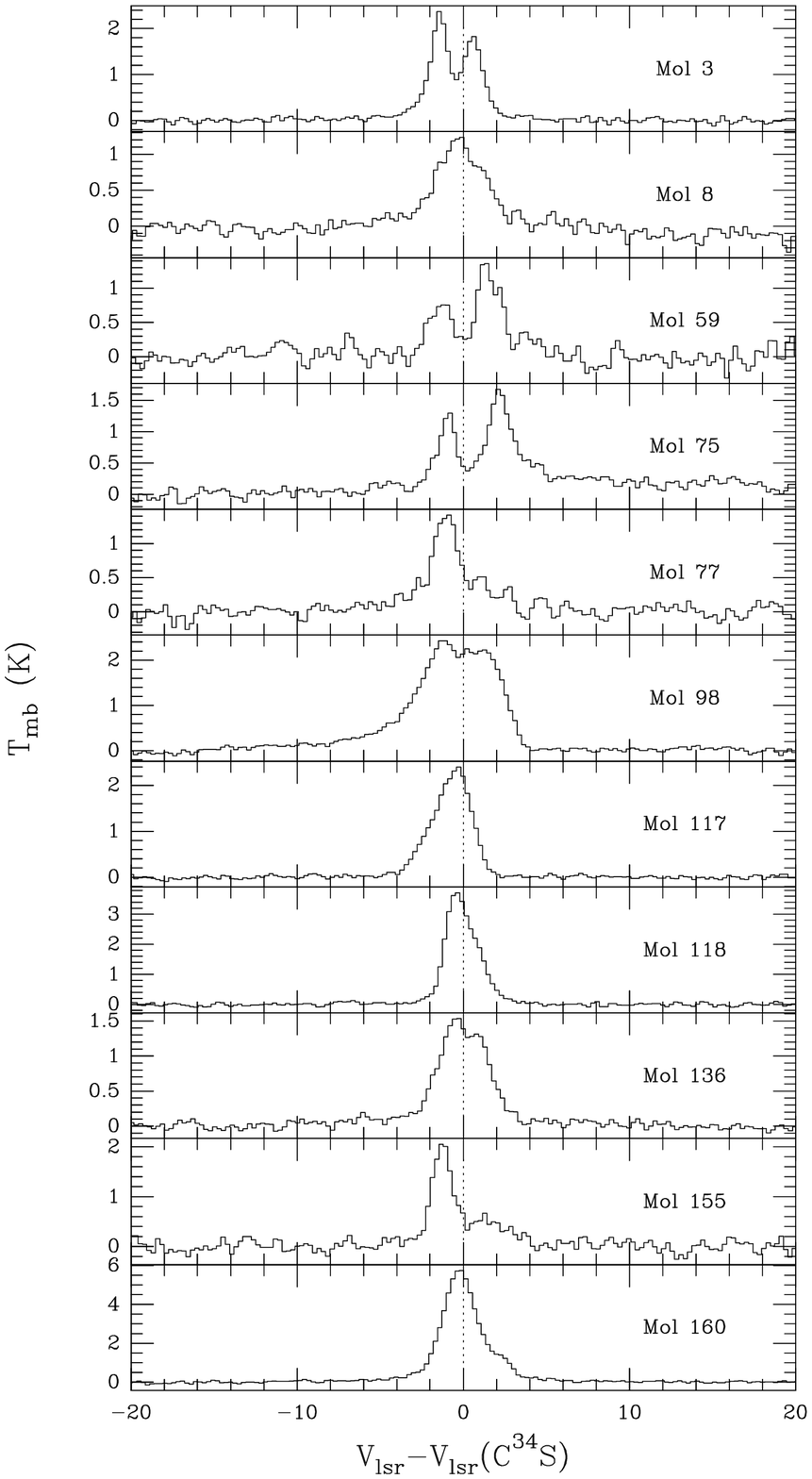}}
 \hfill
 \caption{{\bf b}\ As {\bf a}, but for the HCO$^+$(1$-$0) spectra}
\end{figure}

\addtocounter{figure}{-1}

\clearpage
\begin{figure}
 \resizebox{\hsize}{!}{\includegraphics{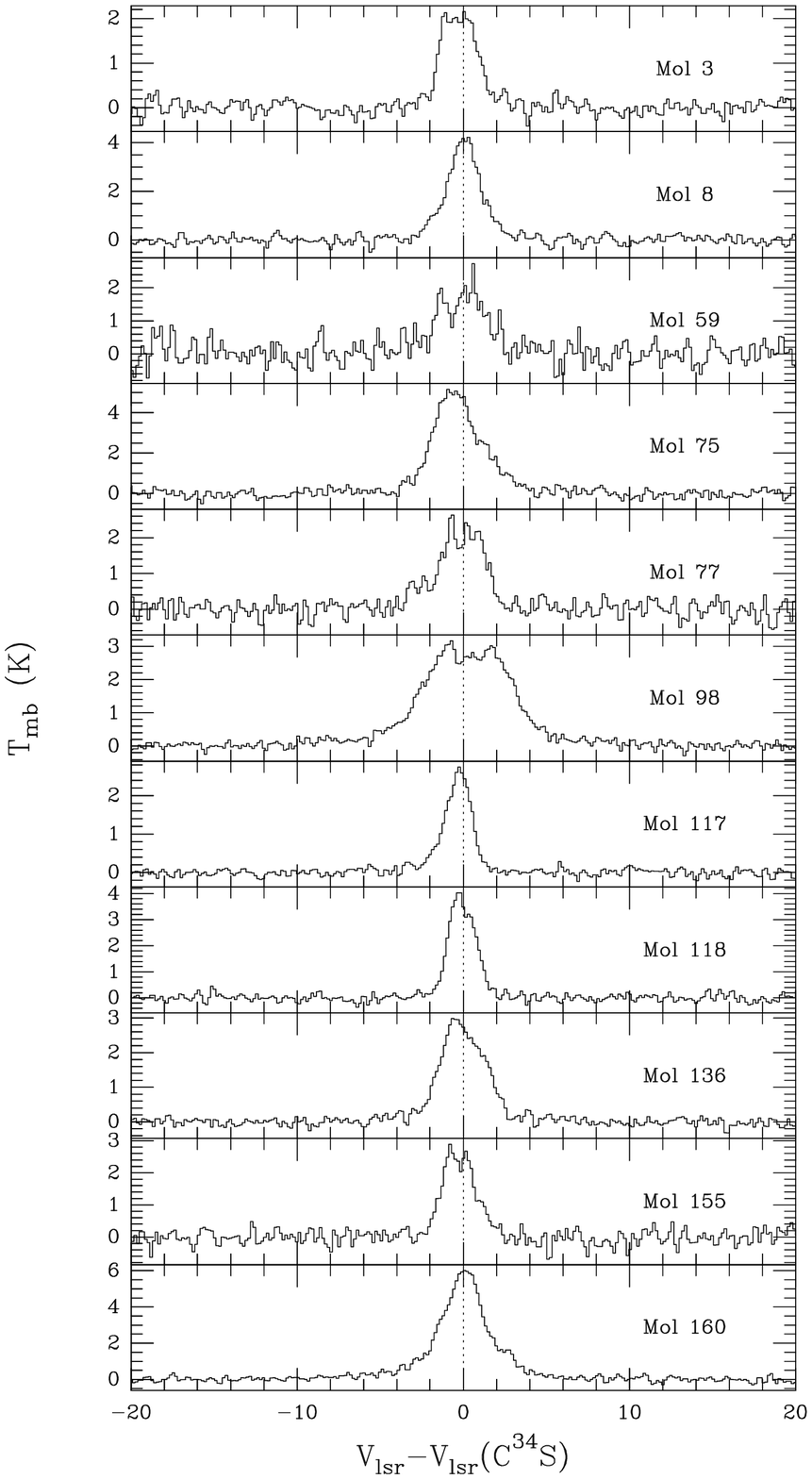}}
 \hfill
 \caption{{\bf c}\ As {\bf a}, but for the CS(3$-$2) spectra}
\end{figure}

\addtocounter{figure}{-1}

\begin{figure}
 \resizebox{\hsize}{!}{\includegraphics{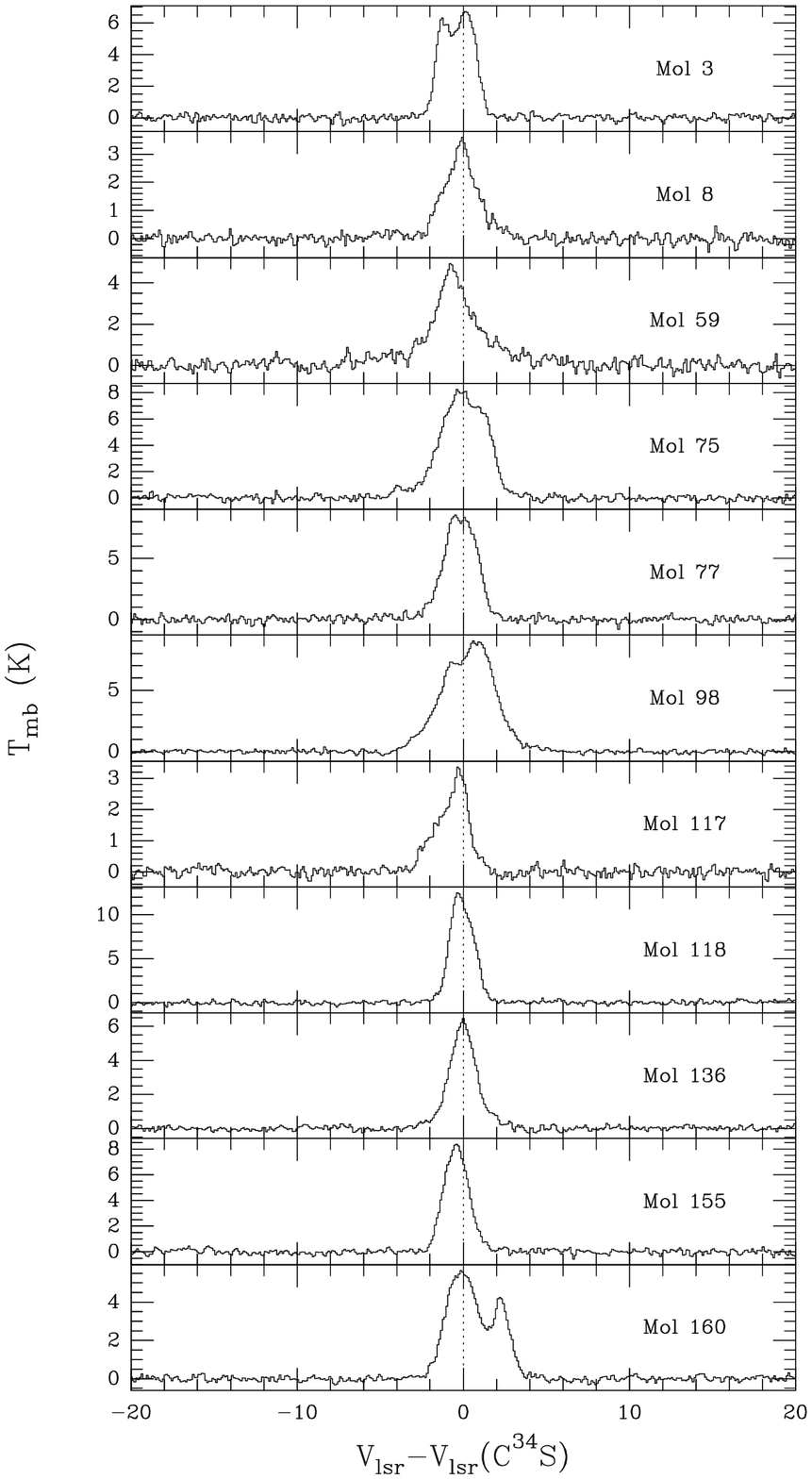}}
 \hfill
 \caption{{\bf d}\ As {\bf a}, but for the C$^{18}$O(2$-$1) spectra}
\end{figure}



\clearpage
\begin{figure}
 \resizebox{\hsize}{!}{\includegraphics{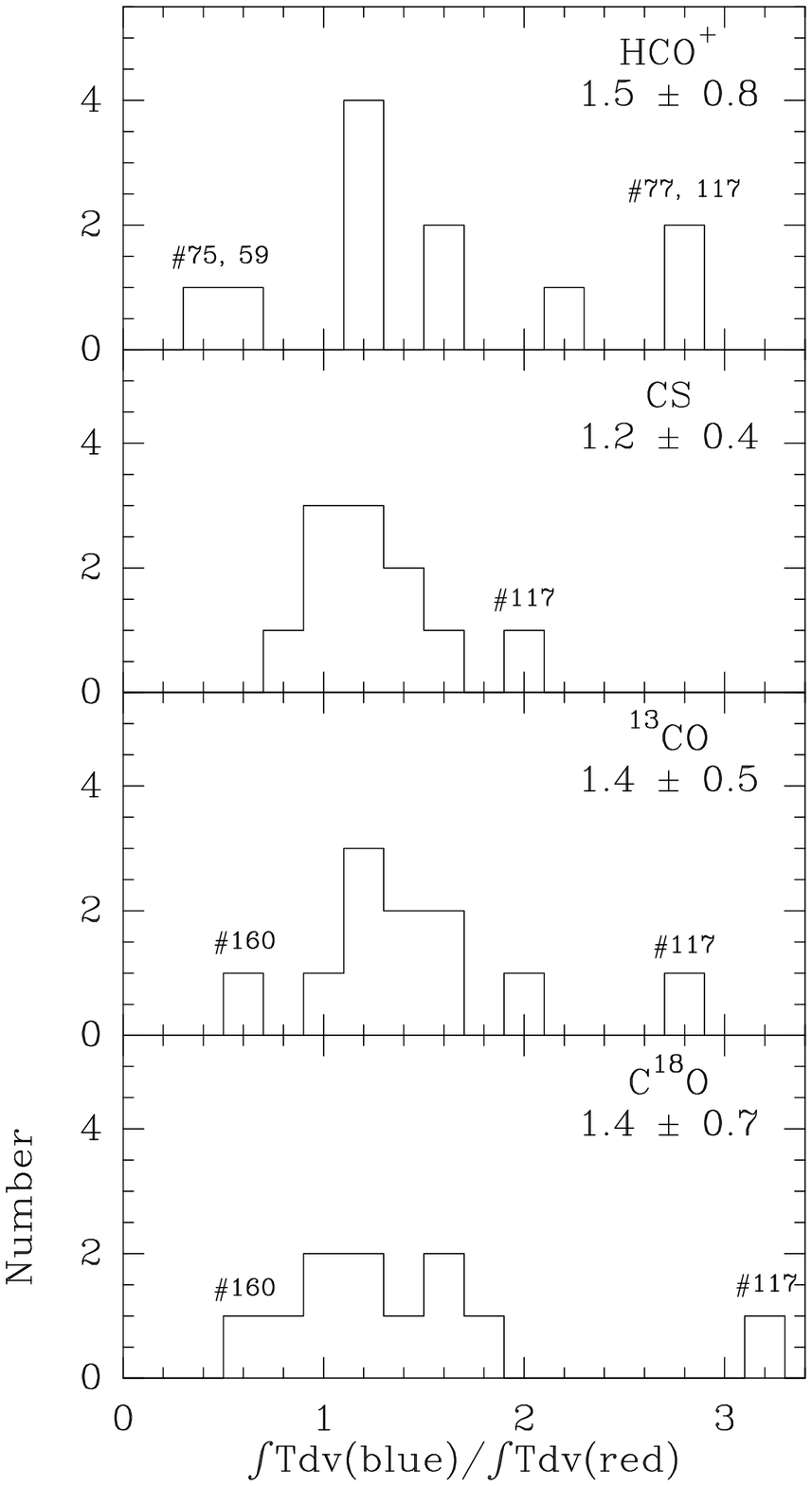}}
 \hfill
 \caption{Distribution of the ratio of $\int Tdv$ over the blue and red
(with respect to the C$^{34}$S velocity) parts of the spectra at the peak
positions, for the 
indicated tracers. For each molecule, the average ratio is indicated, as 
are the source numbers of the extreme values.}
\label{asymm}
\end{figure}



\begin{figure}
 \resizebox{\hsize}{!}{\includegraphics{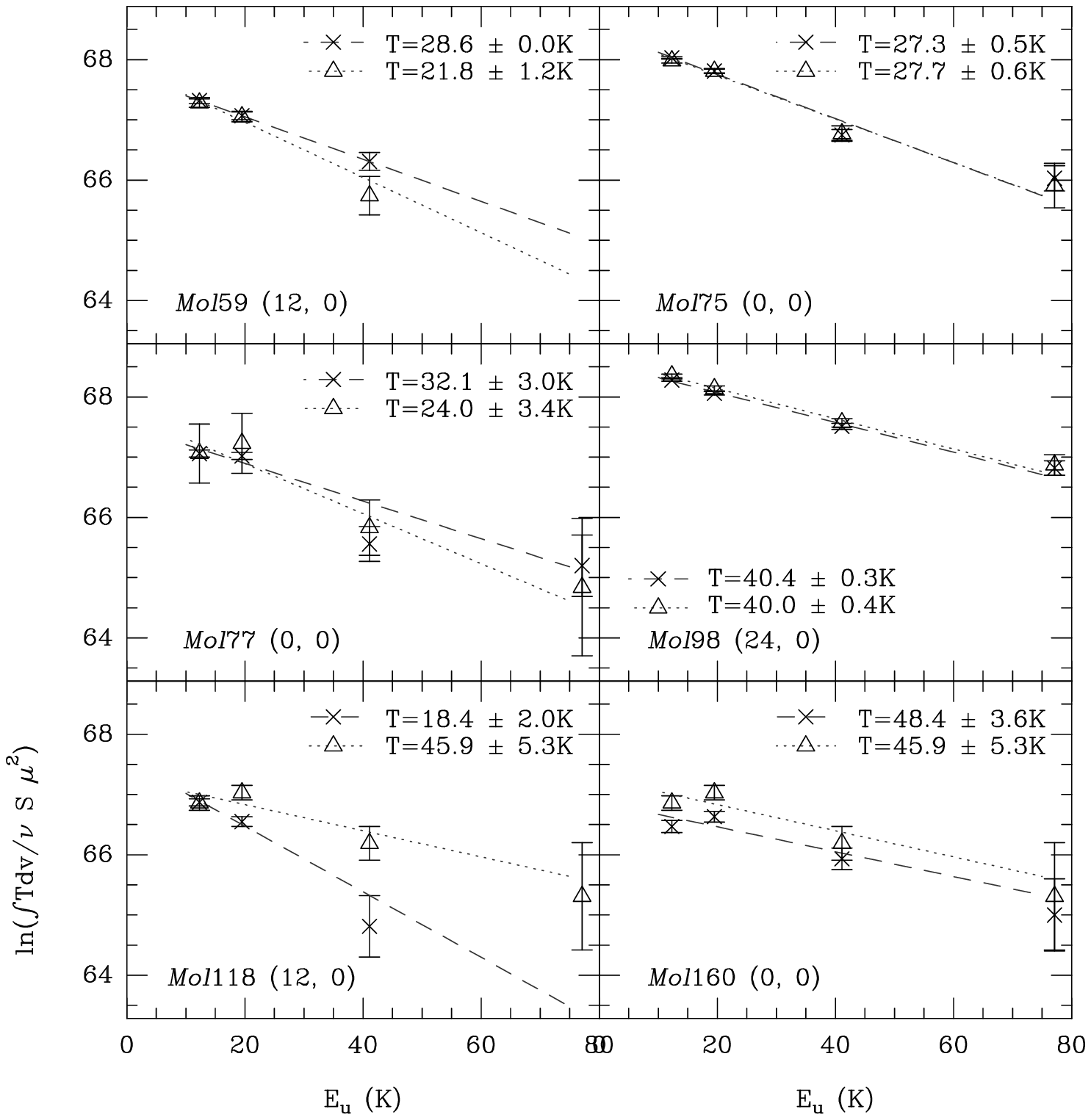}}
 \hfill
 \caption{Boltzmann plots, constructed from the CH$_3$C$_2$H(6--5)
observations. The results from the high- (0.23~kms$^{-1}$) and
low-resolution (2.92~kms$^{-1}$) spectra are shown as crosses and triangles,
respectively. The dashed and dotted lines are weighted least-squares fits 
through the
data points. The sources are identified in the panels, where also the
derived T$_{kin}$ is given. Full results are presented in
Table~\ref{ch3c2h_results}.}
\label{mol_boltz}
\end{figure}



\clearpage
\begin{figure}
 \resizebox{\hsize}{!}{\includegraphics{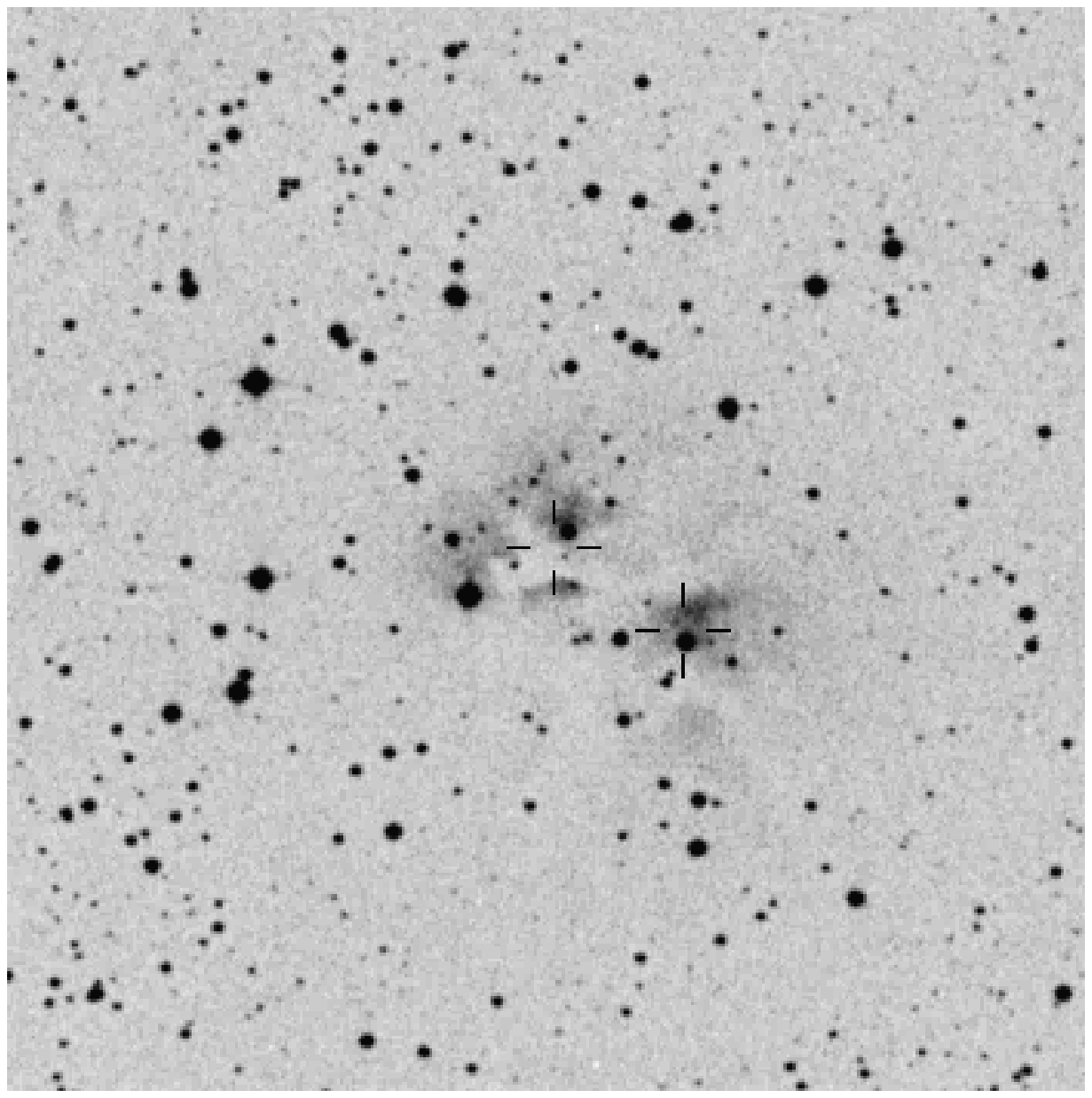}}
 \hfill
 \caption{Optical image (Digital Sky Survey) of a 10\arcmin
$\times$ 10\arcmin\ region around Mol~3. N is up, E is left; the (0,0)
position of the mapped region is indicated by the easternmost cross. The
westernmost cross indicates the location of the 6-cm continuum peak detected
by White \& Gee \cite{white} (GN0042.0$-$2; see text).}
\label{optical3}
\end{figure}



\begin{figure}
 \resizebox{\hsize}{!}{\includegraphics{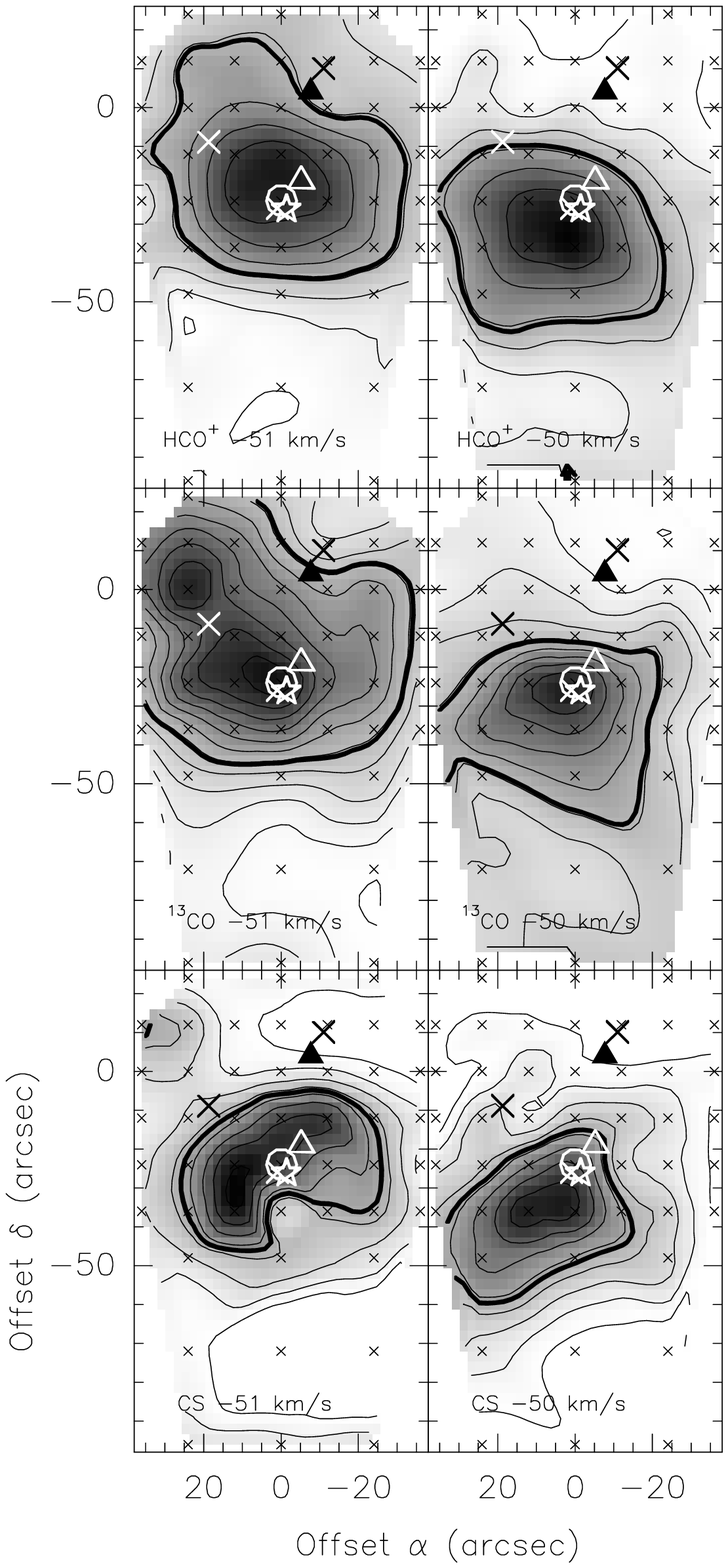}}
 \hfill
 \caption{\ Mol~3 data. Integrated emission of the red and blue emission
(with respect to $-50.85$~kms$^{-1}$) for the indicated molecules. Red and
blue integration limits are $-55, -48$ (HCO$^+$), $-54, -48.5$ ($^{13}$CO), 
and $-54, -48$ (CS). The symbols have the following meaning: filled
triangle --
IRAS source; open circle -- molecular peak; open triangle -- 850~$\mu$m
peak; open star -- H$_2$O maser position; big crosses -- VLA-D array 3.6~cm
peaks. Small crosses indicate the observed positions. Contour values are
0.3(0.5)3.9, 0.2(0.5)3.2~Kkms$^{-1}$ (HCO$^+$ blue, red); 5.5(3)41.5,
2(3)31~Kkms$^{-1}$ ($^{13}$CO blue, red); 0.2(0.4)3.2,
0.2(0.5)4.2~Kkms$^{-1}$ (CS blue, red); the thick contours indicate the FWHM
level.}
\label{mol3bluered}
\end{figure}



\clearpage
\begin{figure*}

\resizebox{\hsize}{!}{\rotatebox{-90}{\includegraphics{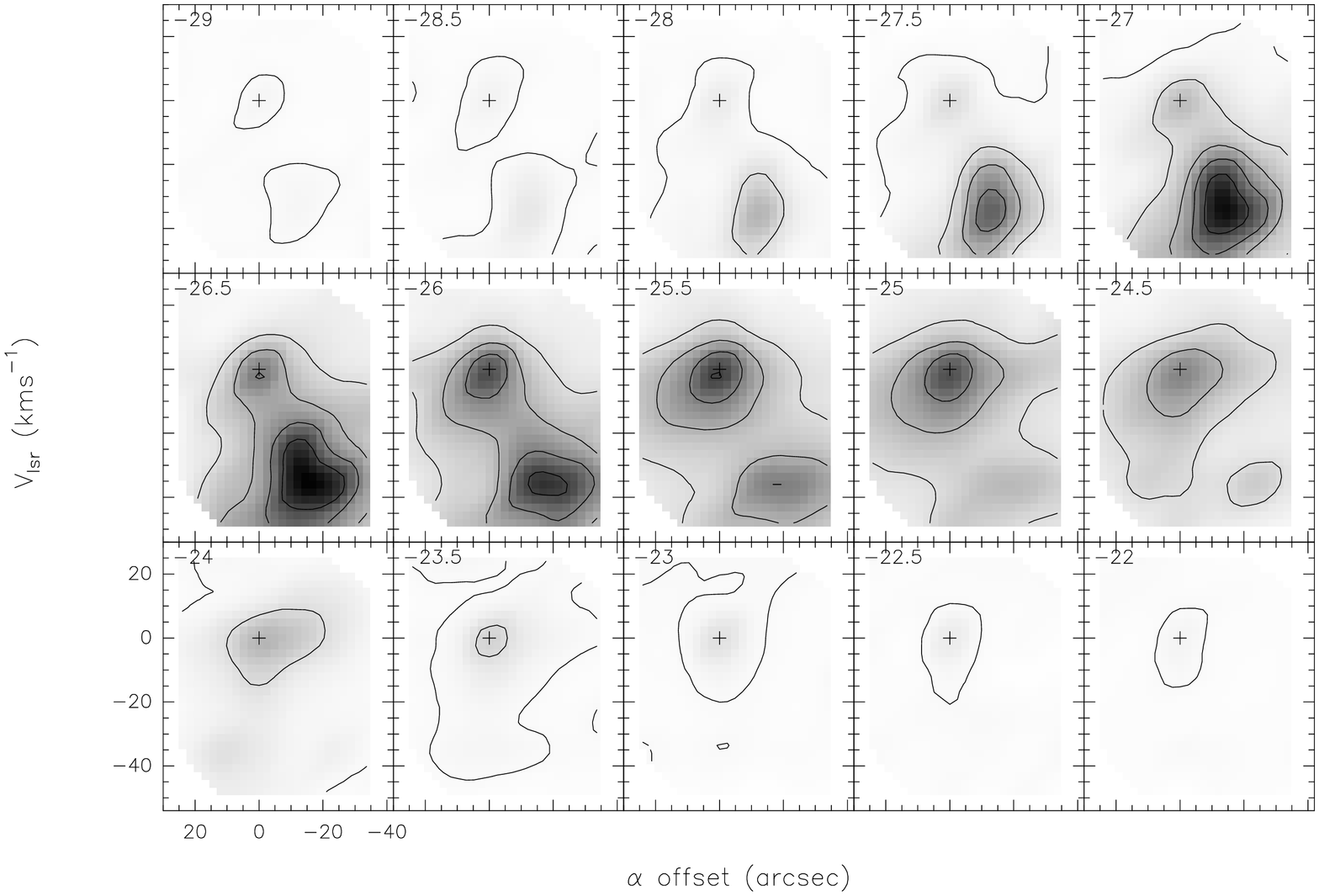}}}
 \hfill
 \caption{{\bf a}\ Mol~8 data. Integrated emission of the $^{13}$CO line,
shown in steps of 0.5~kms$^{-1}$. The cross indicates the position of the
sub-mm peak ($\approx$ IRAS source position). Contours 0.5(4)17~Kkms$^{-1}$.
The HCO$^+$ and CS emission show the same behaviour, and are not shown}
\label{mol8chan}
\end{figure*}



\clearpage
\begin{figure}

\resizebox{\hsize}{!}{\includegraphics{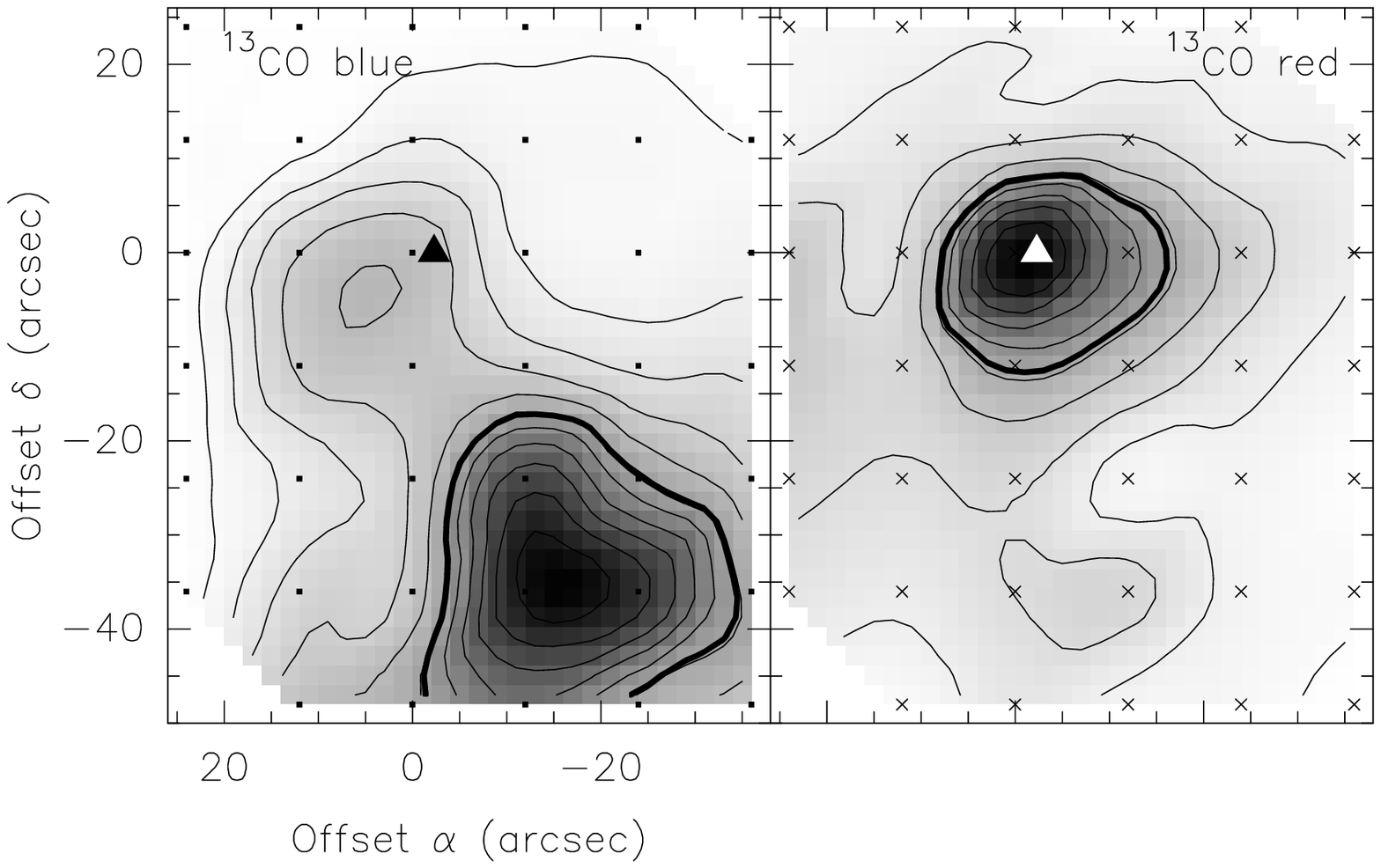}}
 \hfill
 \caption{Mol~8 data. Emission under the $^{13}$CO lines, integrated over the
blue and red components (separated by Gaussian fits). Contour values are
2(4)45~Kkms$^{-1}$ (blue) and 2(4)38~Kkms$^{-1}$ (red), respectively. The
thick contours indicate the FWHM level. The triangle represents the IRAS
position.}
\label{mol8gauss}
\end{figure}



\begin{figure}

\resizebox{\hsize}{!}{\includegraphics{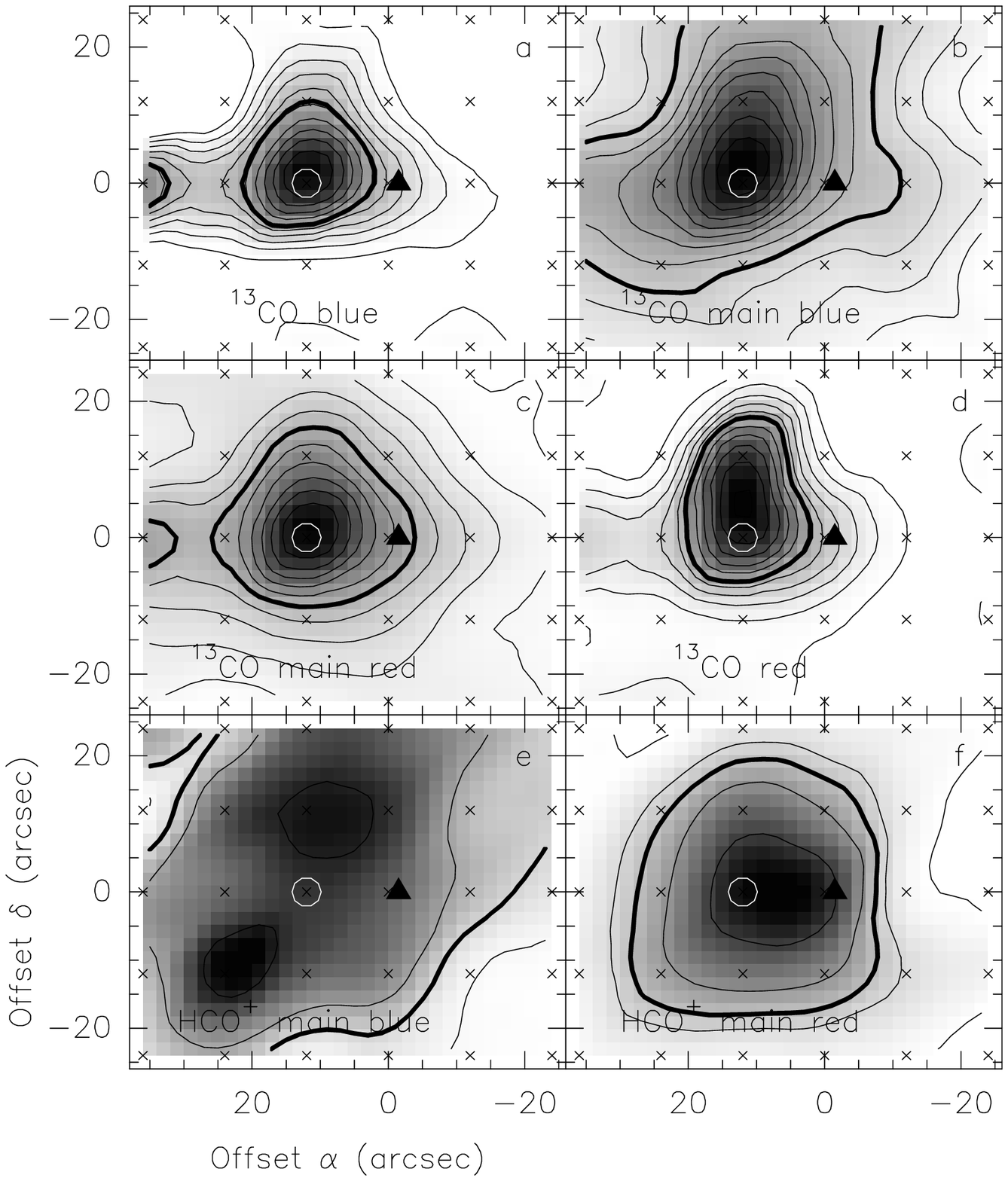}}
 \hfill
 \caption{Mol~59 data. Emission in the $^{13}$CO (top 4 panels) and HCO$^+$
(bottom 2 panels) lines, integrated over different velocity intervals:
{\bf a}: 105$-$111~kms$^{-1}$ [lowest contour 0.5, step 1~Kkms$^{-1}$]; {\bf
b}: 111$-$114.3~kms$^{-1}$ [5.5, 1.5~Kkms$^{-1}$]; {\bf c}:
114.3$-$117~kms$^{-1}$ [1.5, 1.5~Kkms$^{-1}$]; {\bf d}: 117$-$122~kms$^{-1}$
[0.5, 1~Kkms$^{-1}$]; {\bf e}: 111$-$114.3~kms$^{-1}$ [0.5,
0.5~Kkms$^{-1}$]; {\bf f}: 114.3$-$117~kms$^{-1}$ [0.5, 0.5~Kkms$^{-1}$].
Thick contours in each panel indicate the half-maximum level. The IRAS
position is indicated by the triangle, the white circle is the position
where longer integrations were made. Small crosses are the observed positions.}
\label{mol59areas}
\end{figure}



\clearpage
\begin{figure}

\resizebox{\hsize}{!}{\includegraphics{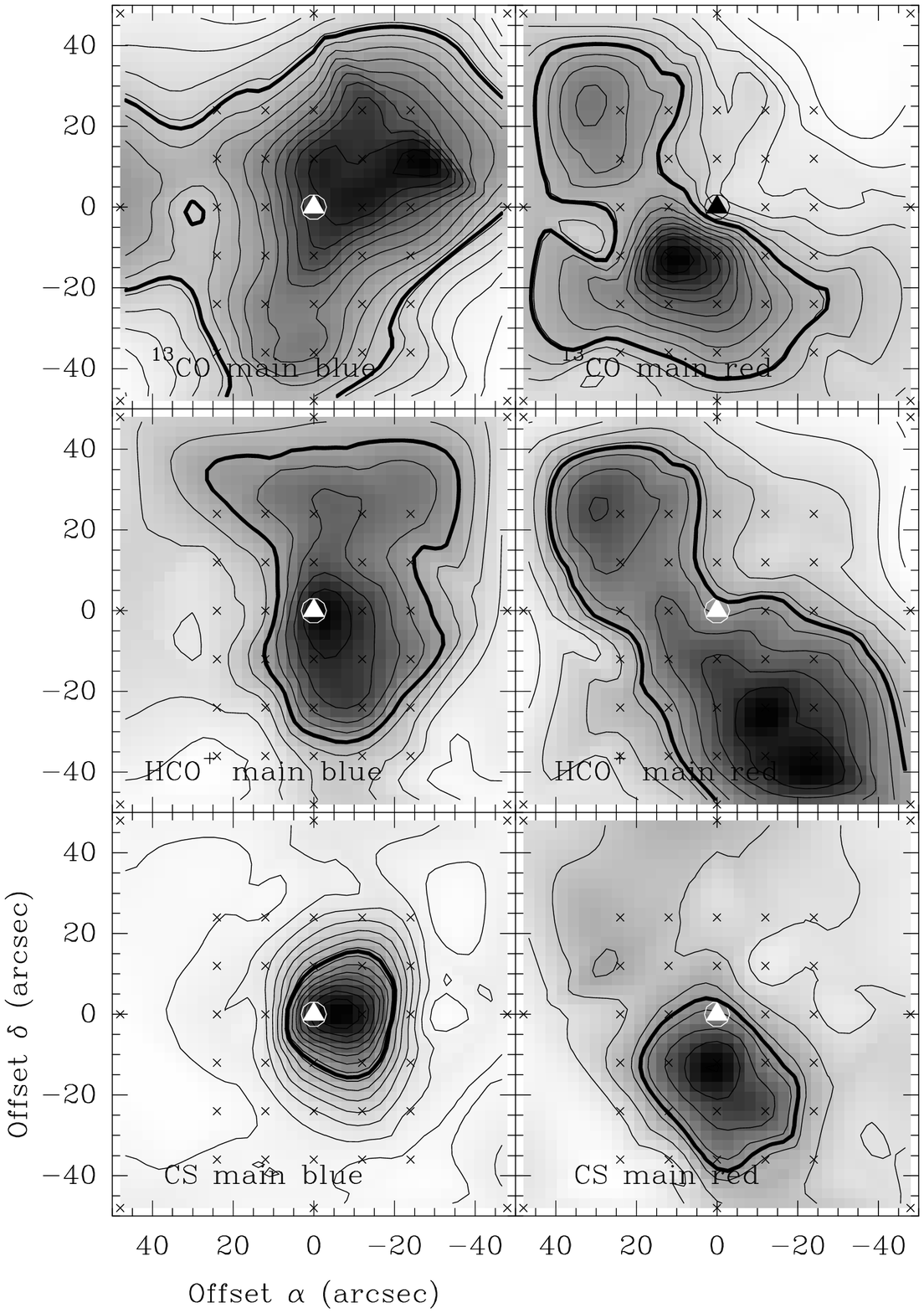}}
 \hfill
 \caption{Mol~75 data. Emission in the $^{13}$CO (top 2 panels), HCO$^+$
(middle 2 panels), and CS (bottom 2 panels) lines, integrated over the blue
and red parts of the emission (see text).
Thick contours in each panel indicate the half-maximum level. The IRAS
position is indicated by the triangle, the white circle is the position
where longer integrations were made. Small crosses are the observed positions.
Contours: $^{13}$CO blue 8(2)37~Kkms$^{-1}$, red 4(2)36; HCO$^+$ blue
0.2(0.2)2, red 0.2(0.8)7.4; CS blue 1(1)15, red 1(1)7.}
\label{mol75bluered}
\end{figure}



\clearpage
\begin{figure*}

\resizebox{\hsize}{!}{\rotatebox{-90}{\includegraphics{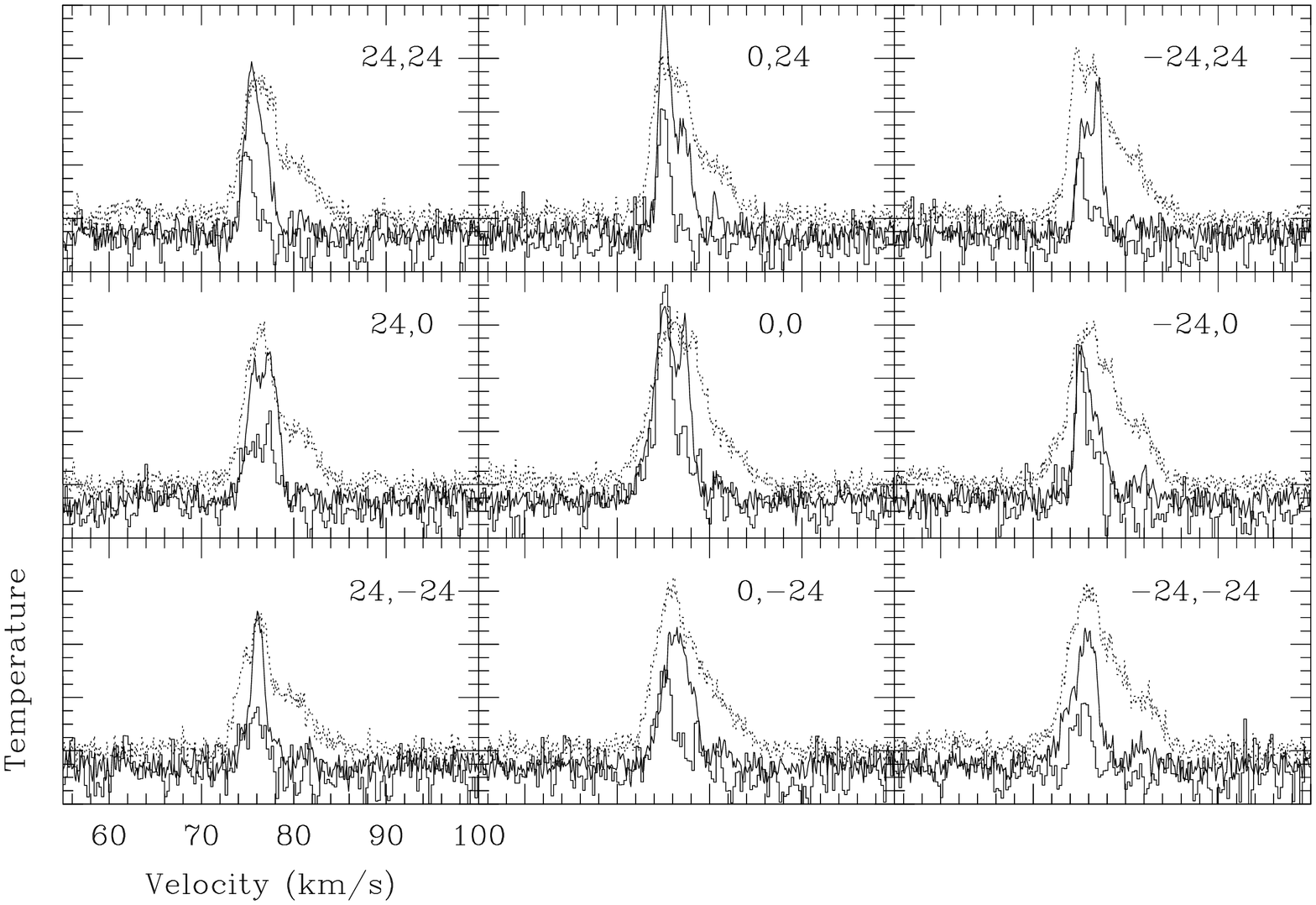}}}
 \hfill
 \caption{Mol~77 data. Comparison of line profiles of $^{12}$CO (dashed),
$^{13}$CO (drawn), and HCO$^+$ (histogram), in the central part of the cloud.
The vertical scale is temperature, but scaled such as to make all lines
clearly visible.
The NRAO 12-m $^{12}$CO spectra were actually taken on a 29\arcsec\ grid, 
rather than 24\arcsec. The beam of the $^{12}$CO observations ($\sim$29\arcsec)
is comparable to that of the IRAM 30-m HCO$^+$ ($\sim$27\arcsec); the
$^{13}$CO observations were made with a $\sim$11\arcsec\ beam. These plots
suggest that the dip/shoulder in the IRAM spectra is probably due to the
superposition in velocity of 2 components.}
\label{mol77grid}
\end{figure*}



\clearpage
\begin{figure}

\resizebox{\hsize}{!}{\includegraphics{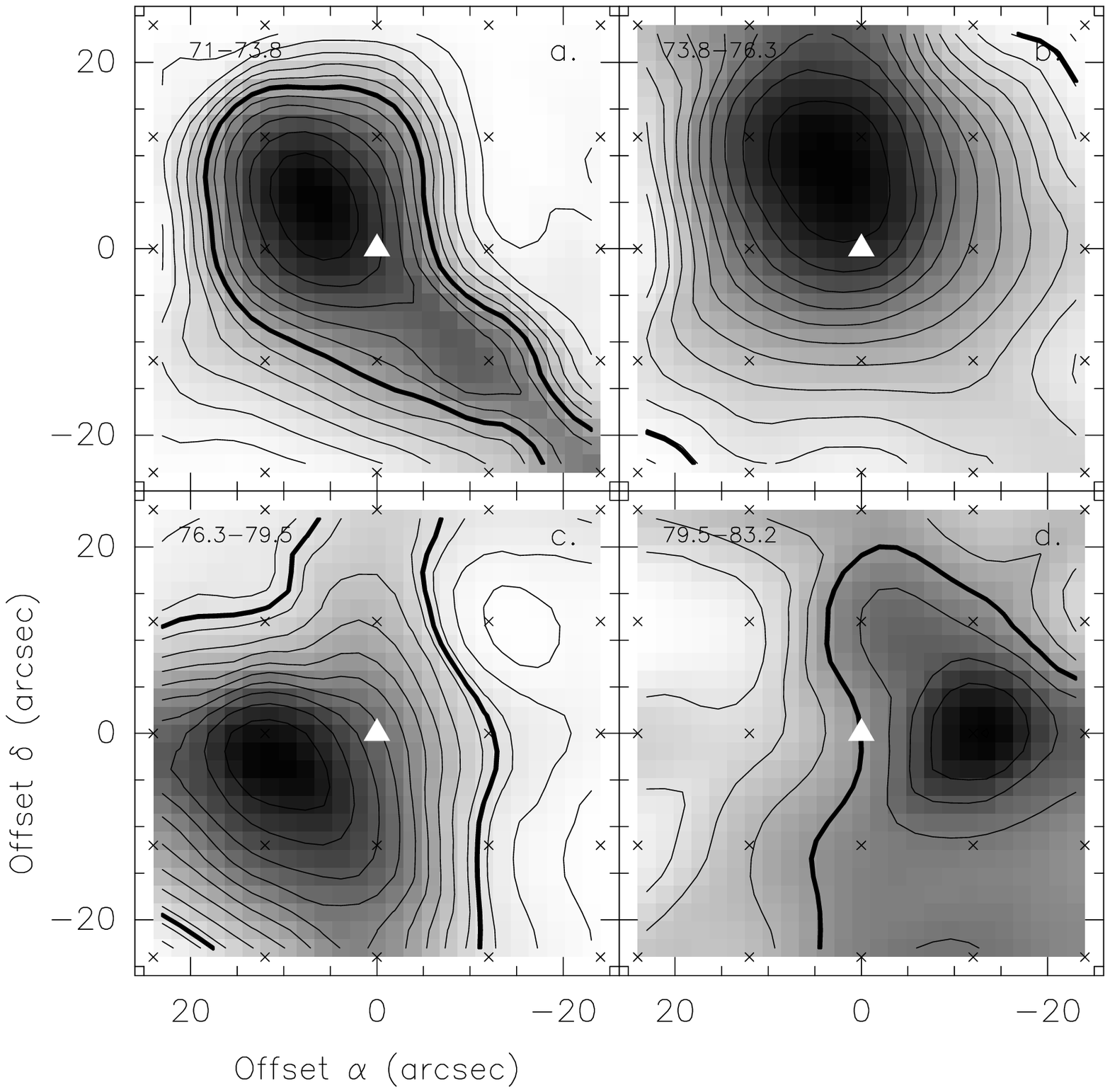}}
 \hfill
 \caption{{\bf a.}\ Distribution of $\int Tdv$ of the $^{13}$CO emission
towards Mol~77,
integrated over the velocity ranges indicated in the panels. In each panel
an individual velocity component is isolated. 
Thick contours in each panel indicate the half-maximum level. The IRAS
position is indicated by the triangle, and is identical to the position
where longer integrations were made. Small crosses are the observed positions.
Contours: 0.5(0.5)6~Kkms$^{-1}$ (panel a.); 10(1)22 (b.); 
6(1)19 (c.); 0.5(0.5)4 (d.)}
\label{mol77areas}
\end{figure}

\addtocounter{figure}{-1}

\begin{figure}

\resizebox{\hsize}{!}{\includegraphics{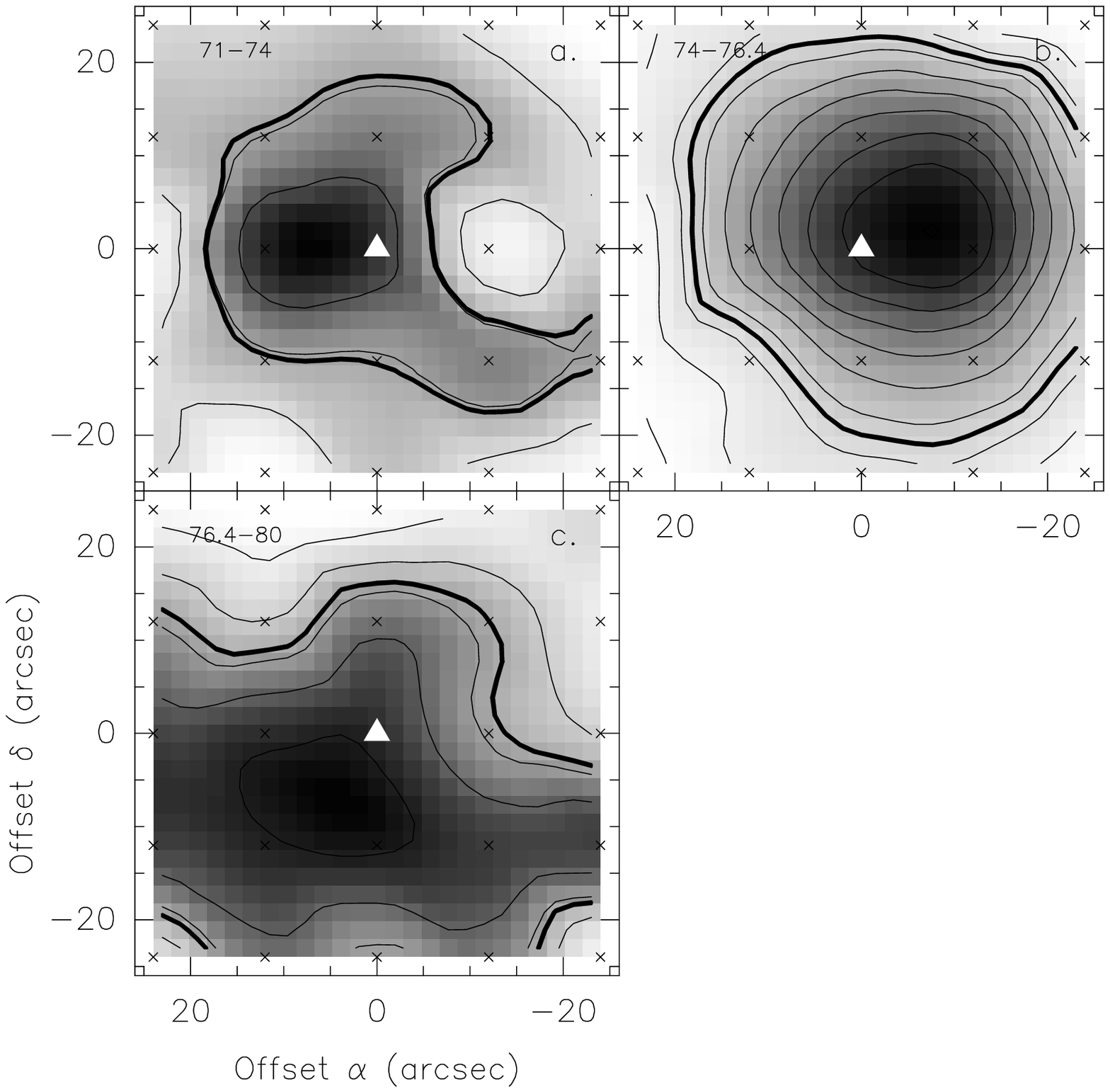}}
 \hfill
 \caption{{\bf b.}\ As {\bf a}, but for HCO$^+$. 
Contours: 0.2(0.2)0.8~Kkms$^{-1}$ (panel a.); 0.8(0.2)2.6 (b.); 
0.2(0.2)1.2 (c.)}
\end{figure}

\addtocounter{figure}{-1}

\begin{figure}

\resizebox{\hsize}{!}{\includegraphics{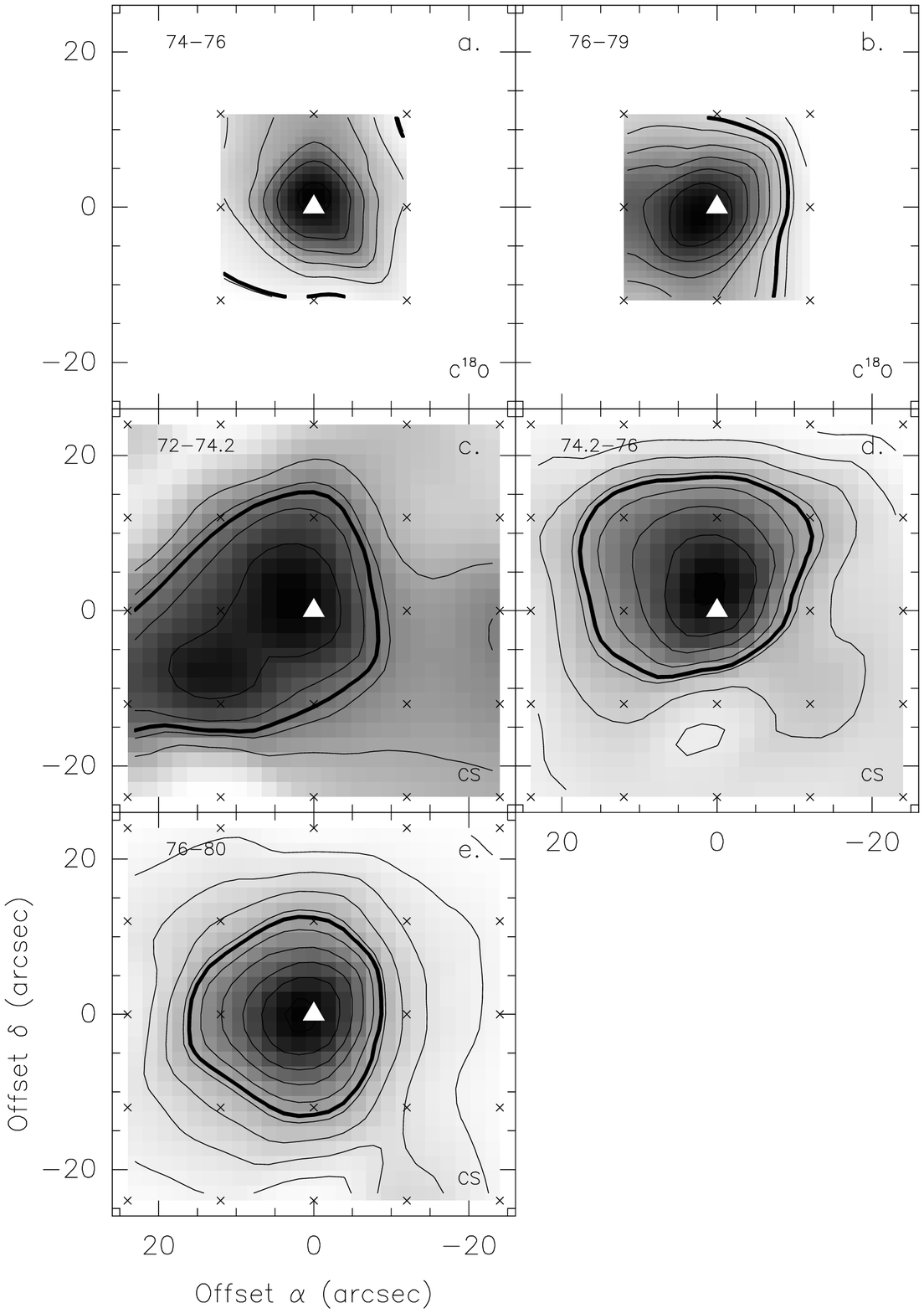}}
 \hfill
 \caption{{\bf c.}\ As {\bf a}, but for C$^{18}$O (panels a, b) and
CS (panels c-e). 
Contours: 5(1)10~Kkms$^{-1}$ (panel a.); 3(1)11 (b.); 
0.2(0.2)1 (c.); 0.2(0.4)3.4 (d.); 0.2(0.4)4 (e.)}
\end{figure}



\clearpage
\begin{figure}

\resizebox{\hsize}{!}{\includegraphics{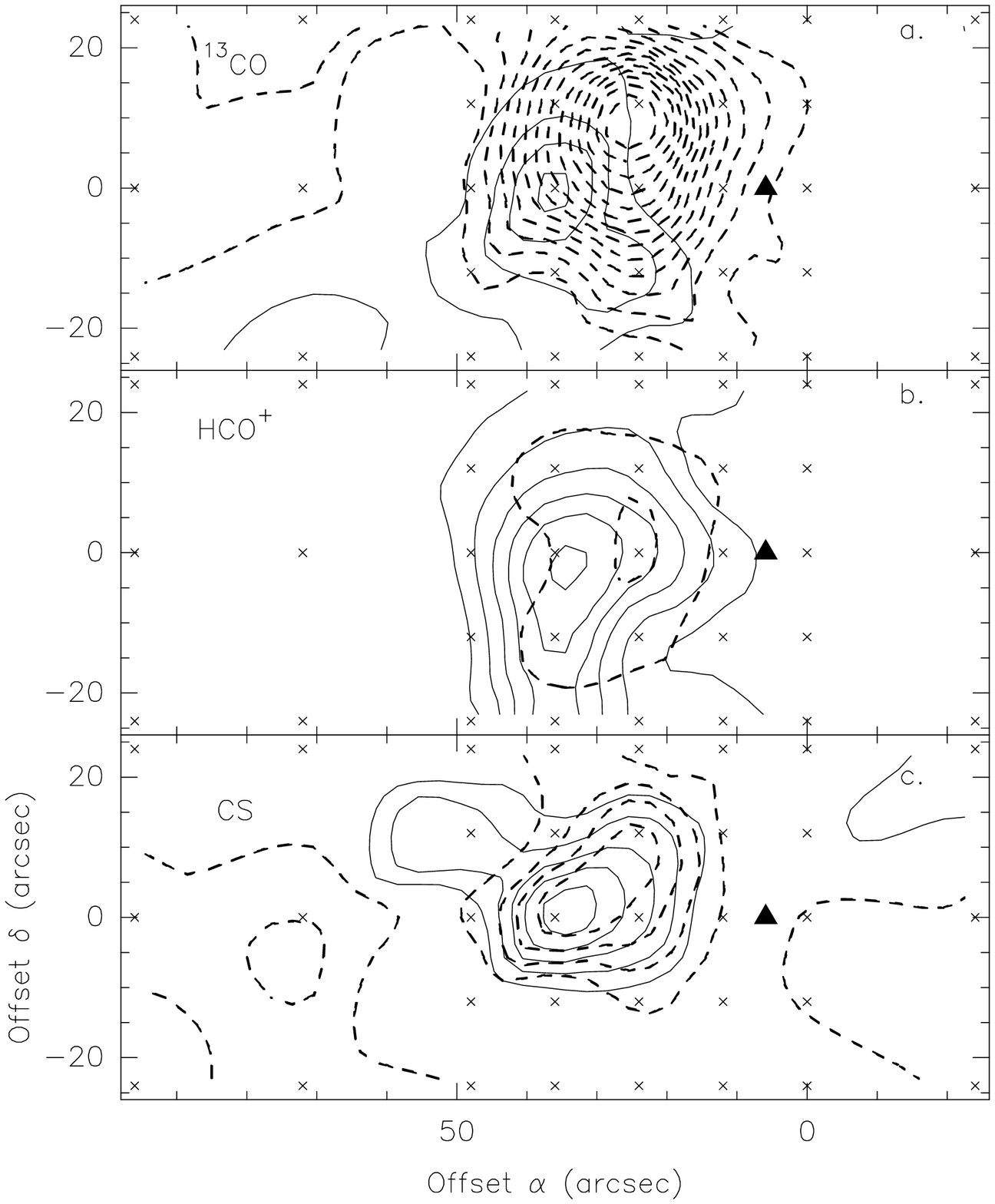}}
 \hfill
 \caption{Mol~98 data. Emission in the wings of the profiles of
$^{13}$CO (a.), HCO$^+$ (b.), and CS (c.). Drawn and dashed contours refer to
blue and red wings, respectively. Integrated velocity interval and contour 
values (blue, red: range, low(step)max) are:
52--54: 1(1)4, 60--65: 1(1)14~Kkms$^{-1}$ (a.); 45--54: 0.5(0.5)3, 60--64: 
0.5(0.5)1~Kkms$^{-1}$ (b.); 48--54: 0.5(0.5)2.5, 61--65: 
0.5(0.5)2.5~Kkms$^{-1}$ (c.)}
\label{mol98wings}
\end{figure}



\begin{figure}

\resizebox{\hsize}{!}{\includegraphics{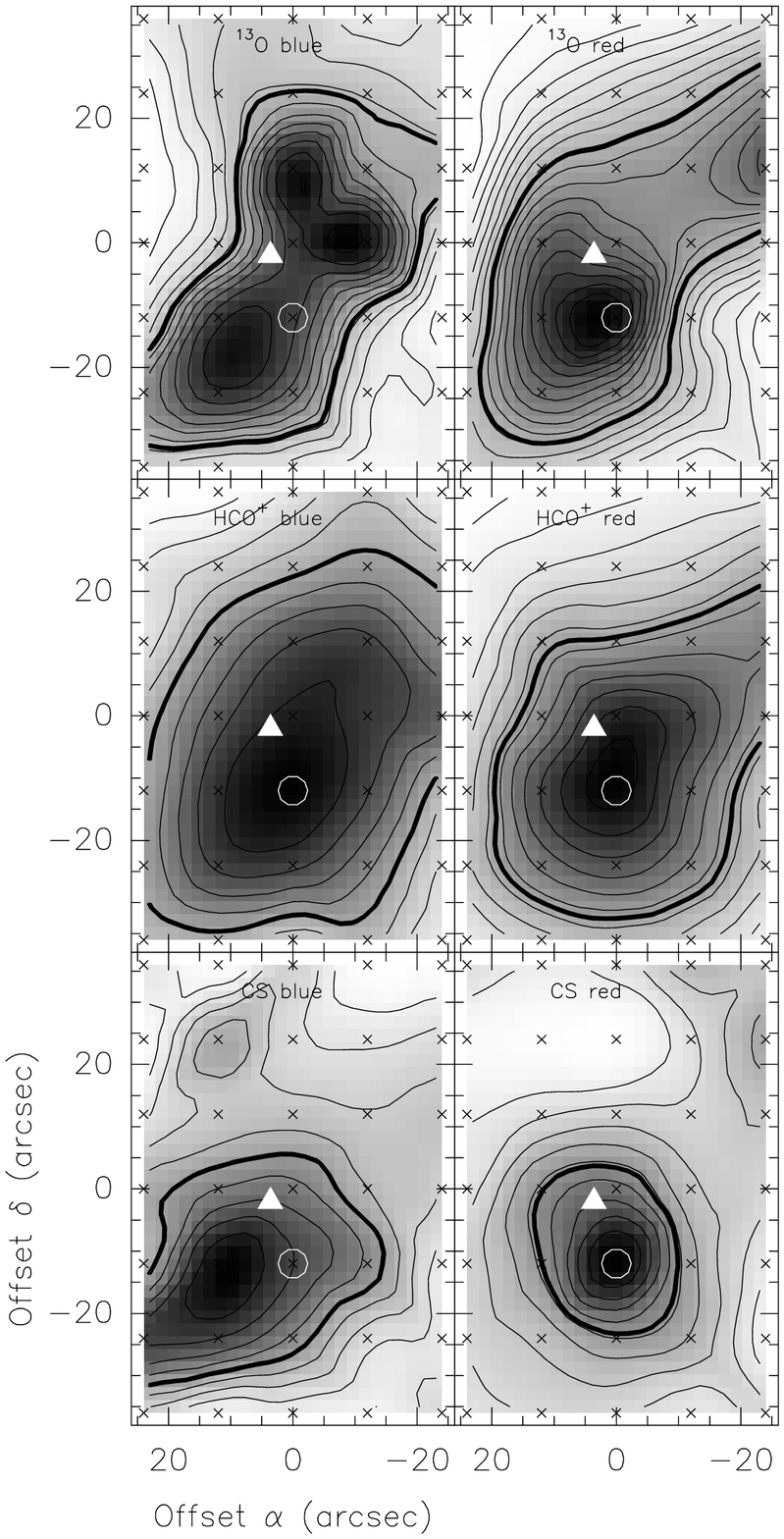}}
 \hfill
 \caption{Mol~117 data. Emission in the $^{13}$CO (top 2 panels), HCO$^+$
(middle 2 panels), and CS (bottom 2 panels) lines, integrated over the blue
and red parts of the emission (see text).
Thick contours in each panel indicate the half-maximum level. The IRAS
position is indicated by the triangle, the white circle is the position
where longer integrations were made. Small crosses are the observed positions.
Contours: $^{13}$CO blue 3.5(1.5)23~Kkms$^{-1}$, red 4.5(2.5)49.5; HCO$^+$ blue
0.2(0.2)2, red 0.2(0.4)4.8; CS blue 0.2(0.2)2, red 0.2(0.4)3.8.}
\label{mol117bluered}
\end{figure}



\clearpage
\begin{figure*}

\resizebox{\hsize}{!}{\rotatebox{-90}{\includegraphics{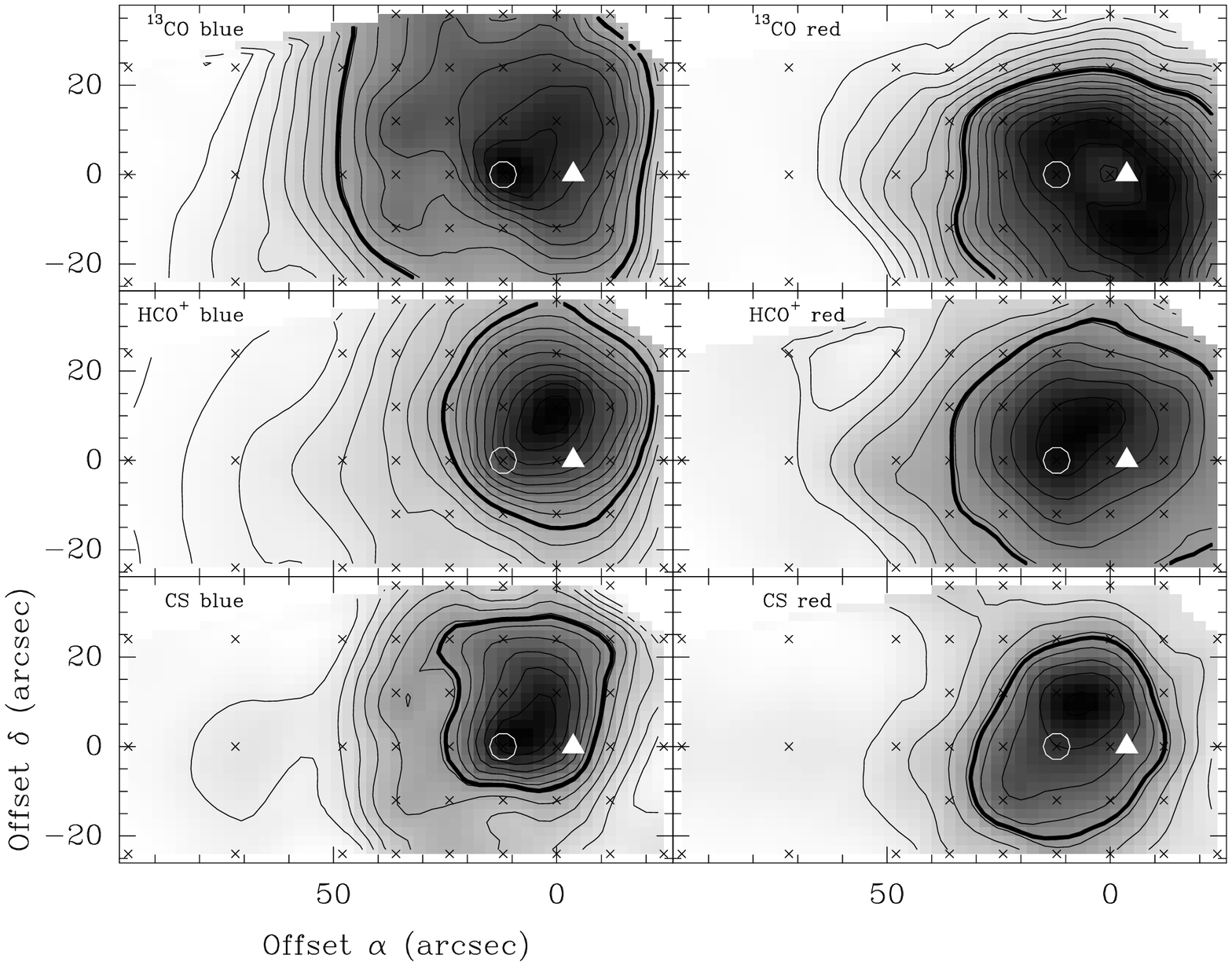}}}
 \hfill
 \caption{Mol~118 data. Emission in the $^{13}$CO (top 2 panels), HCO$^+$
(middle 2 panels), and CS (bottom 2 panels) lines, integrated over the blue
and red parts of the emission (see text).
Thick contours in each panel indicate the half-maximum level. The IRAS
position is indicated by the triangle, the white circle is the position
where longer integrations were made. Small crosses are the observed positions.
Contours: $^{13}$CO blue 2(3)45~Kkms$^{-1}$, red 1(1)15; HCO$^+$ blue
0.2(0.4)6.8, red 0.2(0.2)2; CS blue 0.2(0.4)4.6, red 0.2(0.4)3.4.}
\label{mol118bluered}
\end{figure*}



\clearpage
\begin{figure*}

\resizebox{12cm}{!}{\includegraphics{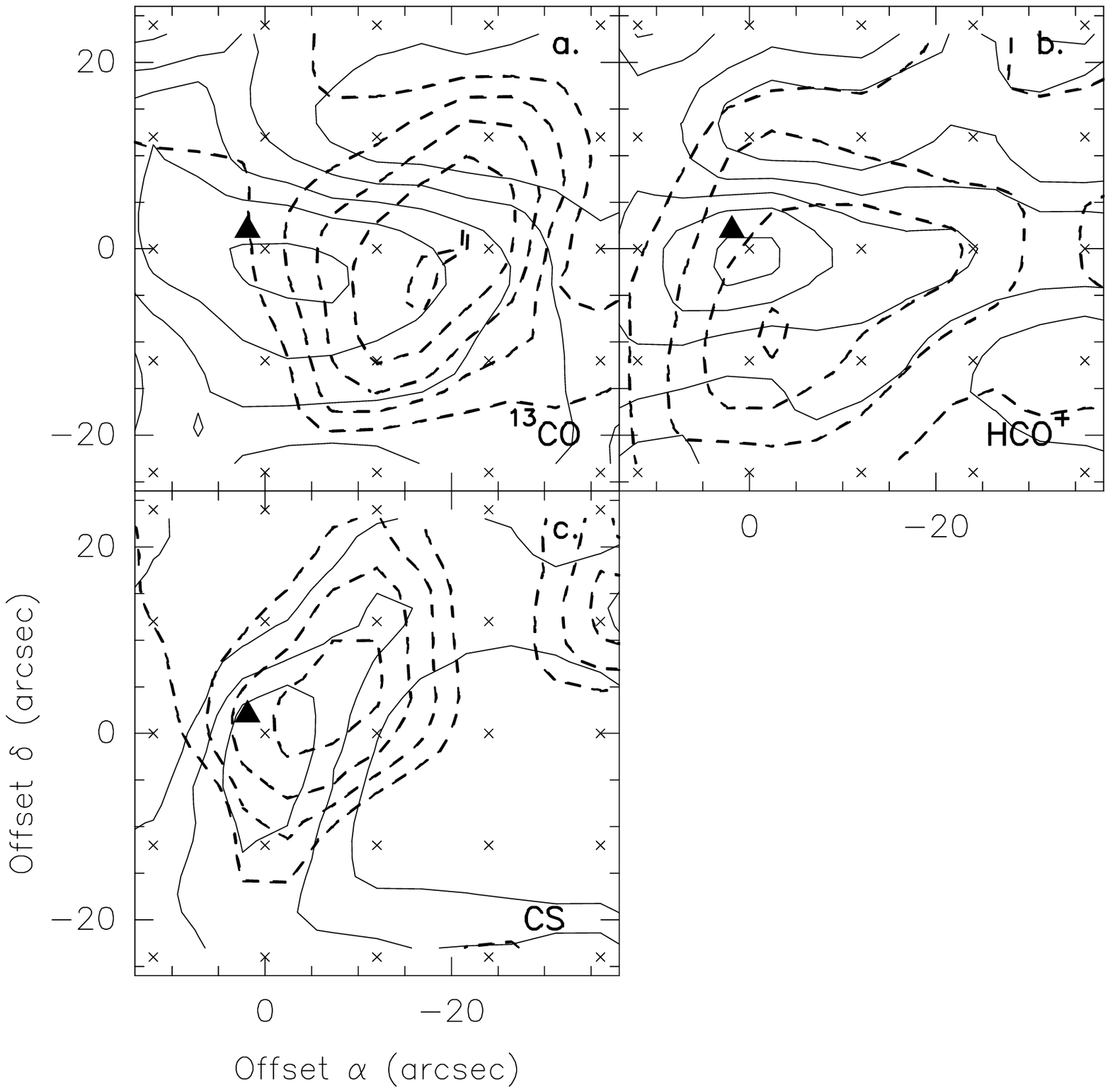}}
 \hfill
 \parbox[b]{55mm}{
 \caption{Mol~136 data. Emission in the wings of the profiles of
$^{13}$CO (a.), HCO$^+$ (b.), and CS (c.). Drawn and dashed contours 
refer to blue and red wings, respectively. Integrated velocity interval for
the blue emission is $-$51 to $-$48; for the red emission it's $-$44.5 to
$-$41 ($-$42 for CS). Contour values (low(step)max) are:
1(1)5~Kkms$^{-1}$ (a.); 0.2(0.2)1~Kkms$^{-1}$ (b., blue), 
0.2(0.2)0.8~Kkms$^{-1}$ (b., red);
0.2(0.3)1.1~Kkms$^{-1}$ (c., blue), and 0.2(0.2)1~Kkms$^{-1}$ (c., red). The
triangle indicates the position of the IRAS source.}
\label{mol136wings}}
\end{figure*}



\clearpage
\begin{figure*}

\resizebox{\hsize}{!}{\includegraphics{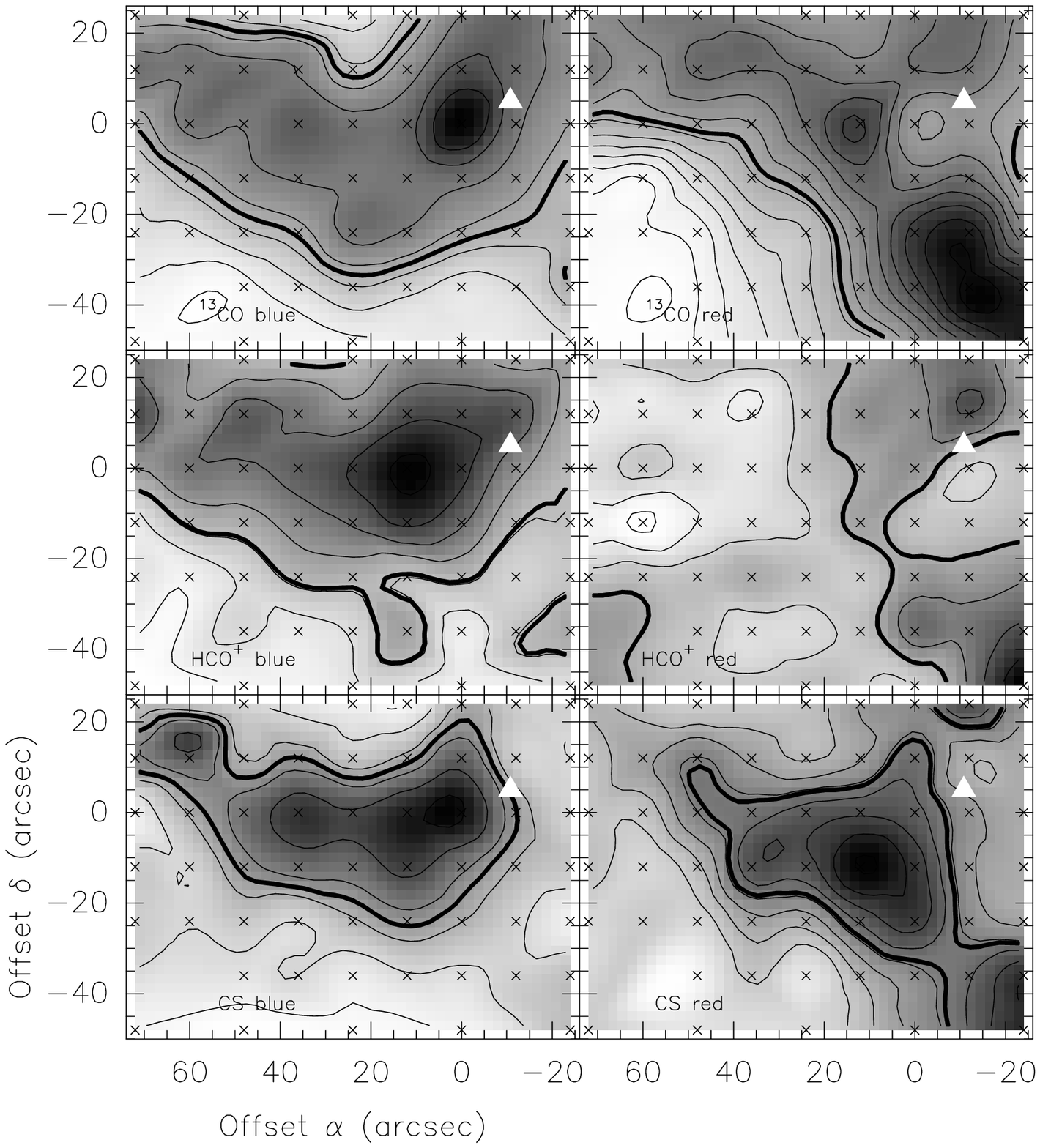}}
 \hfill
 \caption{Mol~155 data. Emission in the $^{13}$CO (top 2 panels), HCO$^+$
(middle 2 panels), and CS (bottom 2 panels) lines, integrated over the blue
and red parts of the emission (see text).
Thick contours in each panel indicate the half-maximum level. The IRAS
position is indicated by the triangle.
Small crosses are the observed positions.
Contours: $^{13}$CO blue 5(5)45~Kkms$^{-1}$, red 3(3)44; HCO$^+$ blue
0.5(0.5)4, red 0.2(0.4)3; CS blue 0.2(0.8)4.4, red 0.2(0.4)3.4.}
\label{mol155bluered}
\end{figure*}



\clearpage
\begin{figure*}

\resizebox{12cm}{!}{\includegraphics{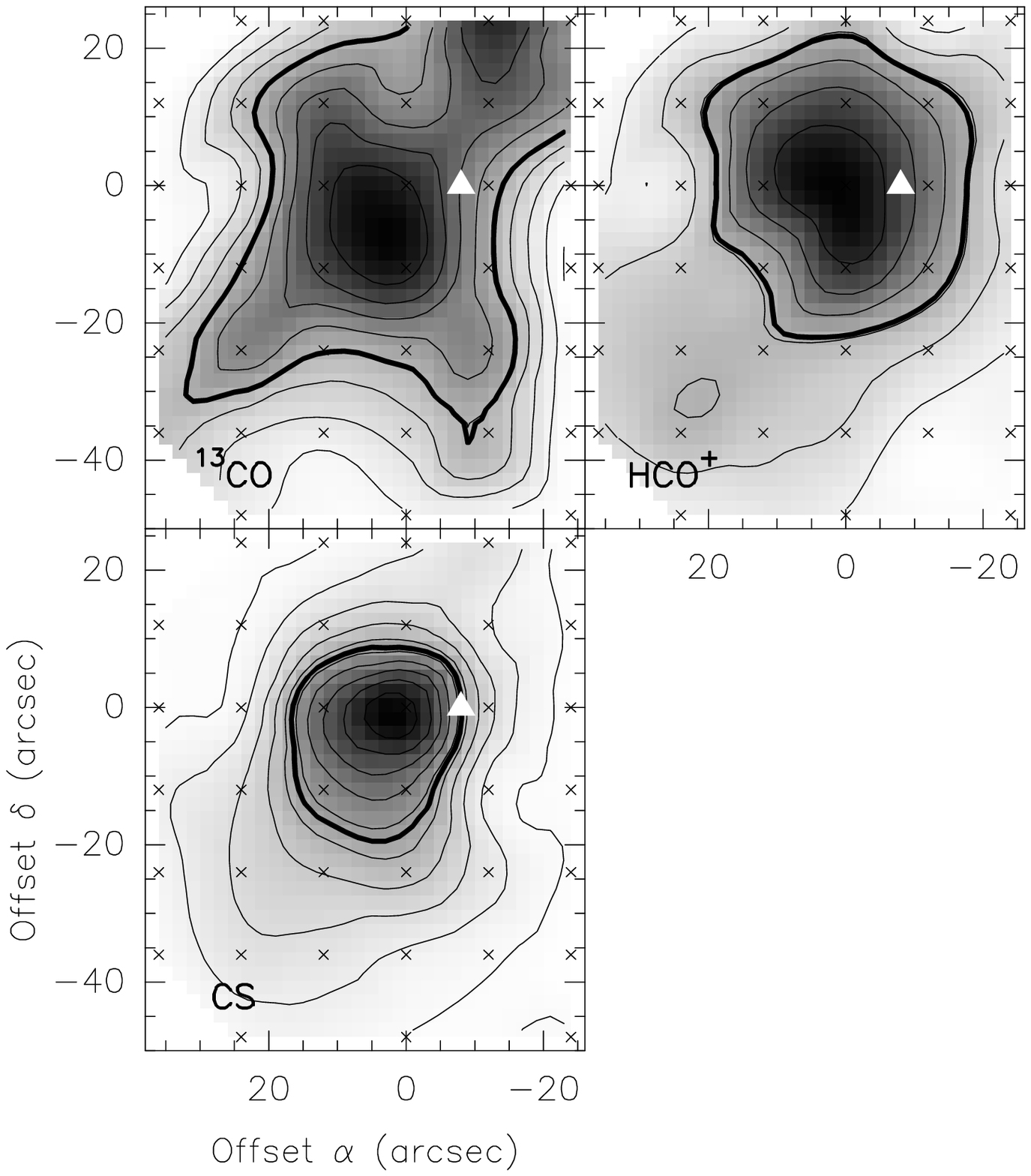}}
 \hfill
 \parbox[b]{55mm}{
 \caption{Mol~160 data. Area under the main component (subtracting
components using Gaussian fits), for the molecules marked in the panels. The
triangle indicates the IRAS position. Contour values are 7(6)61~Kkms$^{-1}$
($^{13}$CO); 1(2)14~Kkms$^{-1}$ (HCO$^+$); 1(2)19~Kkms$^{-1}$ (CS). The
thick contours mark the FWHM values.}
\label{mol160mainarea}}
\end{figure*}



\clearpage
\begin{figure*}

\resizebox{\hsize}{!}{\rotatebox{-90}{\includegraphics{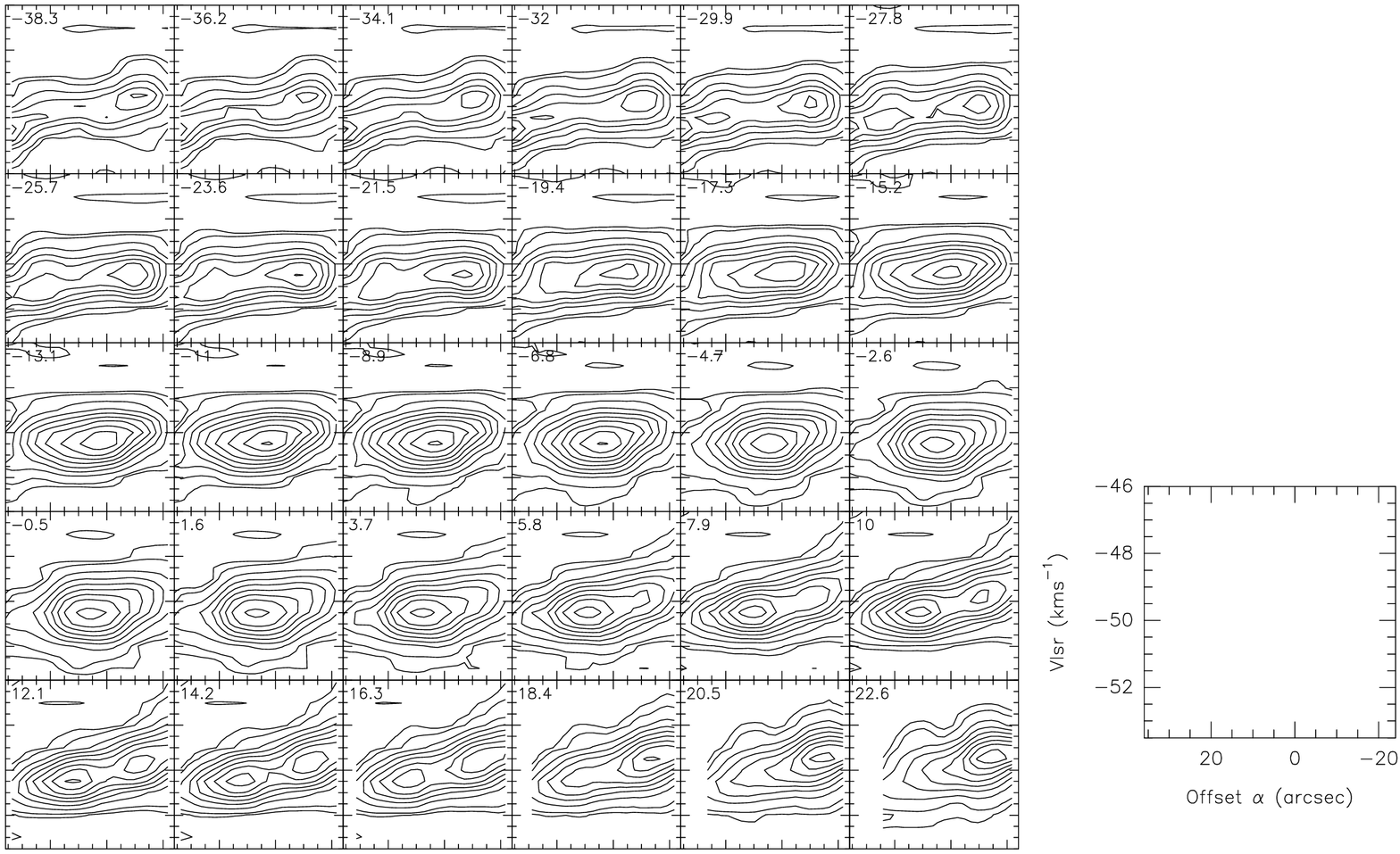}}}
 \hfill
 \caption{Mol~160 data. Right ascension - velocity plots of the 
$^{13}$CO line at the declination offsets marked in each panel. Contour
values are 1, 2(3)27~K.}
\label{mol160rv}
\end{figure*}



\clearpage
\begin{figure}

\resizebox{\hsize}{!}{\includegraphics{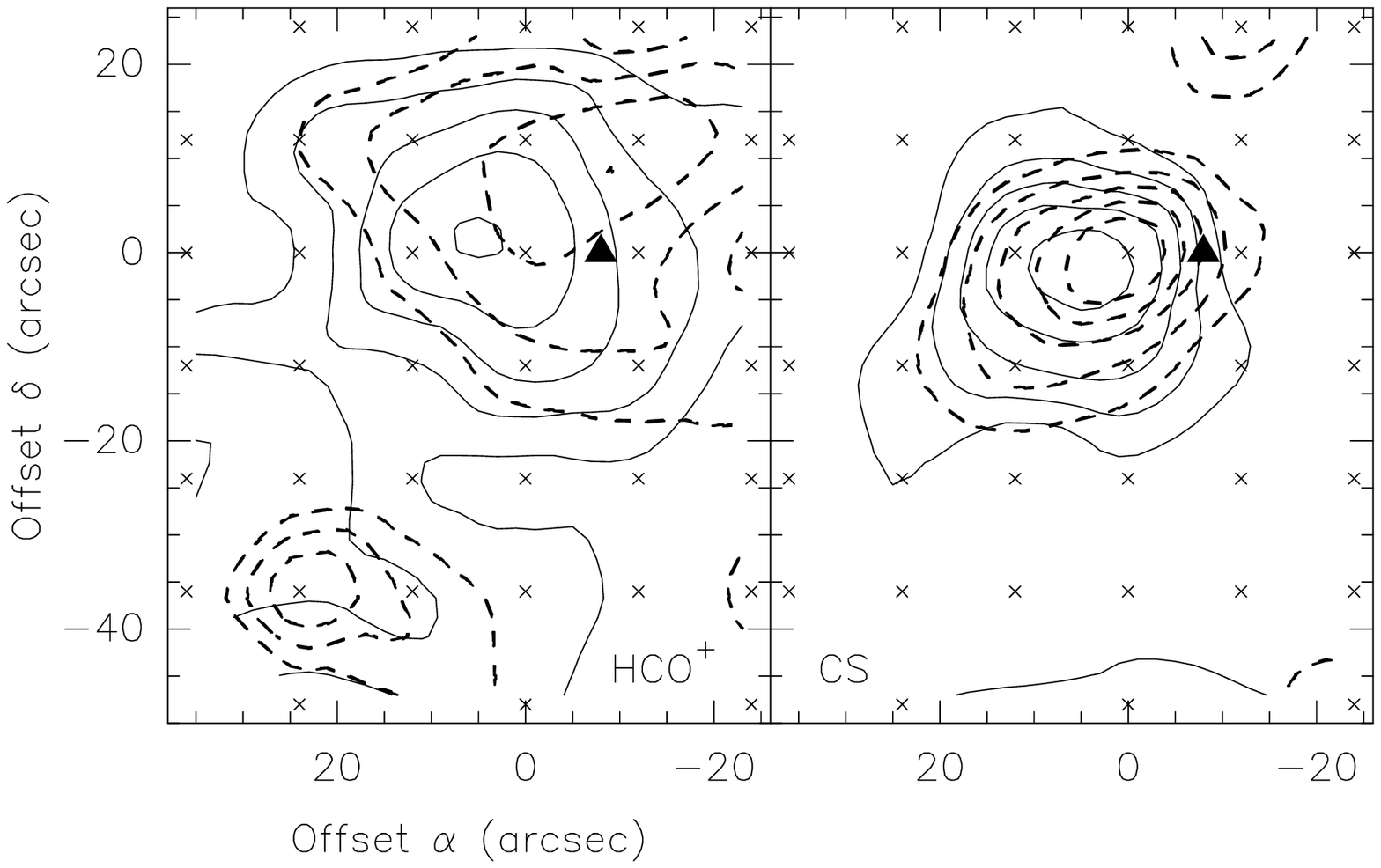}}
 \hfill
 \caption{Mol~160 data. Emission in the wings of the profiles of
HCO$^+$ and CS. Drawn and dashed contours
refer to blue and red wings, respectively. Integrated velocity interval for
the blue emission is $-$53 to $-$52 (HCO$^+$), $-$53 to $-$51.5 (CS); for the
red emission it's $-$48.5 to $-$47 (HCO$^+$), $-$49 to $-$47.5 (CS). Contour 
values (low(step)max) are: 0.2(0.2)1~Kkms$^{-1}$ (HCO$^+$); 
0.5(0.5)3~Kkms$^{-1}$ (CS). The triangle indicates the position of the IRAS 
source.}
\label{mol160wings}
\end{figure}



\begin{figure}

\resizebox{\hsize}{!}{\includegraphics{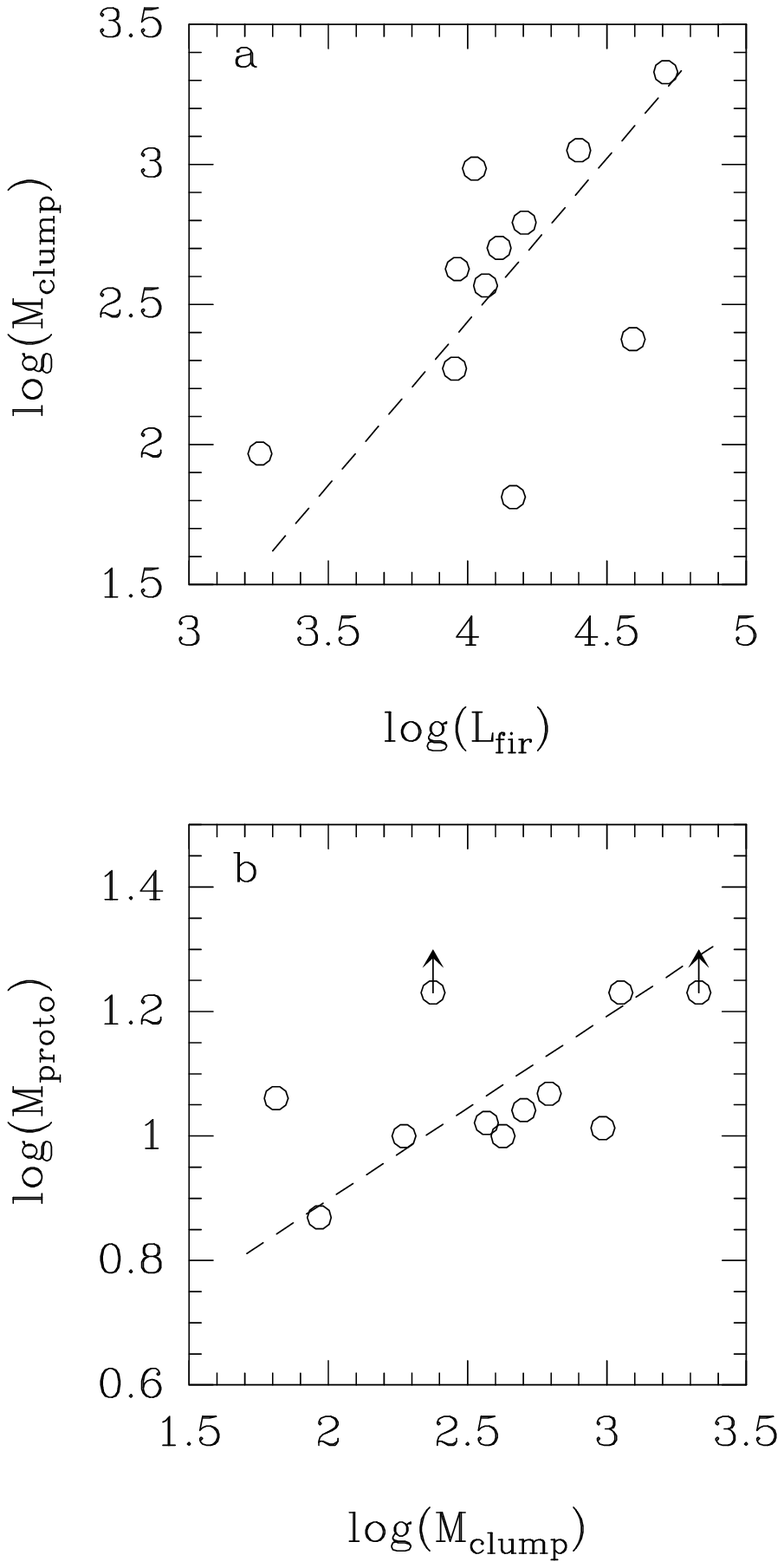}}
 \hfill
 \caption{{\bf a.}\ Relation between clump mass (data from Table~\ref{clmass})
and luminosity (data from Table~\ref{sources}). The least-squares (bisector) 
fit is shown as the dashed line, which has a slope of 1.17 $\pm$ 0.22; 
{\bf b.} Mass of the embedded protostellar object, as function of clump mass. 
The dashed line is the least-squares (bisector) fit to the data, and has a 
slope of 0.30 $\pm$ 0.07.}
\label{masseslumo}
\end{figure}



\end{document}